\documentclass[fleqn,usenatbib]{mnras}

\usepackage{mathptmx}

\usepackage[T1]{fontenc}

\DeclareRobustCommand{\VAN}[3]{#2}
\let\VANthebibliography\thebibliography
\def\thebibliography{\DeclareRobustCommand{\VAN}[3]{##3}\VANthebibliography}

\usepackage{graphicx}	
\usepackage{amsmath}	
\usepackage{amssymb}	

\usepackage{rotating}
\usepackage{longtable}
\usepackage[normalem]{ulem}
\useunder{\uline}{\ul}{}
\usepackage{csquotes}
\usepackage{multirow}
\usepackage{pdflscape}
\usepackage{supertabular}  
\usepackage{color, colortbl}

\definecolor{gray1}{gray}{0.8}
\definecolor{gray2}{gray}{0.8}
\definecolor{gray3}{gray}{0.8}
\cellcolor{gray3}
\cellcolor{white}

\title[]{The morphology of the X-ray afterglows and of the jetted GeV emission in long GRBs}

\author[R. Ruffini et al.]{
R.~Ruffini,$^{1,2,5,7,14}$
R.~Moradi,$^{1,2,15}$\thanks{ruffini@icra.it, rahim.moradi@inaf.it, liang.li@icranet.org, jorge.rueda@icra.it, yu.wang@icranet.org}
J.~A.~Rueda,$^{1,2,4,8,16}$
L.~Li,$^{1,2,15}$
N. Sahakyan,$^{1, 6}$
Y.-C.~Chen,$^{1,2}$
\newauthor
Y.~Wang,$^{1,2,15}$
Y.~Aimuratov,$^{1,2,9}$
L.~Becerra,$^{1,2,17}$
C.~L.~Bianco,$^{1,2,16}$
C.~Cherubini,$^{1,3,11}$
S.~Filippi,$^{1,3,10}$
\newauthor
M.~Karlica,$^{1,2}$
G.~J.~Mathews,$^{1,12}$
M.~Muccino,$^{13}$
G.~B.~Pisani,$^{1,2}$ and
S.~S.~Xue$^{1,2}$
\\
$^{1}$
ICRANet, Piazza della Repubblica 10, I-65122 Pescara, Italy\\
$^{2}$
ICRA, Dipartimento di Fisica, Universit\`a  di Roma ``La Sapienza'', Piazzale Aldo Moro 5, I-00185 Roma, Italy\\
$^{3}$
ICRA, University Campus Bio-Medico of Rome, Via Alvaro del Portillo 21, I-00128 Rome, Italy\\
$^{4}$
ICRANet-Ferrara, Dipartimento di Fisica e Scienze della Terra, Universit\`a degli Studi di Ferrara, Via Saragat 1, I--44122 Ferrara, Italy\\
$^{5}$
ICRANet-Rio, Centro Brasileiro de Pesquisas F\'isicas, Rua Dr. Xavier Sigaud 150, 22290--180 Rio de Janeiro, Brazil\\
$^{6}$
ICRANet-Armenia, Marshall Baghramian Avenue 24a, Yerevan 0019, Republic of Armenia\\
$^{7}$
Universit\'e de Nice Sophia-Antipolis, Grand Ch\^ateau Parc Valrose, Nice, CEDEX 2, France
\\
$^{8}$
Dipartimento di Fisica e Scienze della Terra, Universit\`a degli Studi di Ferrara, Via Saragat 1, I--44122 Ferrara, Italy\\
$^{9}$
Fesenkov Astrophysical Institute, Observatory 23, 050020 Almaty, Kazakhstan
\\
$^{10}$
Department of Engineering, University Campus Bio-Medico of Rome, Nonlinear Physics and Mathematical Modeling Lab, \\Via Alvaro del Portillo 21, 00128 Rome, Italy
\\
$^{11}$
Department of Science and Technology for Humans and the Environment and  Nonlinear Physics and Mathematical Modeling Lab, \\University Campus Bio-Medico of Rome, Via Alvaro del Portillo 21, 00128 Rome, Italy\\
$^{12}$
Center for Astrophysics, Department of Physics, University of Notre Dame, Notre Dame, IN, 46556, USA\\
$^{13}$
Instituto Nazionale di Fisica Nucleare, Laboratori Nazionali di Frascati, I-00044 Frascati, Italy\\
$^{14}$
INAF, Viale del Parco Mellini 84, 00136 Rome, Italy\\
$^{15}$
INAF -- Osservatorio Astronomico d'Abruzzo,Via M. Maggini snc, I-64100, Teramo, Italy\\
$^{16}$
INAF, Istituto di Astrofisica e Planetologia Spaziali, Via Fosso del Cavaliere 100, 00133 Rome, Italy\\
$^{17}$
Instituto de Astrof\'isica, Facultad de F\'isica, Pontificia Universidad Cat\'olica de Chile, Av. Vicu\~na Mackenna 4860, Macul, Santiago, Chile
}

\date{Accepted XXX. Received YYY; in original form ZZZ}

\pubyear{2020}

\begin{document}
\label{firstpage}
\pagerange{\pageref{firstpage}--\pageref{lastpage}}
\maketitle

\begin{abstract}
We recall evidence that long gamma-ray bursts (GRBs) have binary progenitors and give new examples. Binary-driven hypernovae (BdHNe) consist of a carbon-oxygen core (CO$_{\rm core}$) and a neutron star (NS) companion. For binary periods $\sim 5$~min, the CO$_{\rm core}$ collapse originates the subclass BdHN I characterized by: 1) an energetic supernova (the ``\textit{SN-rise}''); 2) a black hole (BH), born from the NS collapse by SN matter accretion, leading to a GeV emission with luminosity $L_{\rm GeV} = A_{\rm GeV}\,t^{-\alpha_{\rm GeV}}$, observed only in some cases; 3) a new NS ($\nu$NS), born from the SN, originating the X-ray afterglow with $L_X = A_{\rm X}\,t^{-\alpha_{\rm X}}$, observed in all BdHN I. We record $378$ sources and present for four prototypes GRBs 130427A, 160509A, 180720B and 190114C: 1) spectra, luminosities, SN-rise duration; 2) $A_X$, $\alpha_X=1.48\pm 0.32$, and 3) the $\nu$NS spin time-evolution. We infer a) $A_{\rm GeV}$, $\alpha_{\rm GeV}=1.19 \pm 0.04$; b) the BdHN I morphology from time-resolved spectral analysis, three-dimensional simulations, and the GeV emission presence/absence in $54$ sources within the Fermi-LAT boresight angle. For $25$ sources, we give the integrated and time-varying GeV emission, $29$ sources have no GeV emission detected and show X/gamma-ray flares previously inferred as observed along the binary plane. The $25/54$ ratio implies the GeV radiation is emitted within a cone of half-opening angle $\approx 60^{\circ}$ from the normal to the orbital plane. We deduce BH masses $2.3$--$8.9~M_\odot$ and spin $0.27$--$0.87$ by explaining the GeV emission from the BH energy extraction, while their time evolution validates the BH mass-energy formula.
\end{abstract}




\section{Introduction}\label{sec:1} 

The year 2021 marks the $50^{th}$ anniversary of the paper ``Introducing the black hole'' \citep{1971PhT....24a..30R} and of the black hole (BH) mass-energy formula \citep{1970PhRvL..25.1596C,1971PhRvD...4.3552C,1971PhRvL..26.1344H,Hawking:1971vc}. Since those days,  interest in BHs has spread worldwide and their study represents one of the most innovative fields of fundamental physics and astrophysics. There has also been an exponential growth of observational and theoretical developments which are finally reaching the {momentous} result of unveiling the process of rotational energy extraction from a rotating Kerr BH.  We indicate the path of this discovery in the present paper. This realization has allowed for the identification of the code of GRBs: one of the most complex sequences of a very large number of non-repetitive classical and quantum events, each of which are characterized by specific spectral and temporal properties. In parallel, a new arena for fundamental physics has been revealed by the dubbed ``\emph{blackholic quantum}'' \citep{2020EPJC...80..300R}. This enormous conceptual progress has not been reached straightforwardly: it has come from an intense dedicated process with continuous feedback between theoretical understanding, unprecedented panchromatic observational progress, and modification of basic interpretation paradigms: they have all been truly essential. We first summarize in this introduction some of the contributions which have led to the start of this most complex inquiry into the  the most powerful energy source in the Universe.

\subsection{The initial ``golden age'' of Relativistic Astrophysics}

\textit{The first breakthrough} in relativistic astrophysics was the discovery of pulsars in $1967$ \citep{1968Natur.217..709H}, and the discovery of a pulsar in the core of the Crab Nebula \citep{1968Sci...162.1481S,1969PhRvL..22..311R}. The identification of the energy source of the pulsar with a fast rotating newly born neutron star (NS); the new NS ($\nu$NS), coincident with the supernova (SN) explosion led to a new paradigm in SN understanding \citep{1969supe.book.....S}. As we show in this paper, we are gaining a deeper understanding of both of SNe and of the role of the $\nu$NS in the binary driven hypernova (BdHN) systems.

\textit{The second breakthrough} came from the launch in $1970$ of the first X-Ray telescope, observing in the $2$--$20$~keV energy band: the Uhuru satellite \citep[see e.g.][]{GiacconiRuffini1978,2003RvMP...75..995G}. Uhuru paved the way for a crucial working method in developing a multiwavelength collaboration with optical and radio observatories. Thanks to the theoretical understanding, this led to the discovery, inside our own galaxy, of a large number of binary X-ray sources composed of a main sequence star and a companion NS (like Hercules X-1 and Centaurus X-3) and a binary system composed of a main sequence star and a BH, which gave the first evidence for the discovery of a BH in our Galaxy \citep[see][for details]{1974asgr.proc..349R,GiacconiRuffini1978}. It was soon realized that these binary X-ray sources would themselves further evolve as the companion main sequence star would undergo a SN explosion on timescales of $10^8$~y \citep{1974asgr.proc..349R}. In view of the limited number of such binary X-ray sources in our Galaxy, the expected observational rate of the final evolution of such binary systems would be on the order of $10^{-8}$ events/y in our Galaxy. The point that was missed at the time was the existence of the process of ``\emph{induced gravitational collapse}'', which was identified years later \citep{2001ApJ...555L.117R, 2012ApJ...758L...7R}. This implies an unprecedented energy emission of $\sim 10^{54}$~erg, making them observable from all galaxies in the entire Universe: if the number of galaxies in our past light-cone is taken into account, the expected observational rate of the final evolution of such binary X-ray sources in the entire Universe is on the order of $10$--$100$~events/y. 

\textit{The third breakthrough} was the introduction in $1971$ of the BH mass-energy formula by Christodoulou, Hawking, Ruffini \citep{1970PhRvL..25.1596C,1971PhRvD...4.3552C,1971PhRvL..26.1344H,Hawking:1971vc}, and the BH extractable energy by reversible and irreversible transformation (in geometric $c=G=1$ units): 
\begin{subequations}
\begin{align}
\label{aone}
M^2 &= \frac{J^2}{4 M^2_{\rm irr}}+M_{\rm irr}^2,\\
S &= 16 \pi M^2_{\rm irr}    \\ 
\delta S &= 32 \pi  M_{\rm irr} \delta M_{\rm irr}  \geq 0 ,
\end{align}
\end{subequations}
where  $J$, $M$, $M_{\rm irr}$ and $S$ are the  angular momentum, mass, irreducible mass and horizon surface area of the BH. 

Again in this article, we indicate the path to observe for the first time the BH extractable energy process, which can be as high as $29\%$ of the BH mass for an extreme Kerr BH. We measure as well the BH mass and spin in selected BdHN.

Just at the end of this ``initial golden age of relativistic astrophysics'', the discovery of GRBs was publicly announced in February $1974$ at the annual meeting of the American Association for the Advancement of Science, in San Francisco \citep[see details in][]{GurskyRuffini1975}. In that meeting, observations by the Vela $5$ and Vela $6$ satellites were presented. These satellites operated in the $3$--$12$~keV X-ray energy band and, for the first time, in the $150$--$750$~keV (Vela $5$) and $300$--$1500$~keV (Vela $6$) gamma-ray energy bands. Tens of gamma-ray events per year of unknown origin, lasting for a few seconds, and originating outside the Solar System,  were named ``Gamma-Ray Bursts'' \citep[details in][]{1973ApJ...182L..85K,1975ASSL...48...47S}.

What has became clear only recently, and further clarified in this article, is that precisely the late catastrophic evolution of the binary X-ray sources leads to the BdHNe: the progenitors of a class of long gamma-ray bursts (GRBs). Indeed, these highest luminosity energy sources in the Universe are observed to occur at a rate of $10$--$100$ events/y, consistent with the order of magnitude estimate given above.

We proceed to focus on the most recent developments, selecting crucial observational milestones, theoretical developments, and define the interpretation paradigms which have recently led to a unified understanding of the GRBs.  

\subsection{The largest ever multi-wavelength observational efforts}

The earliest evidence for high-energy radiation above $100$~MeV from GRBs were the observations by the Energetic Gamma-Ray Experiment Telescope (\textit{EGRET)}, operating in the energy range $\sim$ $20$~MeV--$30$~GeV, onboard of the Compton Gamma-Ray Observatory (\textit{CGRO}, $1991$--$2000$). The detection was triggered by the Burst And Transient Source Experiment (\textit{BATSE}), operating in energy range of $\sim 20$--$2000$~keV. \textit{EGRET} has detected five GRBs which, from our understanding today, were long duration bursts: GRB~910503, GRB~910601, GRB~930131, GRB~940217, and GRB~940301 \citep[see e.g.][and references therein]{1996MmSAI..67..161K}. Unfortunately, no redshift was known at the time.

A new epoch started with the launch of the Beppo-Sax satellite in $1996$, joining the expertise of the X-ray and gamma-ray communities. Its gamma-ray burst monitor (GRBM) operating in the $40$–-$700$~keV energy band determined the trigger of the GRB, and two wide-field cameras operating in the $2$–-$30$~keV X-ray energy band allowed the localization of the source to within arc-minutes resolution. This enabled a follow-up with the narrow-field instruments (NFI) in the $2$--$10$~keV energy band. 

Beppo-SAX achieved three major results:
\begin{enumerate}
    \item  
    The discovery of the X-ray afterglow (GRB~970228, \citet{Costa1997}), characterized by an X-ray luminosity decreasing with a  power-law with index of $\alpha_{X}=-1.48\pm 0.32$ (see \citealp{DePasquale2006}, as well as \citealp{Li2015a,2016ApJ...833..159P,Li2018a}). In this article, we specifically address the astrophysical origin of the afterglow.
    \item 
    The determination of the accurate positions by the NFI, transmitted to the optical \citep{vanParadjis1997} and radio telescopes \citep{1997Natur.389..261F}, allowed the determination of the GRB cosmological redshifts. The first redshift was measured for GRB 970508 \citep{1997Natur.387..878M}, using the LRIS instrument of the Keck II telescope \citep{1995PASP..107..375O}. The derived distances of $\approx5$--$10$~Gpc confirmed their cosmological origin and their unprecedented energetics, $\approx 10^{50}$--$10^{54}$~erg, thus validating our hypothesis derived from first principles \citep{1975PhRvL..35..463D,1998bhhe.conf..167R}.
    \item  
    The discovery of the temporal and spatial coincidence of GRB 980425 with SN 1998bw \citep{1998Natur.395..670G}, which suggested the connection between GRBs and SNe, was soon supported by many additional events \citep[see e.g.][]{2006ARA&A..44..507W,2011IJMPD..20.1745D,2012grb..book..169H, Li2012, Li2018}. The astrophysical origin of this coincidence is addressed in this article within the BdHN approach.
\end{enumerate}

The Neil Gehrels \emph{Swift} Observatory (hereafter indicated as \textit{Swift}) followed in 2004. It was conceived as a panchromatic space observatory dedicated to the observations of GRBs. The GRB trigger is detected by the large field of view of its Burst Alert Telescope (BAT) \citep{2005SSRv..120..143B}, operating in the hard X-ray band. This is followed up by the fast and automatic observations of the onboard narrow fields instruments XRT \citep{2005SSRv..120..165B} and UVOT \citep{2005SSRv..120...95R} operating in the soft/medium X-ray and in the optical/UV bands, respectively. The BAT telescope operates in the  $15$--$150$~keV energy band and can detect the GRB prompt emission while accurately determining its position in the sky within $3$~arcmin. Within $90$~s, Swift can repoint the XRT telescope, operating in the $0.3$--$10$~keV energy range, and relay promptly the burst position to the ground. Unfortunately, this does not allow the establishment of the initial \textit{Swift}-XRT detection prior to the \textit{Swift}-BAT trigger, as later explained in this article.

Thanks to the Swift satellite, the number of detected GRBs increased rapidly to $1300$ sources with known redshifts \citep[see, e.g.,][]{2020arXiv200305153G}. By analyzing the light-curve of some long GRBs, \citet{Nousek2006} and \citet{Zhang2006} discovered three power-law segments in the XRT flux light curves prior to the afterglow emission\citep[see, also,][]{Li2015a,Li2018}. We refer in this article to these segments as the ``Nousek-Zhang power-laws''. All the X-ray afterglow observations considered in this article refer to \textit{Swift}-XRT observation.

The high-energy astrophysics era of GRB observations started with the launch of \textit{AGILE} in $2007$ \citep{2009A&A...502..995T} with the onboard Gamma-Ray Imaging Detector (\textit{GRID}) operating in the $30$~MeV--$50$~GeV energy range. \textit{AGILE} was soon followed by the launch in June~$2008$ of the \textit{Fermi} satellite, having onboard the Gamma-ray Burst Monitor (GBM) operating in the $8$~keV--$40$~MeV energy range \citep{2009ApJ...702..791M} and the Large Area Telescope (LAT) operating in the $20$~MeV--$300$~GeV energy range \citep{2009ApJ...697.1071A}.

\textit{AGILE-GRID} detected the first long GRB with emission above $100$~MeV and with a photometric redshift of $z=1.8$, GRB~080514B \citep{2008A&A...491L..25G}. It was followed four months later by the detection of GRB~080916C \citep{2009A&A...498...89G} by \textit{Fermi} with one of the largest isotropic energies ever detected, $E_{\rm iso}= (4.07 \pm 0.86) \times 10^{54}$~erg, and a photometric redshift of $z= 4.35$. These were followed by a large number of long GRBs observed by LAT with both GeV emission and with a well-defined $z$. All the high-energy long GRBs
considered in this article are based on the first and second Fermi-LAT GRB catalogs \citep{2013ApJS..209...11A, 2019ApJ...878...52A}.

\begin{figure*}
\centering
\includegraphics[width=1.0\hsize,clip]{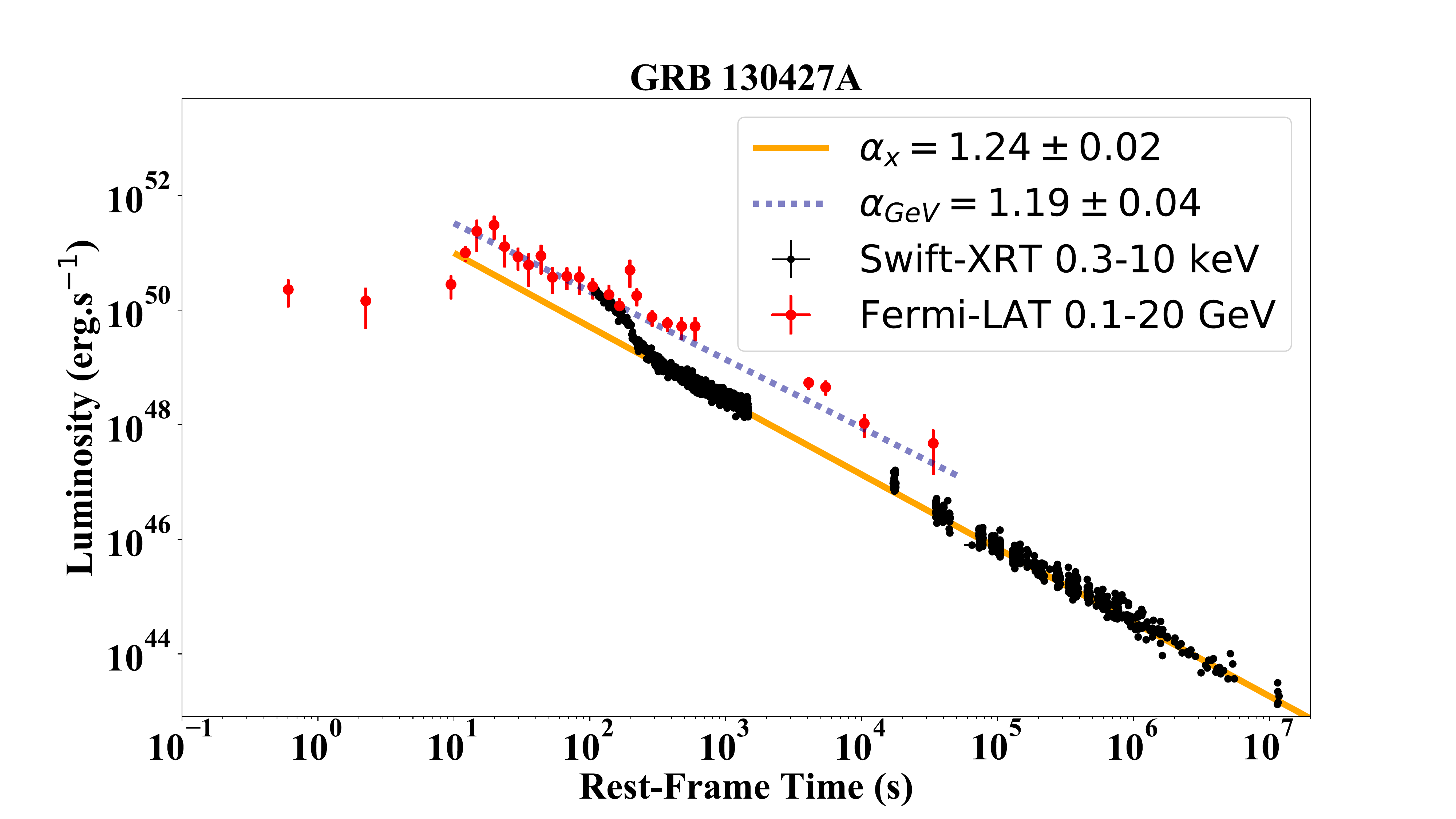} 
\caption{Luminosity of BdHN I 130427A: the black data points represent the rest-frame $0.3$--$10$~keV luminosity obtained from \textit{Swift}-XRT. It follows a decaying power-law with amplitude $(3.65\pm 0.63)\times 10^{52}$~erg~s$^{-1}$ and index $\alpha_X=1.24\pm 0.02$. The red data points show the rest-frame in $0.1$--$20$~GeV luminosity observed by \textit{Fermi}-LAT. It follows a decaying power-law with an amplitude of $(5.1\pm 0.2)\times 10^{52}$~erg~s$^{-1}$ and index $\alpha_{\rm GeV}=1.19\pm 0.04$. Details are given in Sections~\ref{sec:xrayafterglow}, \ref{sec:4} and \ref{sec:8}.}\label{fig:130427AF} 
\end{figure*}

\begin{figure*}
\centering
\includegraphics[width=1.0\hsize,clip]{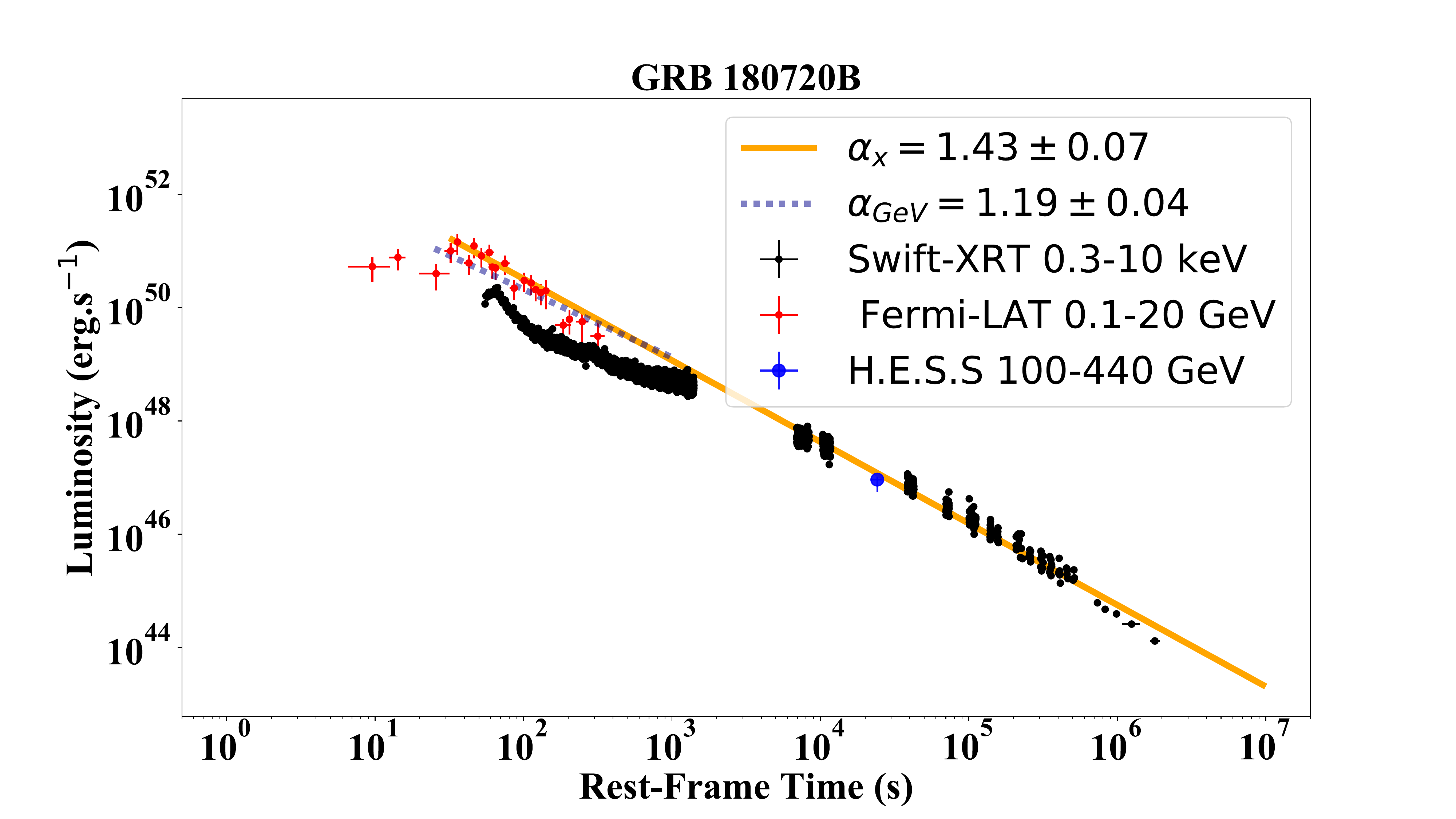} 
\caption{Luminosity of BdHN I 180720B: the black data points represent the rest-frame $0.3$--$10$~keV luminosity obtained from \textit{Swift}-XRT. It follows a decaying power-law with index $\alpha_X=1.43\pm 0.07$. The blue data point shows the rest-frame $100$--$440$~GeV luminosity observed by H.E.S.S. The red data points show the rest-frame $0.1$--$20$~GeV luminosity observed by \textit{Fermi}-LAT. It follows a decaying power-law with amplitude $(5.4\pm 0.6)\times 10^{52}$~erg~s$^{-1}$ and index $\alpha_{\rm GeV}=1.19\pm 0.04$. Details are given in Sections~\ref{sec:xrayafterglow}, \ref{sec:4} and  \ref{sec:8}.}\label{fig:hess} 
\end{figure*} 

\begin{figure*}
\centering
\includegraphics[width=1.0\hsize,clip]{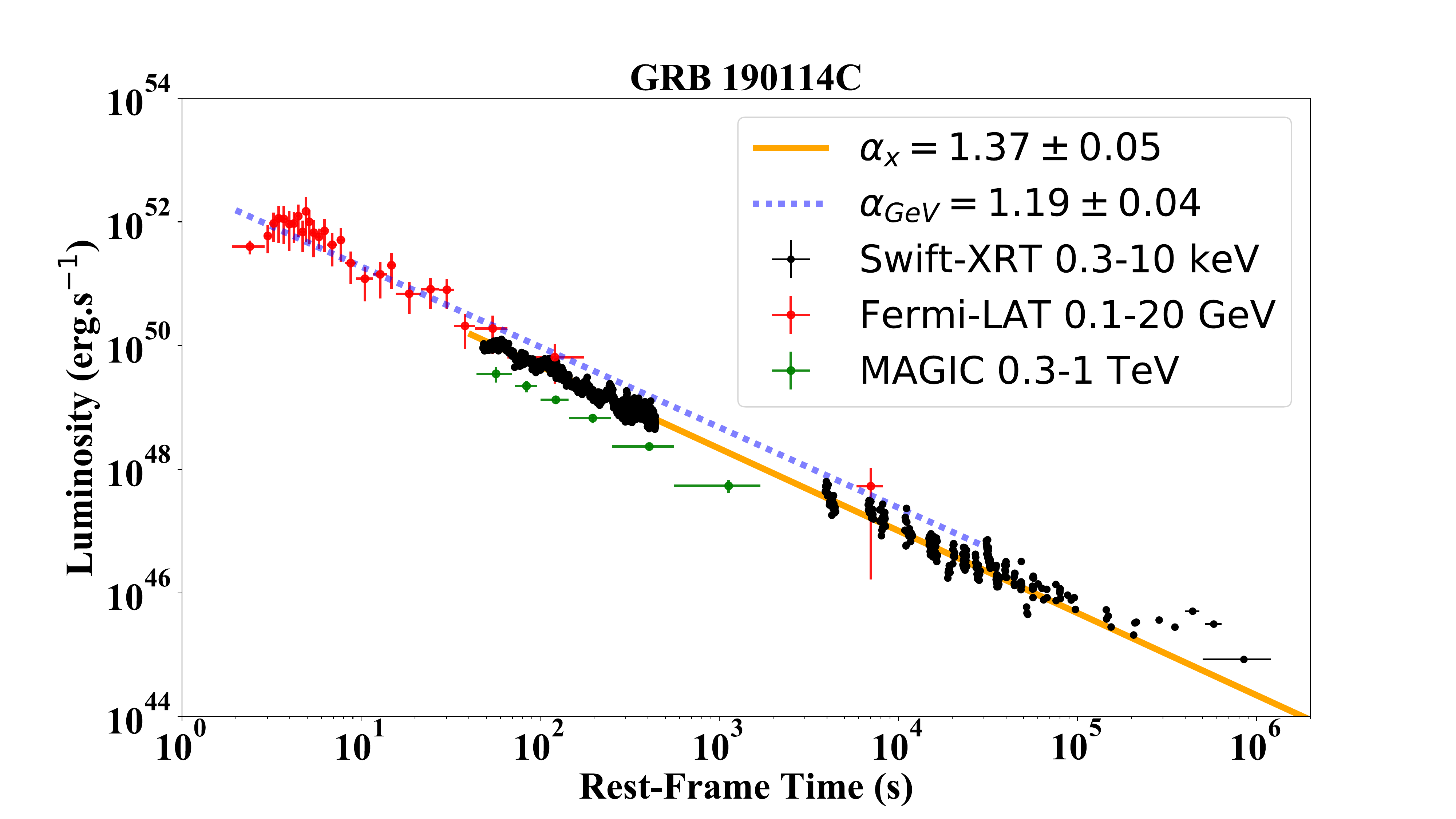} 
\caption{Luminosity of BdHN I 190114C: the black data points represent the rest-frame $0.3$--$10$~keV luminosity obtained from \textit{Swift}-XRT. It follows a decaying power-law with index $\alpha_X=1.37\pm 0.05$. The red data points show the rest-frame $0.1$--$20$~GeV luminosity observed by \textit{Fermi}-LAT. It follows a decaying power-law with amplitude $(4.6\pm 0.6)\times 10^{52}$~erg~s$^{-1}$ and index $\alpha_{\rm GeV}=1.19\pm 0.04$. The green data points show the rest-frame $0.3$--$1$~TeV luminosity obtained from MAGIC. Details are given in Sections~\ref{sec:xrayafterglow}, \ref{sec:4} and \ref{sec:8}. }\label{fig:tev190114C} 
\end{figure*} 

The leading observations from space observatories were followed by a multitude of equally essential observations from ground-based observatories spanning the globe. The leading role was taken by the largest optical telescopes, e.g. the VLT from ESO with its X-shooter instrument \citep{2011A&A...536A.105V}, and radio telescopes. This vastest ever multiwavelength observational campaign has been recently further extended to the very-high-energy (VHE) domain with the GRB detection by observatories on the ground. This is the case of the observations of GRB 190114C by the Imaging Atmospheric Cherenkov Telescopes MAGIC (see Fig.~\ref{fig:tev190114C} and \citealp{2019Natur.575..455M}), designed to detect VHE gamma-rays from $30$~GeV to more than $50$~TeV \citep[see e.g.][]{2016APh....72...61A,2016APh....72...76A}, the observations of GRB 180720B by H.E.S.S (see Fig.~\ref{fig:hess} and  \citealp{2019Natur.575..464A}), operating in the energy range from tens of GeV to tens of TeV \citep[see e.g.][]{2006A&A...457..899A}, as well as GRB 190829A \citep{2020arXiv200100648C}, which we also address in this article.

\subsection{The short GRBs with binary NS progenitors}\label{sec:1.3}

One of the main results of the observations of the CGRO satellite \citep{2000eaa..bookE4537.} was the isotropic distribution of the GRBs when expressed in galactic coordinates \citep{1992Natur.355..143M}. This result gave the first preliminary indication of the cosmological nature of GRBs. This was later confirmed by irrefutable evidence from the observations of Beppo-Sax, as mentioned above. An additional result was the clear indication of the existence of two different classes of GRBs: the short and the long GRBs \citep{1993ApJ...413L.101K}. This classification has been confirmed and further extended as we recall in Section~\ref{sec:2}, now duly expressing all quantities, after Beppo-Sax, in the rest frame of the source.

The first proposal of successfully relating a GRB to an astrophysical cosmological source came from the vision of {Bohdan} Paczynski and collaborators, who identified the progenitors of short GRBs with merging NS binaries  \citep[see, e.g.,][]{1986ApJ...308L..43P,Eichler:1989jb,1991ApJ...379L..17N,1992ApJ...395L..83N,1992ApJ...388L..45M,1992ApJ...395L..83N}. This result was later confirmed by Beppo-Sax \citep{1998ApJ...507L..59L,2000ApJ...534L.197L,2006MNRAS.366..219L,2014ARA&A..52...43B}. Complementary information came from the localization of short GRBs at large off-sets from their host galaxies and with no star formation evidence \citep[see, e.g.,][]{Fox2005,Gehrels2005,2014ARA&A..52...43B}. The following fundamental discovery came from the identification of the first short GRB in the GeV band by \textit{AGILE}. The first observation of a short GRB  was done by \textit{AGILE} who detected GRB~090510A at a  spectroscopic redshift of $z=0.903$, with $E_{\rm iso}= (3.95 \pm 0.21) \times 10^{52}$~erg, and a significant GeV emission $E_{\rm LAT}= (5.78 \pm 0.60) \times 10^{52}$~erg. On the basis of the observed energetics of this source, and its spectral properties, we proposed that in this short GRB we witness the birth of a BH, which we associate with the onset of the GeV emission: the signature of this event \citep{2016ApJ...831..178R}.

This identification further evolved with the introduction of the two sub-classes of short bursts \citep{2015ApJ...798...10R,2016ApJ...831..178R,2016ApJ...832..136R,2017ApJ...844...83A}. The first sub-class corresponds to short bursts with isotropic energies $E_{\rm iso}<10^{52}$~erg (in the rest-frame $1$--$10^{4}$~keV energy band) and rest-frame spectral peak energies $E_{\rm p,i}<2$~MeV.  These are expected to originate when the NS-NS merger leads to a single massive NS (M-NS) with a mass below the NS critical mass. We have called these sources short gamma-ray flashes (S-GRFs).

The second sub-class corresponds to short bursts with $E_{\rm iso}\gtrsim10^{52}$~erg and $E_{\rm p,i}\gtrsim2$~MeV. It was assumed that these sources, in analogy with the prototype GRB~090510, originate from a NS-NS merger in which the merged core overcomes the NS critical mass and gravitationally collapses to form a BH. We have called these sources genuine short GRBs (S-GRBs); see  \citet{2016ApJ...831..178R,2019ApJ...886...82R}; six of such short GRBs have been identified, all emitting GeV emission with a decaying luminosity of index $\alpha_{\rm GeV,short}=-1.29\pm 0.06$ \citep{2019ApJ...886...82R}; see Fig.~\ref{fig:02} in Section~\ref{sec:8}.

We show how, by following these pathfinding works on short GRBs, we have progressed in formulating the theory of the BdHNe: the theory of long GRBS based on binary progenitors. Before this, however, we summarize the traditional long GRB models based upon a single progenitor. 

\subsection{Long GRBs in the traditional model}\label{sec:1.4}

A review of  the traditional long GRB model is facilitated by the extensive book by Bing Zhang and many references therein  \citep{2018pgrb.book.....Z}. As recounted there, the papers by \citet{1992MNRAS.258P..41R,1538-4357-482-1-L29}, and \citet{1993ApJ...405..273W} have characterized this traditional model. \citet{1992MNRAS.258P..41R}  proposed a single BH
as the origin of GRBs emitting an ultrarelativistic blast wave, whose expansion follows the Blandford-McKee self-similar solution \citep{1976PhFl...19.1130B}.
\citet{1993ApJ...405..273W} linked the GRB origin to a Kerr BH emitting an ultrarelativistic jet originating from the accretion of toroidal material onto the BH. The BH was assumed to be  produced from the direct collapse of a massive star, a ``failed'' SN leading to a large BH of approximately $5 M_{\odot}$, possibly as high as $10 M_{\odot}$, a ``\emph{collapsar}''. We will address this interesting idea within our BdHN model in section~\ref{sec:5}. 

In these ultrarelativistic blast wave models, the afterglow is
explained by the synchrotron/synchrotron self-Compton (SSC)
emission from accelerated electrons when the
blast wave of $\Gamma \sim 1000$ is slowed down by the circumburst medium \citep{1994ApJ...433L..85W,1997MNRAS.288L..51W,1995ApJ...455L.143S,1997ApJ...489L..37S,1998ApJ...497L..17S}.

As pointed out by \citet{2018pgrb.book.....Z}, these ultrarelativistic blast wave models have been applied to explain a vast number of observations:
\begin{enumerate}
    \item 
    The X-ray afterglow as well as the steep and shallow decay in the ``Nousek-Zhang'' phase, the X-ray and the gamma-ray flares.
    \item 
    The optical and the radio emissions.
    \item 
    The high-energy emission in the GeV band observed in some long GRBs by Fermi-LAT.
\end{enumerate}

An example of this method is the recent case of GRB 190114C, in which the traditional approach has been applied: 
\begin{enumerate}
\item 
to jointly explain the emissions in the TeV observed recently by MAGIC \citep{GCN23701, 2019Natur.575..455M, 2019Natur.575..459M}; see Fig. \ref{fig:tev190114C};
\item 
to explain the emission in the MeV and GeV bands observed by the Fermi GBM and LAT satellites in the jetted emission;
\item  
To explain the emission in the MeV and keV bands observed by \textit{Swift} including the emission in the optical and radio.
\end{enumerate}

In the traditional model, all of these emissions occur jointly using the kinetic energy of an ultrarelativistic blast wave with Lorentz factor  Gamma $\sim 10^3$, emitting at distances of $\sim 10^{16}$~cm-$10^{18}$~cm, implying  total energies reaching $10^{55}$~erg. 

This approach, however,  encounters some contradictions with model-independent constraints. Moreover, there is no requirement that these different emission processes be explained by a single origin, i.e. the kinetic energy of a blast wave. As we are going to show in this article, each one of the above mentioned emissions finds its reason for existence in different specific processes originating in different specific episodes during the BdHN evolution. Each episode implies a different process and less demanding  energy requirements.

\begin{figure*}
\centering
\includegraphics[width=0.49\hsize,clip]{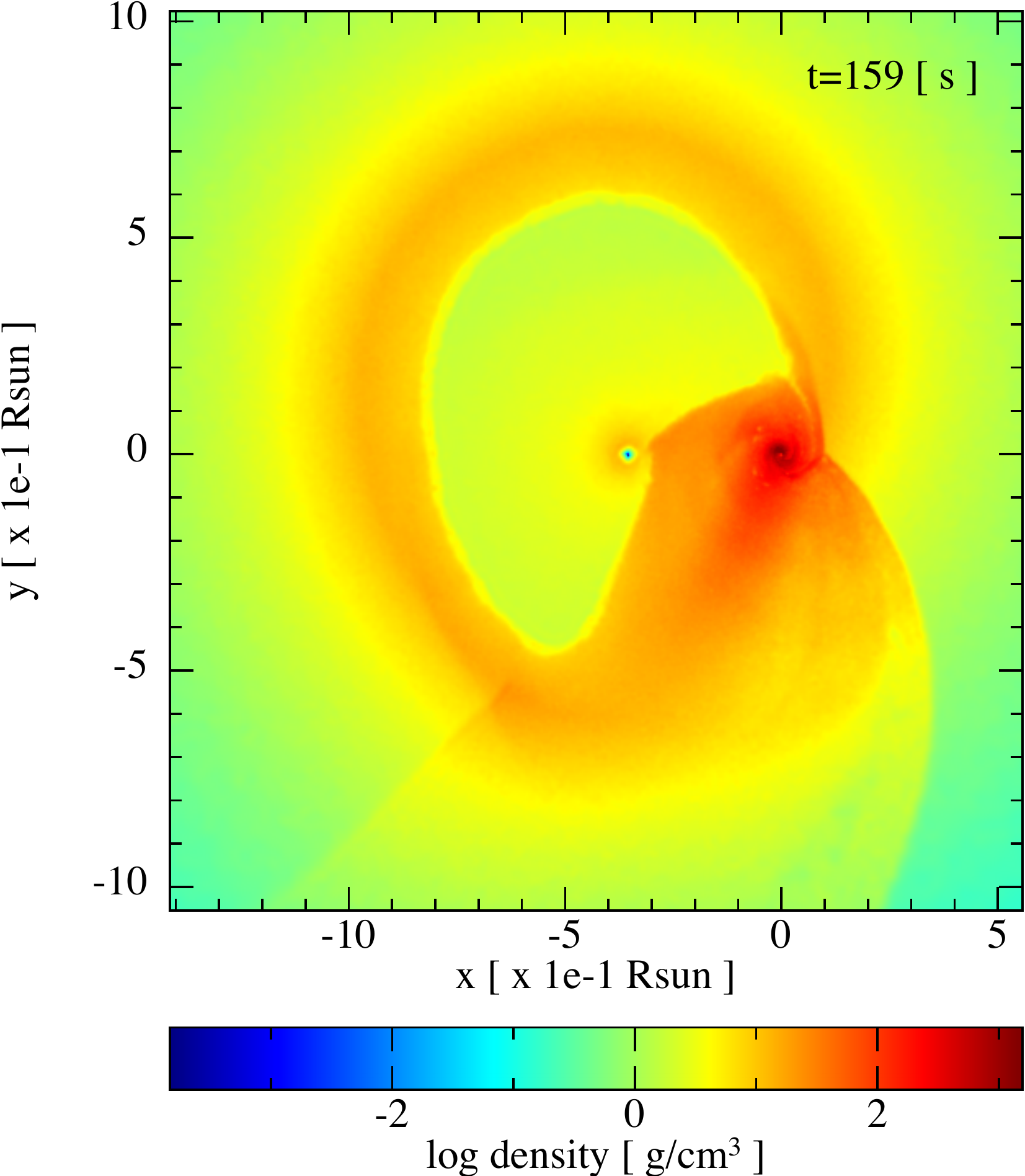}
\includegraphics[width=0.49\hsize,clip]{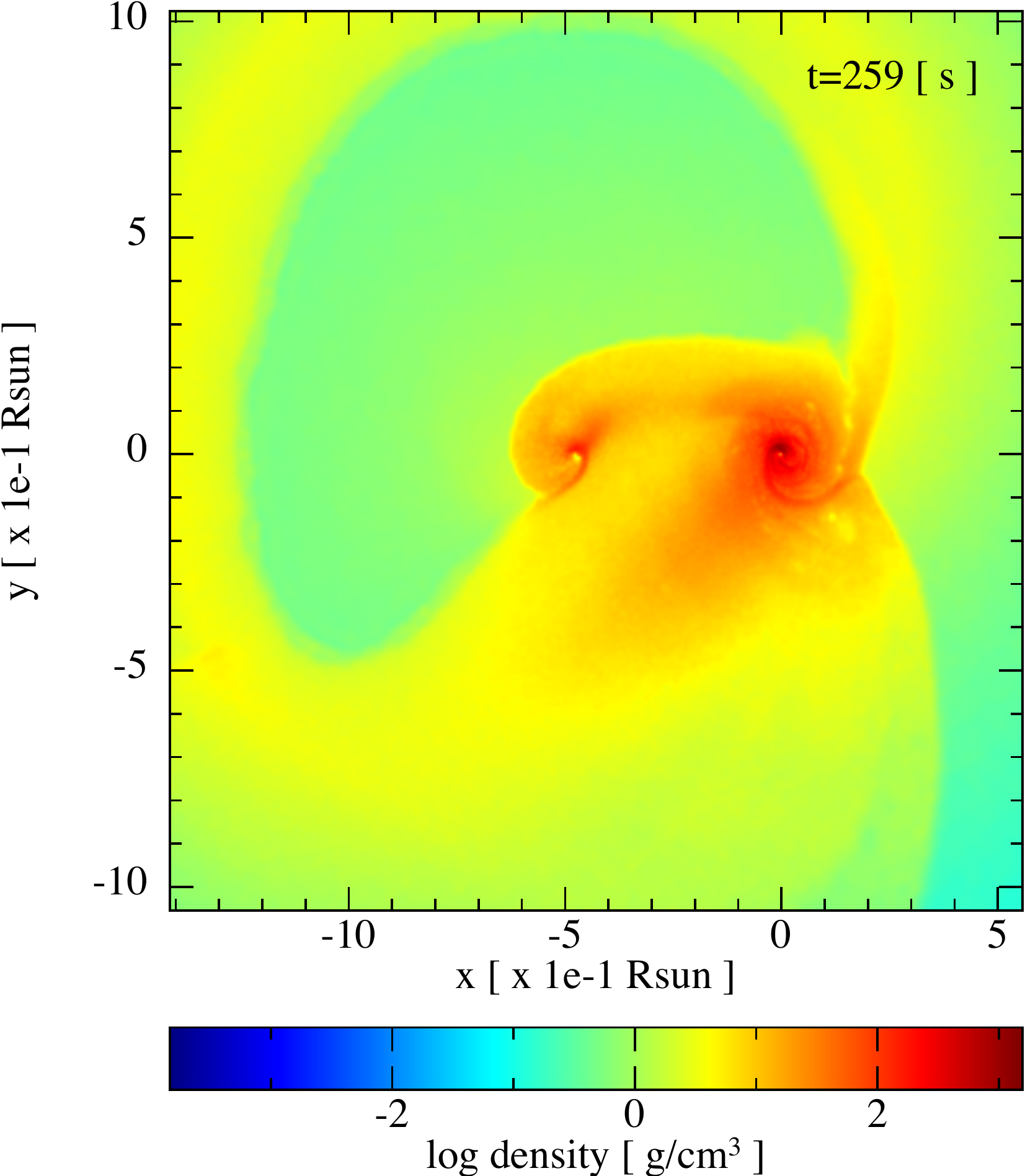}
\includegraphics[width=0.49\hsize,clip]{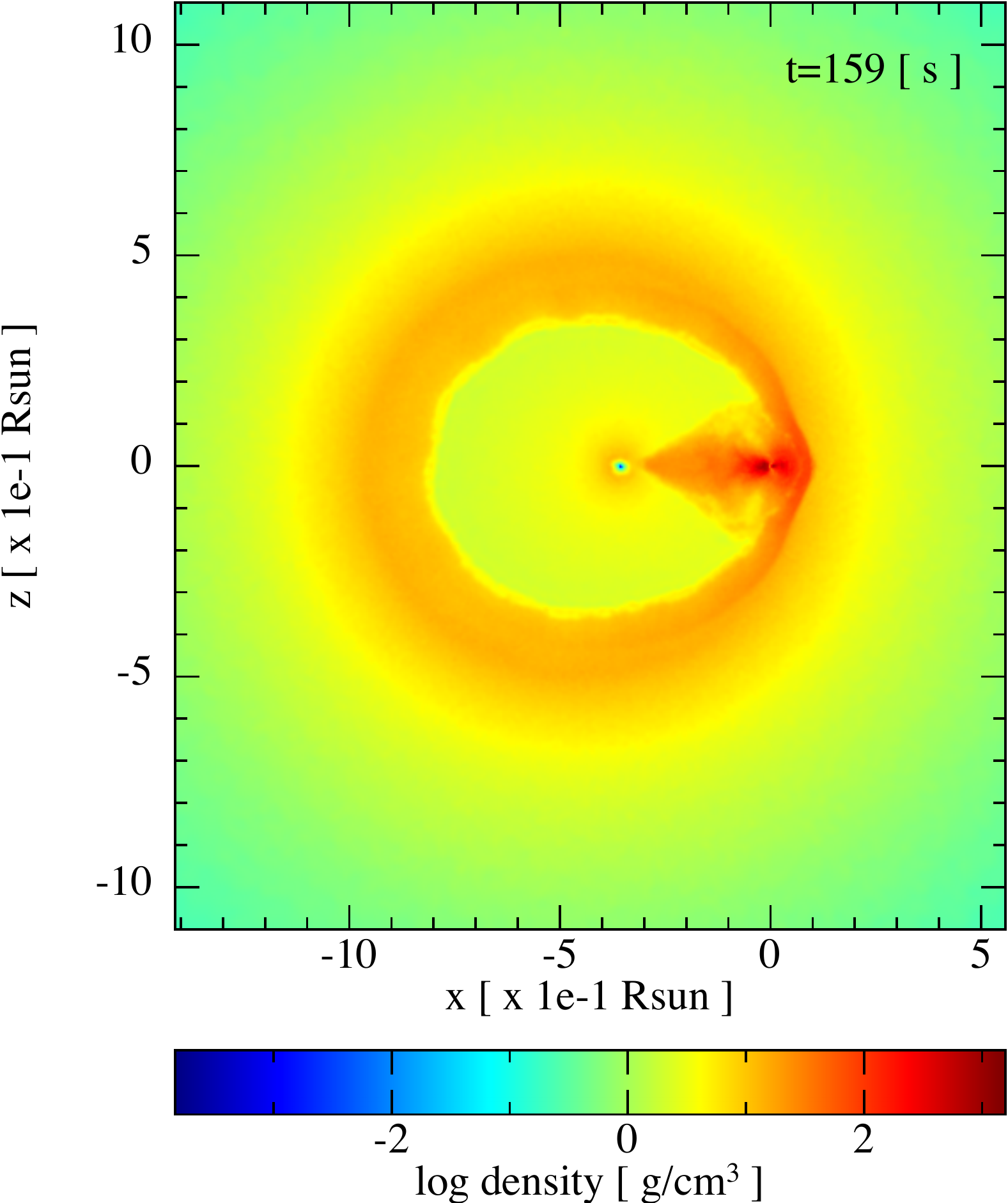}
\includegraphics[width=0.49\hsize,clip]{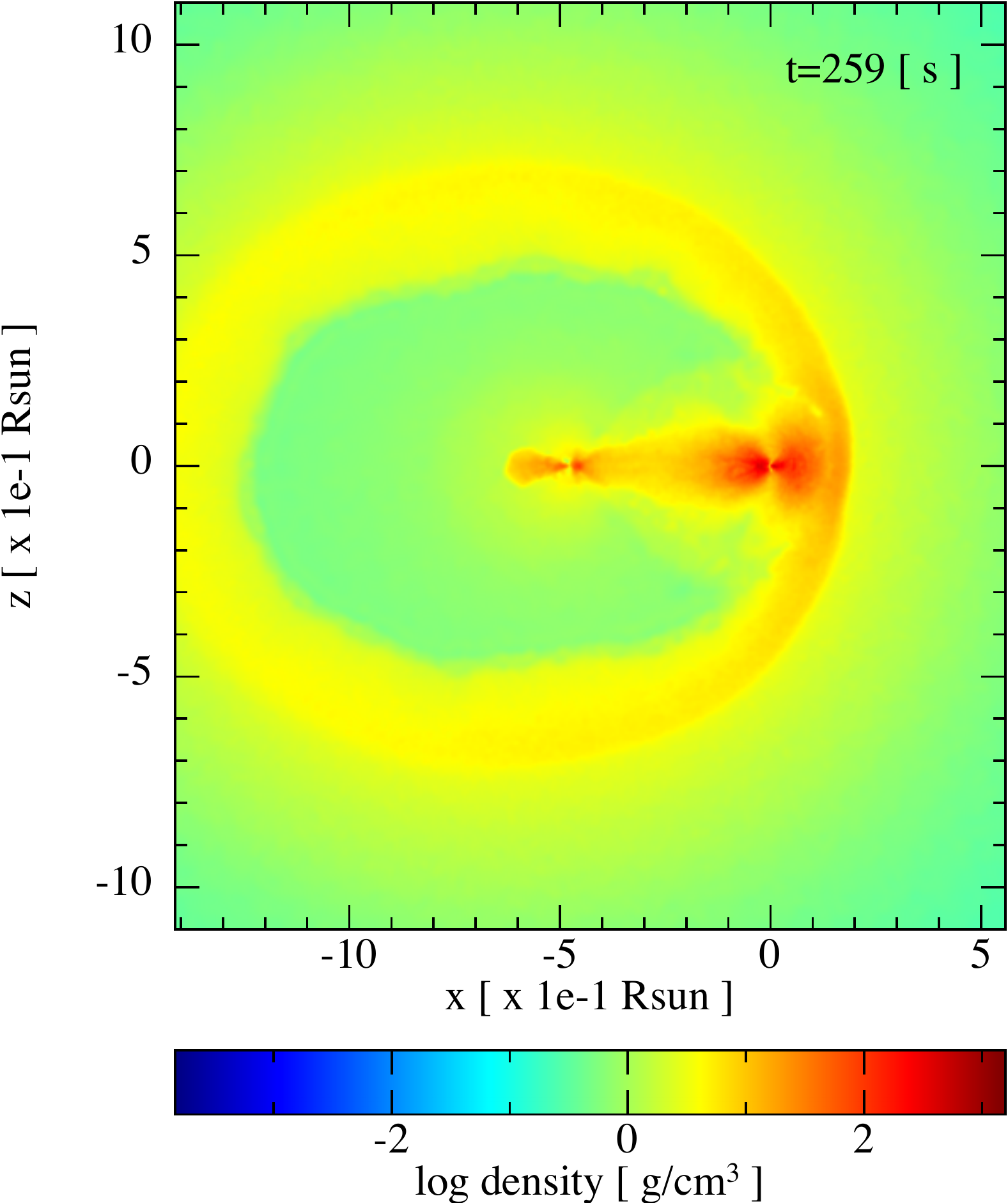}
\caption{A SPH simulation from \citet{2019ApJ...871...14B} of the exploding CO-star as the SN in the presence of a companion NS: Model ``25m1p08E'' (see table~2 therein). The CO-star is obtained from the evolution of a $25~M_\odot$ zero-age main-sequence (ZAMS) progenitor which leads to a pre-SN CO-star mass $M_{\rm CO}=6.85~M_\odot$. The initial mass of the $\nu$NS (formed at the center of the SN) is $1.85 M_\odot$ and the one of the NS companion is $M_{\rm NS}=2~M_\odot$. The initial orbital period is $4.8$~min. The upper panels show the mass density on the binary equatorial plane and the lower ones correspond to the plane orthogonal to it, at two selected times from the SN explosion ($t=0$ of the simulation), $159$~s and $259$~s. The reference system is rotated and translated so that the x-axis is along the line that joins the $\nu$NS and the NS, and the axis origin $(0,0)$ is located at the NS position. For this simulation, the NS collapses reaching the secular axisymmetric instability point with a mass $2.26 M_\odot$ and angular momentum $1.24 G M^2_\odot/c$, while the $\nu$NS is stable with mass and angular momentum, respectively, $2.04 M_\odot$ and $1.24 G M^2_\odot/c$. Up to the final simulation time, the binary system kept bound although the binary orbit widens, reaching an orbital period of $16.5$~min and an eccentricity of $\epsilon = 0.6$. The collapse of the NS to the newly-formed BH, characteristic of a BdHN I, occurs at $t=21.6$~min. 
}
\label{fig:SPHsimulation}
\end{figure*}

\subsection{The role of binary systems as progenitors of long GRBs}\label{sec:1.5}

The role of binary systems as progenitors of long GRBs in our approach involves three assumptions: 

A) That all long GRBs, not only the short GRBs, originate from binary systems. These binaries are composed of different combinations of CO$_{\rm core}$, NS, white dwarfs (WDs), BH and $\nu$NS; see Table~\ref{acronyms}.  We classify all GRBs in nine different subclasses on the basis of their energetics, their spectra and their duration expressed in the rest-frame of the source. Only in \textit{some} of these subclasses the presence of a BH occurs \citep[see e.g.][]{2016ApJ...832..136R,2018ApJ...869..151R,2019ApJ...874...39W}; detail in section~\ref{sec:2}.  

B) We focus on BdHNe with a binary progenitor composed of a CO-star and a companion binary NS. As the CO$_{\rm core}$ gravitationally collapses, it gives origin to a SN and its iron core collapses to form a {newborn NS ($\nu$NS)}. The hypercritical accretion of the SN ejecta on the companion NS leads, for binary periods $\lesssim 5$~min, to the formation of a BH. This happens when the NS critical mass is reached and overcome \citep{2016ApJ...833..107B}. We denote these systems  as BdHNe I in which a BH is formed. The BdHNe I are characterised by an isotropic energy, estimated by the Fermi-GBM, in the range $10^{52}<E_{\rm iso}<10^{54}$~erg. In the opposite case, i.e. for longer binary periods, a more massive NS (MNS) originates from the SN hypercritical accretion process \citep{2019ApJ...874...39W}. These BdHNe II are characterised by $10^{50}<E_{\rm iso}<10^{52}$~erg \citep{2016ApJ...832..136R}. The BdHNe III are characterized by binaries with even longer periods, so with more widely separated components, leading to an even weaker energy emission with $10^{48}<E_{\rm iso}<10^{50}$~erg.

C) We make use of recent theoretical results in the study of the hypercritical accretion of the SN ejecta both on the companion NS and the $\nu$NS \citep[see e.g.][]{2016ApJ...832..136R, 2016ApJ...833..107B, 2018ApJ...852...53R, 2019ApJ...871...14B, 2020ApJ...893..148R}. We rely on the three-dimensional (3D) simulations performed with a new
Smoothed-Particle-Hydrodynamics (SPH) code developed in collaboration with Los Alamos National laboratory \citep[see e.g.][and reference therein]{2019ApJ...871...14B}. We here give special attention to this procedure in order to reconstruct the morphology of the BdHNe, which has a strong dependence on the viewing angle as a result of the binary nature of the progenitor. We use the observations of the GeV emission observed by Fermi-LAT present only in \textit{some} BdHN to infer their morphology and visualize its nature by SPH simulations; see section~\ref{sec:morphology} and section~\ref{sec:11} and Fig.~\ref{fig:SPHsimulation}. 

\subsection{{The role of the binary progenitor in the SN associated with long GRBs}}

{
Contrary to the case of short GRBs, the necessity of a binary progenitor in long GRBs did not arise from the very beginning, and possibly the most important observational piece of evidence of this need can be identified in the temporal and spatial coincidence of GRB 980425 \citep{2000ApJ...536..778P} and SN 1998bw \citep{1998Natur.395..670G}, and the subsequent systematic spectroscopic analysis of additional GRB-SN associations \citep[see][for a review]{2017AdAst2017E...5C}.
}

{
There are two key observational aspects of the SNe associated with GRBs pointing to a relevant role of binary interactions: 1) they are of type Ic, namely both hydrogen and helium lack in their spectra, and 2) the spectral lines are broad-lined implying their ejecta expand at very high expansion velocities of the order of $10^4$~km~s$^{-1}$, implying kinetic energies of up to $10^{52}$~erg, reason for which they have been dubbed HN \citep{2017AdAst2017E...5C}.}

{
The first feature, namely that these SNe are of type IC implies that they possibly originate from helium stars, CO$_{\rm core}$, or Wolf-Rayet stars which have rid of their outermost layers \citep[see, e.g.,][]{2011MNRAS.415..773S}. Indeed, it has been recognized that the a binary companion would most efficiently help in stripping off the pre-SN star outermost layers by tidal effects, multiple mass-transfer and common-envelope episodes \citep[see, e.g.,][]{1988PhR...163...13N,1994ApJ...437L.115I,2007PASP..119.1211F,2010ApJ...725..940Y,2011MNRAS.415..773S}.
}

{
The second feature, namely the observed high-expansion velocities of the SN ejecta, is more delicate and less straightforward to account for in theoretical models. In the BdHN model, numerical simulations in \citet{2018ApJ...852...53R} have shown that the explosion of the GRB within the SN might transfer to it sufficient energy and momentum to convert an initial ordinary SN into a HN. Therefore, broad-lined SNe or HNe in the BdHN model does not necessarily need to be born as such, instead they can be the outcome of the GRB feedback into the SN \citep[see also][]{2019ApJ...871...14B}. Evidence of such a transition from a SN into a HN in a BdHN has been observationally identified in GRB 151027A \citep[see][for details]{2018ApJ...869..151R}.}

{
In addition, binary interactions may enforce corotation of the pre-SN star (i.e the CO$_{\rm core}$) thereby spinning it up to high rotation rates. For BdHN I, this implies a rotation period of the CO$_{\rm core}$ of the order of minutes, so a rotational energy $\sim 10^{50}$~erg \citep{2019ApJ...874...39W}. Of course, this can not explain directly an observed kinetic energy of $10^{52}$~erg. The core-collapse of the iron core of this rotating CO$_{\rm core}$, by angular-momentum conservation, implies the birth of a millisecond period $\nu$NS, which may well power the SN by injecting into it energies of the order of $10^{52}$~erg \citep[see][for more details]{2019ApJ...874...39W, 2020ApJ...893..148R}. It may also happen that binary interactions spin up the CO$_{\rm core}$ beyond corotation bringing it to even to higher rotation rates $\sim 1$~rad~s$^{-1}$ \citep[see, e.g.,][]{2014ApJ...793...45N, 2018MNRAS.474.2419G, 2019ApJ...872..155F}, which would imply a much larger rotational energy of a few $10^{52}$~erg, ready to be used in the SN event.
}

{
There is increasing observational evidence on the high energetics of up to $10^{52}$~erg and the complex nature of the SN from the X- and gamma-ray precursors to the prompt radiation in long GRBs \citep[see, e.g.,][]{2019ApJ...874...39W}. In order to account for such a complexity, we have dubbed these early phases of the BdHN as ``\textit{SN-rise}'' \citep{2019arXiv191012615L}. The \textit{SN-rise} triggers the entire BdHN, so it includes the SN explosion as well as the feedback of the hypercritical accretion onto the $\nu$NS and onto the binary companion NS. We dedicate section~\ref{sec:snrise} to their analysis giving examples in the case of BdHN I and II.
}

{
We can conclude that the binary progenitor of the BdHN model provides a natural explanation of the observational features of the SN associated with long GRBs. Having said this, it is now appropriate to discuss the formation of the CO$_{\rm core}$-NS binary progenitors of the BdHN from the stellar-evolution viewpoint.
}

{
It is well-known from the stellar evolution theory and observations that massive binaries might evolve to form binaries composed of compact objects, e.g. WD-WD, NS-WD, NS-NS and NS-BH. Leaving aside specific technical details, traditional evolutionary paths lead the compact remnant of the more massive star, after undergoing SN, to common-envelope phase with the companion, and after the collapse of the companion star leading to the second SN, the system forms a compact-object binary providing it keeps bound \citep{1999ApJ...526..152F, 2012ApJ...759...52D, 2014LRR....17....3P}. It is very interesting that alternative evolutionary scenarios have been recently proposed in the X-ray binary and SN community leading to the so-called ``\textit{ultra-stripped}'' binaries used to explain NS-NS and low-luminosity SNe \citep[see e.g.][for details]{2013ApJ...778L..23T, 2015MNRAS.451.2123T}. The binary in these cases, after first SN, experiences multiple mass-transfer phases leading to the expulsion of the hydrogen and helium shells of the secondary. As proposed in \citet{2015ApJ...812..100B, 2015PhRvL.115w1102F}, these evolutionary scenarios are a plausible path to form CO$_{\rm core}$-NS binary progenitors of BdHN.
}

{
From the above descends the question of whether such a population of binaries might or not include the progenitors of the BdHN. The orbital periods of the binary at the end of the evolution in these population synthesis codes are $50$--$5000$~h \citep{2013ApJ...778L..23T}. They have been used as a main channel to form NS-NS, but the formation of NS-BH binaries, which are the final outcome left by BdHN I, have not been up to now considered in population synthesis numerical codes. One of the main reasons for this is that the physical processes involved in a BdHN I, occurring when shorter orbital periods of the order of minutes are allowed, lead to BH formation and they have not accounted for yet in these numerical codes. This is certainly a major research which deserves to be pursued in the near future. 
}

{
We refer to \citet{2015PhRvL.115w1102F} for additional details on the following estimation of the BdHN progenitor population. Ultra-stripped binaries are expected to be $0.1$--$1\%$ of the total SN \citep{2013ApJ...778L..23T}, which is estimated to be $2\times 10^4$~Gpc$^{-3}$~y$^{-1}$ \citep[see e.g.][]{2007ApJ...657L..73G}. The population densities of BdHN II/III and BdHN I have been estimated to be $\sim 100$~Gpc$^{-3}$~y$^{-1}$ and $\sim 1$~Gpc$^{-3}$~y$^{-1}$, respectively~\citep{2016ApJ...832..136R}. The above numbers imply, for instance, that BdHN I would comprise only the $0.5\%$ of the ultra-stripped binaries. These estimates confirm, in passing, the rareness of the GRB phenomenon.
}

Since 2018, our research on BdHN has acquired a different status by promoting technical progress in the visualization and in the data analysis, as well as in the introduction of new theoretical paradigms and identification of new astrophysical regimes which we further extend in this article. We start with a specific example of BdHN simulation.

\subsection{A specific BdHN I SPH simulation}

In Fig.~\ref{fig:SPHsimulation}, we show the results of a specific SPH simulation of a BdHN I from  \citet{2019ApJ...871...14B}. It represents the implosion of a CO$_{\rm core}$ of $6.85 M_{\odot}$ giving origin to the explosion of a SN in presence of a binary 
companion NS of $M_{\rm NS}=2 M_{\odot}$. An additional NS of $1.85 M_{\odot}$ originates from the collapse of the Fe-core within the CO$_{\rm core}$ (the green dot at the center of the SN in the two left figures). We indicate as $\nu$NS this newborn neutron star, in order to differentiate it from the binary companion NS. The two upper panels correspond to the mass density in the binary equatorial plane of the binary progenitor, which we label for short as ``seen in the orbital plane''. The lower panels correspond to  viewing in a plane orthogonal to the  equatorial plane of the binary progenitor, indicated for short as ``seen from the top''. This figure well summarizes the central role of the SN in triggering the BDHN1 Phenomenon: by first creating the $\nu$NS and the accreting SN ejecta both on the $\nu$NS and the binary NS companion. The sequence of the accretion process is followed in these Figures 159 s and 259 s. Following the hypercritical accretion process the $\nu$NS reaches a mass and angular momentum,$2.04 M_\odot$ and $1.24 G M^2_\odot/c$, respectively. Up to the final simulation time. Similarly the binary NS companion collapses reaching the secular axisymmetric instability point with a mass of $2.26 M_\odot$ and angular momentum $1.24 G M^2_\odot/c$ at $t = 21.6$ min. In this model the initial binary period of the circular orbit is $4.8$ min. The binary orbit then widens, reaching an orbital period of $16.5$~min and an eccentricity of $\epsilon = 0.6$. 
We are going to give specific examples in selected GRBs of this process in section~\ref{sec:9} with the determination of the mass and spin of the newborn BH. This figure is also essential in emphasizing the implications of the different viewing angles implied by the binary nature of the progenitors, which have been also neglected in the traditional approach.

We further exemplify, in the next two sections, the large amount of results inferred on the BdHN nature utilizing the two above viewing angles. 

\subsection{The upper limits on the Lorentz $\Gamma$ factor and nature of the afterglow}

The observations of BdHN I ``seen in the orbital plane'' have
been addressed in a series of articles based essentially on the X-ray observations made with the XRT detector in Swift \citep[see e.g.,][and references therein]{2018ApJ...852...53R}. They have
been essential in identifying model-independent upper limits on
the Lorenz $\Gamma$ factors of the emission regions during the
gamma-ray flare, the X-ray flares phase, the flare-plateau and
the early afterglow phases (the Nousek-Zhang phase), following the initial ultra-relativistic prompt radiation phase.

The traditional approach had shown that gamma-ray spikes in the prompt emission occur at $\sim 10^{15}$--$10^{17}$~cm with Lorentz gamma factor $\Gamma\sim10^{2}$--$10^{3}$ \citep[e.g.,][]{Li2020}. Using a novel data analysis we have shown that the time of occurrence, duration, luminosity and total energy of the X-ray flares correlate with $E_{\rm iso}$. A crucial feature has been identified in the observation of thermal emission in the X-ray flares that we have shown occurs at radii $\sim10^{12}$~cm with $\Gamma\lesssim 4$. The upper limit of Lorentz factor, $\Gamma \lesssim 2$, has been there established in the analysis of the X-ray flares. Equally, an upper limit $\Gamma \lesssim 3$ has been set in the transition from a SN to a Hypernova (HN) in GRB 151027A \citep{2018ApJ...869..151R}. Finally, the limit $\Gamma \lesssim 2$ has been established in the thermal emission in the early part of the afterglow phase of GRB 130427A \citep{2018ApJ...869..101R}.

The enormous kinetic energy of an ultra-relativistic blast wave needed in the traditional approach to explain the energy source of the afterglow has been therefore superseded: the above mentioned stringent upper limits on the $\Gamma$ factors exclude any ultra-relativistic motion. 

The origin of the afterglow of long GRBs and these mildly-relativistic processes have been successfully identified in the synchrotron emission produced by relativistic electrons in the SN ejecta, powered by the hypercritical accretion of the SN into the spinning $\nu$NS of $1.5~M_{\odot}$ and its pulsar-like emission \citep{2018ApJ...869..101R, 2019ApJ...874...39W, 2020ApJ...893..148R}. From the amplitude of their decaying X-ray luminosities observed by \textit{Swift}-XRT \citep{2016ApJ...833..159P} the spin of the $\nu$NS and the strength and structure of its magnetic field in specific BdHN I and II have recently been obtained 
\citep{2020ApJ...893..148R}.

It is important that the synchrotron process occurring in the interaction of the SN ejecta with the $\nu$NS, requires a much smaller energy to explain the nature of the afterglow in our present approach based on the hypercritical accretion of from the SN onto the $\nu$NS \citep{2019ApJ...874...39W,2020ApJ...893..148R} than the ones purported in the ultrarelativistic blast waves.
\begin{table}
\centering
\begin{tabular}{lc}
\hline\hline
Extended wording & Acronym \\
\hline
Binary-driven hypernova & BdHN \\
Black hole                    & BH \\
Carbon-oxygen star      & CO-star \\ 
fallback-powered kilonova      & FB-KN \\ 
Gamma-ray burst         & GRB \\
Gamma-ray flash          & GRF \\
gamma-ray flash kilonovae & GR-K \\
Massive neutron star     & M-NS \\
Neutron star                & NS \\
New neutron star          & $\nu$NS \\
Short gamma-ray burst  & S-GRB \\
Short gamma-ray flash  & S-GRF \\
Supernova                  & SN \\
Supernova rise              &SN-rise\\
Ultrashort gamma-ray burst & U-GRB \\ 
White dwarf                & WD \\
X-ray flash                  & XRF \\
\hline
\end{tabular}
\caption{Alphabetic ordered list of the acronyms used in this work.}
\label{acronyms}
\end{table}

\subsection{The ``\emph{inner engine}'' of  BdHN I} 

The observations of the BdHN I ``seen from the top'' are the
main topic of this article. They lead to an identification of the morphology of BdHN I, to the origin of the MeV, GeV, and TeV emissions observed by the GBM and LAT instruments {on board} the Fermi satellite, the MAGIC and the H.E.S.S telescopes, {as well as to a contribution to ultrahigh-energy cosmic rays (UHECRs) from GRBs \citep[see, e.g.,][]{2020EPJC...80..300R}}. Particularly important has been the recent identification of the physical process occurring in the ``\textit{inner engine}'' originating the GeV emission as ``seen from the top'' in GRB 130427A, also confirmed in three additional BdHN I GRB 160509A, GRB 160625B and GRB 190114C \citep{2019ApJ...886...82R,2019arXiv191012615L}.

In these works: 
\begin{enumerate}
    \item 
    We have proposed that the \textit{inner engine} of a BdHN I is composed of a Kerr BH in a non-stationary state, embedded in a uniform magnetic field $B_0$ aligned with the BH rotation axis, as modeled by the Papapetrou-Wald solution of the Einstein-Maxwell equations \citep{1966AIHPA...4...83P, 1974PhRvD..10.1680W}, and surrounded by an extremely-low density ionized plasma of $10^{-14}$~g~cm$^{-3}$. Using GRB 130427A as a prototype, we have shown that this \textit{inner engine} acts in a sequence of \textit{elementary impulses} emitting ``\textit{blackholic quanta}'' \citep{2020EPJC...80..300R}. The repetition time of the emission of each ``\textit{blackholic quantum}'' of energy ${\cal E} \sim 10^{37}$~erg, is $\sim 10^{-14}$~s at the beginning of the process. Then, it slowly increases with the time evolution. Electrons are accelerated to ultra-relativistic energy near the BH horizon and, propagate along the polar axis, ${\theta =0}$.  They can reach energies of $\sim 10^{18}$~eV, and partially contribute to ultra-high energy cosmic rays (UHECRs). When propagating along ${\theta \neq 0}$ through the magnetic field $B_0$ they give rise to the synchrotron emission of GeV and TeV photons. The \textit{inner engine} operates within a ``cavity'' formed during  the hypercritical accretion of the SN ejecta onto the NS binary companion, and during the BH formation \citep{2019ApJ...883..191R}. This result is the first step toward identifying the BdHN I morphology, presented in this article.
    \item 
    It has been shown that the multiwavelength emissions corresponding to the above acceleration process leading to synchrotron radiation occur in a jet with a half-opening angle of 60$^{\circ}$ from the normal to the binary plane. The jetted emission occurs in selected energy bands in the MeV, GeV, TeV and UHECR.
    \item 
    This result has been applied to GRB 130427A, and we here show that it applies generally to all BdHN I as a consequence of the novel morphology identified in the present article.
    \item 
    We have evaluated the total GeV emission in GRB 130427A and identified its decaying luminosity in the GeV range with a power-law  index of $\alpha_{\rm GeV}=-1.19\pm 0.04$, using the first and the second \textit{Fermi}-GRB catalogs \citep{2013ApJS..209...11A, 2019ApJ...878...52A}.  In this article we generalize this result to all BdHN I emitting GeV radiation.
\end{enumerate}

\subsection{On the measure of the BH mass and spin in BdHN I}
 
For the first time, in \citet{2019ApJ...886...82R} it was shown how to extract the rotational energy of a Kerr BH in an astrophysical system, using  GRB 130427A as a prototype. This was made possible making use of the the mass-energy formula
of the Kerr BH \citep{1970PhRvL..25.1596C, 1971PhRvD...4.3552C, 1971PhRvL..26.1344H, Hawking:1971vc}, given in Eq.~(\ref{aone}). There, it was shown how through the ``\emph{inner engine}'' activity the energetics of the GeV emission could originate near the BH horizon and be explained using the extractable energy of the BH, keeping constant the BH \emph{irriducible mass}. In turn, this has led to the first measure of the initial mass and spin of the BH at its moment of formation: $M=2.3 M_\odot$, its spin, $\alpha = a/M = 0.47$. The present article is dedicated to extend this classic result to all BdHN I, where sufficient GeV emission data are available. This same procedure will be soon extended to active galactic nuclei with BH masses up to $10^{10}M_\odot$. 

\subsection{Structure of the article}

We first give in section~\ref{sec:2} an outline of the nine GRB subclasses presented in \citet{2016ApJ...832..136R}, with a brief summary of their initial states (\textit{in-state}), their final state (out-state), their energetics and spectral properties in the gamma-rays {both in the MeV} and in the GeV emissions. We also recall the binary mergers which include the NS-NS binaries leading to the two classes of short GRBs.

In section~\ref{sec:snrise}, we summarize the previous results \citep{2019arXiv191012615L} on the analysis of the SN-rise of BdHNe I and II obtained from \textit{Fermi}-GBM, and present their relation with the X-ray afterglow observed by \textit{Swift}-XRT.

In section~\ref{sec:xrayafterglow}, following our previous works \citep{2018ApJ...869..101R, 2019ApJ...874...39W, 2020ApJ...893..148R}, we study  properties of the X-ray afterglow of BdHNe and we determine the spin of the $\nu$NS in two BdHNe I, two BdHNe II and one BdHN III system.

In section~\ref{sec:4}, we analyze the properties of the GeV emission in BdHNe I updated following the second GRB catalog presented by Fermi-LAT, which covers the first $10$ years of its operations, from 2008 August 4 to 2018 August 4 \citep{2019ApJ...878...52A}. We address the $378$ BdHNe I with known cosmological redshift; see the list of BdHNe I in \citet{2016ApJ...833..159P,2018ApJ...852...53R} and also the updated list in Appendix~\ref{updated}. We then consider only the $54$ BdHN I with the boresight angle of \textit{Fermi}-LAT smaller than $75^{\circ}$ at the trigger time. We give the details of the $25$ BdHNe I {with observed GeV radiation}, out of the 54. For each of them, we list in Table~\ref{tab:cb} the cosmological redshift, the $E_{\rm p,i}$ of the spectrum, the $E_{\rm \gamma, iso}$ of the source, the \textit{Fermi} GCN, the boresight angle, the $E_{\rm LAT}$, the likelihood test statistic (TS), and some additional distinguishing properties.  
In Table \ref{tab:BdHNe_No_GeV} for the $29$ BdHNe I, we then give the cosmological redshift, the $E_{\rm p,i}$ of the spectrum, the $E_{\rm \gamma, iso}$ of the source, the Fermi GCN, the boresight angle and some distinguishing properties of the associated X-ray emissions. 

In section~\ref{sec:morphology}, we explain the nature of the these BdHNe in terms of a novel morphology of the binary system. The BdHN I have a conical structure normal to the equatorial plane of the binary progenitor.  When the observations are made with a viewing angle lying in the orbital plane of the binary progenitor then the GeV emission is not observable. In this case, only the gamma-ray flare, the X-ray flares and the X-ray plateau remain observable. From the ratio $N_{\rm LAT}/N_{\rm tot} = 25/54$, we infer the presence in the BdHN I of a conical structure of approximately $60^{\circ}$ around the normal to the plane of the binary progenitors. Within this cone all emissions are observable, namely the X-ray, the gamma-ray, the GeV and TEV emission also UHECRs. For larger inclination angle as confirmed theoretically in \citet{2018ApJ...869..151R, 2019ApJ...886...82R}, the GeV radiation is not observable and only flaring activities are observed following the prompt radiation phase.
 
In section~\ref{sec:11}, we show that this novel geometry is indeed present in the recent three-dimensional SPH numerical simulations at the moment of BH formation in a BdHN \citep{2019ApJ...871...14B}.

In section~\ref{sec:8}, for each of the $25$ BdHNe I, we provide the $0.1$--$10$~GeV luminosity light-curves as a function of the time in the rest-frame of the source. We obtain a power-law fit $L_n=A_n t^{-1.19\pm0.04}$~erg~s$^{-1}$ and report the  amplitude $A_n$ and the luminosity at $10$~s from the beginning of the prompt radiation, $L_{\rm 10s}$, with their associated uncertainties. We also provide a correlation between $L_{\rm 10s}$ and $E_{\rm \gamma, iso}$.

In section~\ref{sec:5}, we determine the values of the mass and spin of the BH and the strength of the  magnetic field surrounding the BH in the ``\emph{inner engine}'' of the selected BdHNe I. We also show the process of hypercritical accretion of the SN on a companion NS gives in all cases origin to the newborn BH. 

In section~\ref{sec:9}, we confirm 1) the central role of the SN in giving rise to its hypercritical accretion  on the $\nu$NS and the newly born BH, to the afterglow observed by SWIFT and to the high energy GeV and TeV  emission observed by Fermi-LAT, 2) that the MeV-GeV energetic range is explainable  by extractable rotational energy of a Kerr BH operating n the ``\emph{inner engine}'' and this result allows the determination of the initial mass and spin of the BH, 3) the power-law evolution of the $0.1$--$100$~GeV luminosity after the prompt phase, arises from  the slowing down rate of the BH spin, keeping constant the irreducible mass $M_{\rm irr}$ of the BH.

We finally proceed to the general conclusions in section~\ref{sec:12}. Before proceeding, we indicate in Table~\ref{acronyms} the alphabetic ordered list of acronyms used in this work.

\section{Subclasses of GRBs and definitions of BdHN}\label{sec:2}

We address the specific role of the X-ray emission observed
by the \textit{Swift} satellite as well as the MeV-GeV radiation observed by the Fermi satellite in order to further characterize the nine subclasses of GRBs presented in \citet{2016ApJ...832..136R} and updated in \citet{2018ApJ...852...53R, 2019ApJ...874...39W}, and here further updated in section~\ref{sec:xrayafterglow} and Appendix\ref{updated}. In Table~\ref{tab:a}, we summarize for each GRB subclass their name, the number of
observed sources with cosmological redshift, and their progenitors
characterizing their ``in-state''.

\begin{table*}
\scriptsize
\centering
\begin{tabular}{c|cccccccc}
\hline
Class &   Type & Number  & \emph{In-state}  & \emph{Out-state} & $E_{\rm p,i}$ &  $E_{\rm \gamma, iso}$  &  $E_{\rm iso,Gev}$  \\
& & & & & (MeV) & (erg) &  (erg) 	\\	
\hline
Binary Driven & I  & $378$ &CO star-NS  & $\nu$NS-BH & $\sim0.2$--$2$ &  $\sim 10^{52}$--$10^{54}$ &    $\gtrsim 10^{52}$ \\
Hypernova & II & $(49)$ &CO star-NS    & $\nu$NS-NS & $\sim 0.01$--$0.2$  &  $\sim 10^{50}$--$10^{52}$ &    $-$ \\
(BdHN) & III  & $ (19) $ &CO star-NS    & $\nu$NS-NS & $\sim 0.01$  &  $\sim 10^{48}$--$10^{50}$ &    $-$ \\
& IV   & $0$ & CO star-NS  & BH & -- &  $>10^{54}$ &   $\gtrsim 10^{53}$   \\
\hline
 & I & $18$ &NS-NS & MNS & $\sim0.2$--$2$ &  $\sim 10^{49}$--$10^{52}$  &  $-$ \\
Binary & II  & $6$ &NS-NS & BH & $\sim2$--$8$ &  $\sim 10^{52}$--$10^{53}$ &   $\gtrsim 10^{52}$\\
Merger& III  & $(1)$ &NS-WD & MNS & $\sim0.2$--$2$ &  $\sim 10^{49}$--$10^{52}$ & $-$\\
(BM)& IV  & $(1)$ &WD-WD & NS/MWD & $ < 0.2$ &  $< 10^{51}$  & $-$\\
& V & $(0)$ &NS-BH & direct BH & $\gtrsim2$ &  $>10^{52}$ & $-$ \\
\hline
\end{tabular}
\caption{Summary of the GRB subclasses. In addition to the subclass name, we report the number of GRBs for each subclass. We recall as well the ``in-state'' representing the progenitors and the ``out-state'' as well as the $ E_{\rm p,i}$ and $E_{\rm \gamma, iso}$ for each subclass. The GeV emission is indicated in the last column: for long GRBs it appears only in BdHN I and BdHN IV (BH-SN) while, for short bursts, it appears only for S-GRBs. In all sources with GeV emission, it is $\gtrsim 10^{52}$~erg.}\label{table:taxonomy}
\label{tab:a}
\end{table*}

In all cases the GRB progenitors are binary systems composed
of various combinations of CO$_{\rm core}$, of NSs, of WDs, and of BHs. The ``out-state'' of the corresponding mergers
or accretion processes have been represented in Fig.~7 in
\citet{2016ApJ...832..136R} where we also presented the interesting possibility that ``out-states'' of the GRB subclasses  can  become the ``in-states'' of  new GRB subclasses. In particular, we indicate an example in which the ``out-state'' of a BdHN I can become the ``in-state'' of a short GRB.

In this article, we focus only on long GRBs with BdHN progenitors \citep{2016ApJ...832..136R}:  binary systems composed of a CO$_{\rm core}$, exploding as SN Ic, and a NS binary companion. The presence of such a NS binary companion in close orbit can explain the removing of the outer layers of Hydrogen and Helium of the massive star leading to the CO$_{\rm core}$ \citep[see, e.g.,][]{Ruffini2001c,2012ApJ...758L...7R,2014ApJ...793L..36F}.

As noted in the introduction, when the CO$_{\rm core}$ gravitationally collapses, it gives origin to a SN and its Fe core collapses to form a $\nu$NS. The entire dynamics and evolution of the BdHN is essentially based on these three different components and their interplay: the SN explosion (SN-rise), the $\nu$NS undergoing an overcritical accretion process of the SN ejecta, the binary companion NS also undergoes an overcritical accretion process of the SN ejecta which monotonically increases the binary NS companion mass. In compact binary systems, this accretion causes the NS to obtain its critical mass leading to  the formation of a newborn BH \citep{2015ApJ...812..100B,2016ApJ...833..107B}; see also Fig.~\ref{fig:SPHsimulation}.

We first address the SN hypercritical accretion onto the binary NS companion: the outcome is a strong function of the compactness of the binary system and its binary orbital period.

When the orbital period is as short as $5$~min, the hypercritical accretion proceeds at higher rates and the companion NS reaches its critical mass leading to:
\begin{enumerate}
    \item 
    the formation of a BH and consequently a formation of a new binary system composed of a BH and a $\nu$NS \citep{2014ApJ...793L..36F};
    \item 
    the emission of a very energetic GRB in the range of $10^{52} \lesssim E_{\rm iso}\lesssim 10^{54}$~erg and, peak energy in the range of $0.2$~MeV$<E_{\rm p,i}<2$~MeV lasting a few seconds known as the ultra relativistic prompt emission phase (UPE);
   \item
   the onset of the prolonged power-law GeV emission, triggered by the formation of the newborn BH, with a luminosity described in the rest frame of the source
   \begin{equation}
       L_{\rm GeV}= A_{\rm GeV} \left(\frac{t}{1\,\rm s}\right)^{-\alpha_{\rm GeV}} ~~,
   \end{equation}
   with $\alpha_{\rm GeV}=1.19 \pm 0.04$. One of the main results in this paper is to show that this radiation is present only in a subset of BdHN and the explanation of this result will lead to the determination of the conical BdHN morphology, see section~\ref{sec:8}.
   
\end{enumerate}
 
These systems have been indicated as BdHN I \citep{2015ApJ...798...10R,2015ApJ...812..100B,2016ApJ...833..107B,2016ApJ...832..136R,2019ApJ...874...39W,2019ApJ...886...82R}.

\begin{figure*}
\centering
(a)\includegraphics[width=0.65\hsize,clip]{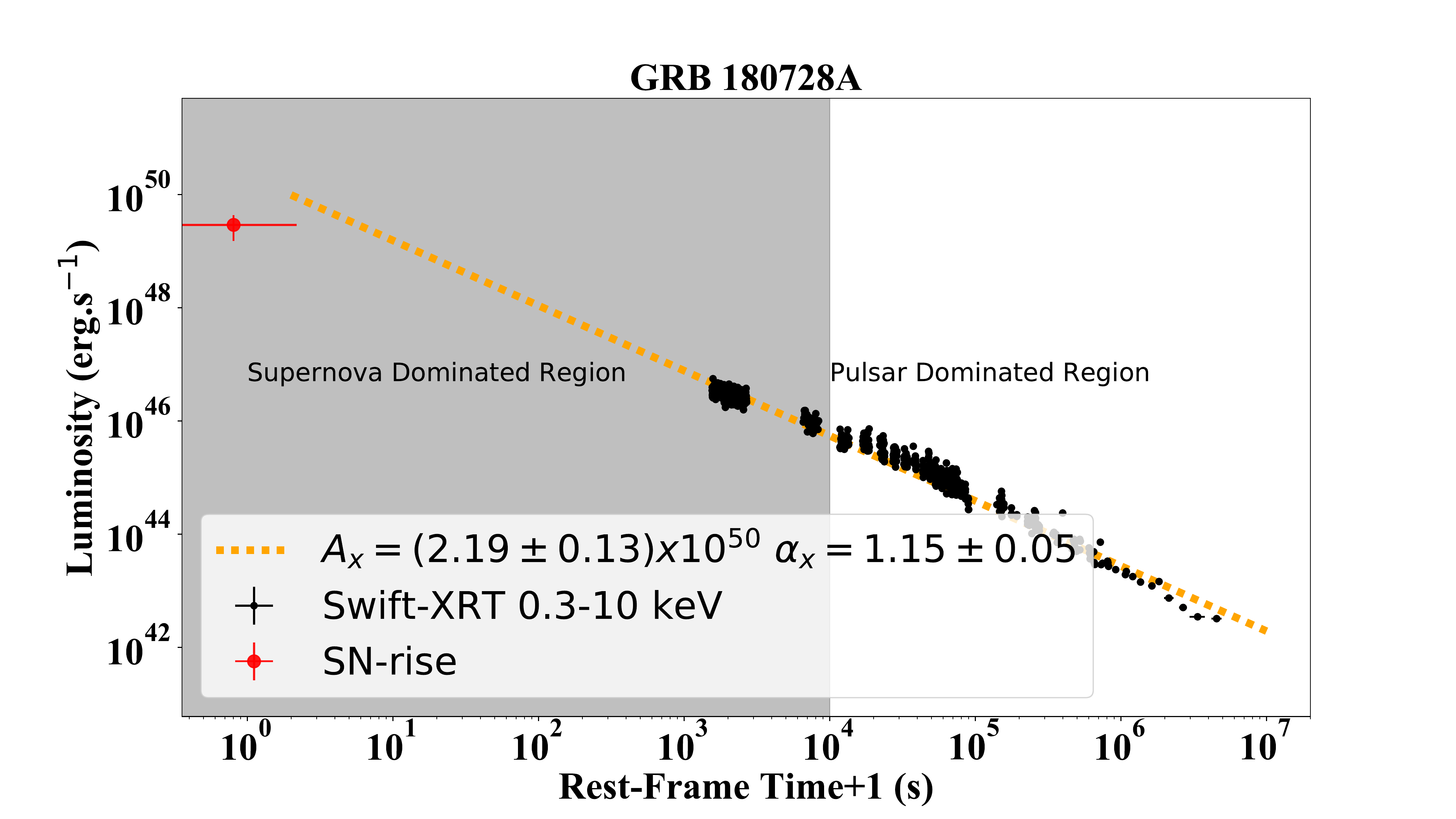}
(b)\includegraphics[width=0.65\hsize,clip]{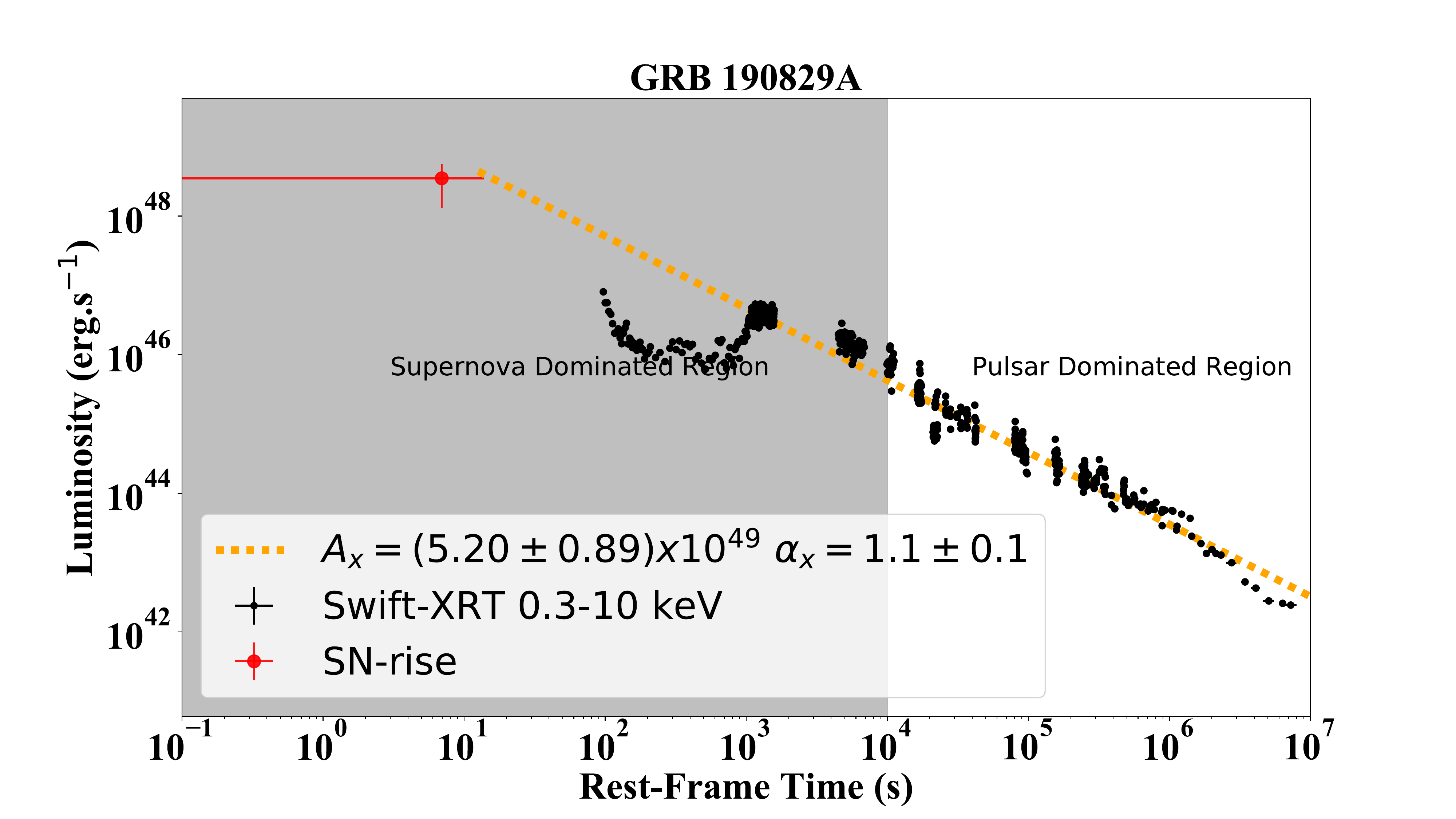}
(c)\includegraphics[width=0.65\hsize,clip]{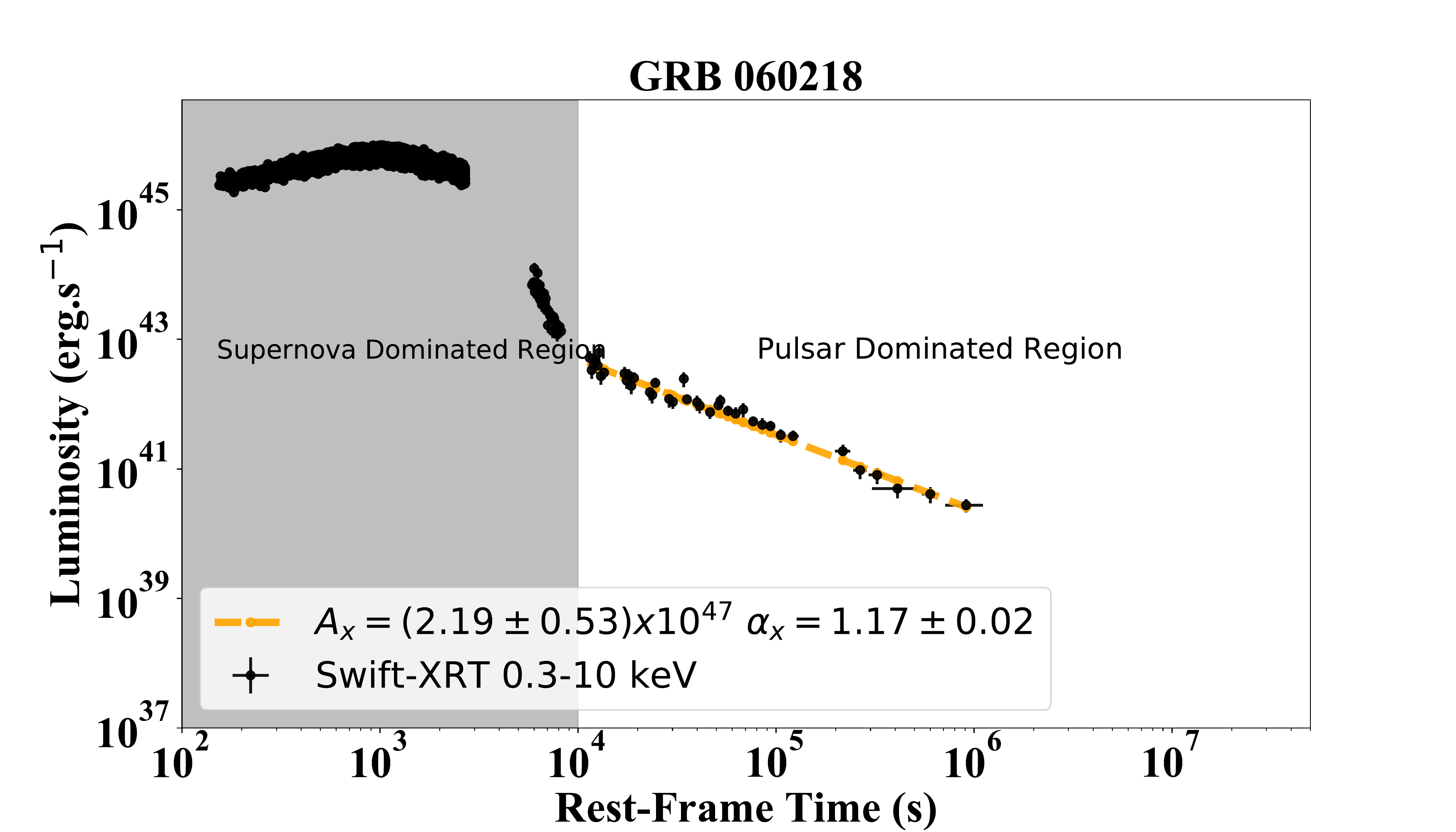}
\caption{The X-ray afterglow luminosity observed by \textit{Swift}--XRT which follow a decaying power-law:  \textbf{a}: GRB 180728A (BdHN II) with amplitude $(2.19\pm 0.13)\times 10^{50}$~erg~s$^{-1}$ and power-law index $\alpha_X=1.15\pm 0.05$. \textbf{b}: GRB 190829A (BdHN II) with amplitude $(5.20\pm 0.89)\times 10^{49}$~erg~s$^{-1}$ and power-law index $\alpha_X=1.1\pm 0.1$. \textbf{c} GRB 060218 (BdHN III) with amplitude $(2.19\pm 0.53)\times 10^{47}$~erg~s$^{-1}$ and power-law index $\alpha_X=1.17\pm 0.02$. The fallback material of the SN on the $\nu$NS produce this X-ray afterglow emission \citep{2020ApJ...893..148R}. In section~\ref{sec:nuns1}, we report the result of the simultaneous fit of the X-ray afterglow of all types of BdHN in order to find the universal power-law index. As shown in \citet{2018ApJ...852...53R, 2018ApJ...869..151R}, until $\sim 10^4$~s the gamma/X-ray afterglow is mainly produced by the SN kinetic energy (\textit{SN dominated region}) and its interaction with the magnetic field of the $\nu$NS. After $ 10^4$~s, as shown by \citet{2018ApJ...869..101R}, the role of $\nu$NS becomes prominent (\textit{pulsar dominated region}).}
\label{fig:2bdhn2}
\end{figure*}

\begin{figure*}[ht!]
\centering
\includegraphics[width=0.49\hsize,clip]{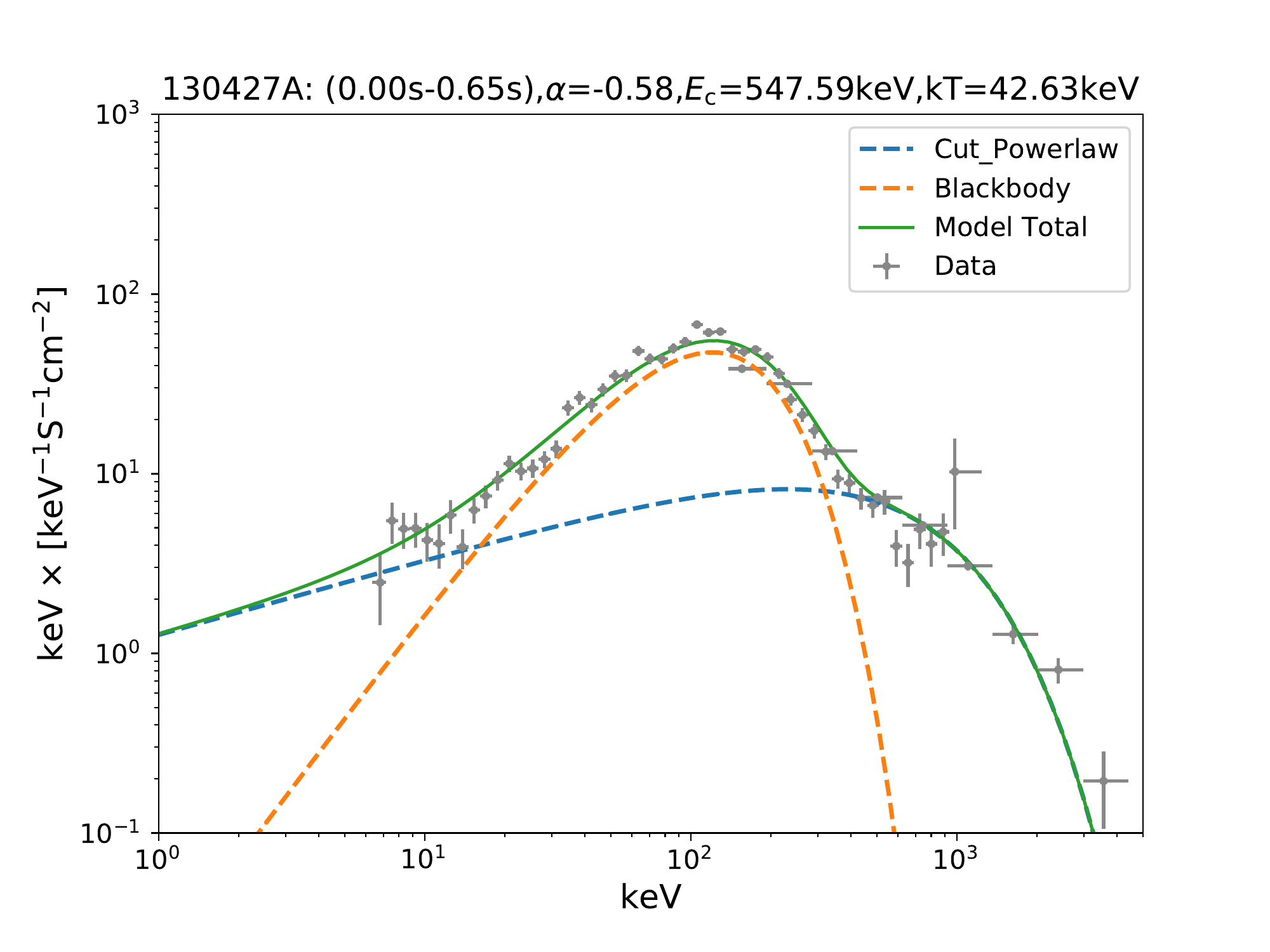}
\includegraphics[width=0.49\hsize,clip]{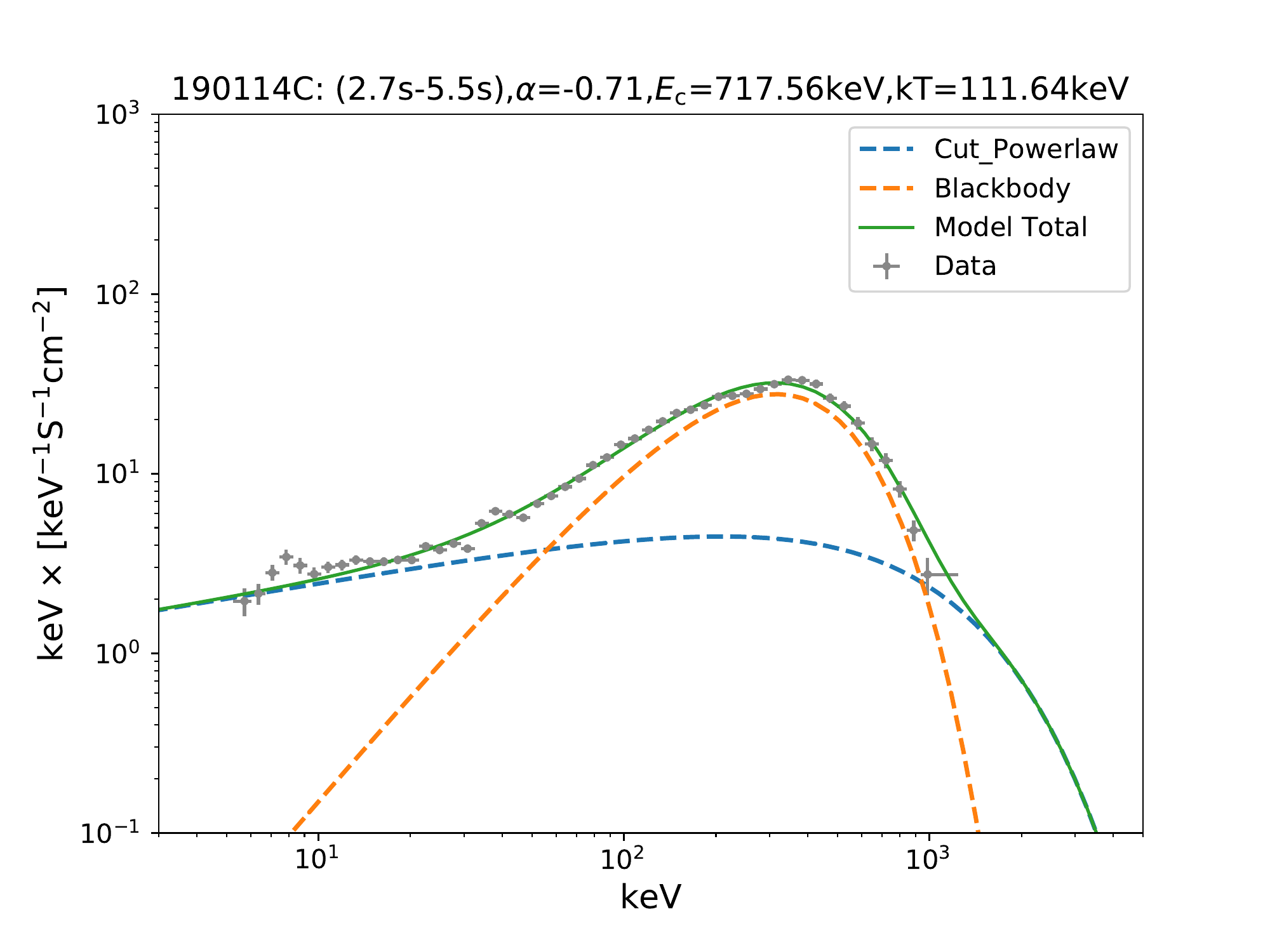}
\includegraphics[angle=0, scale=0.43]{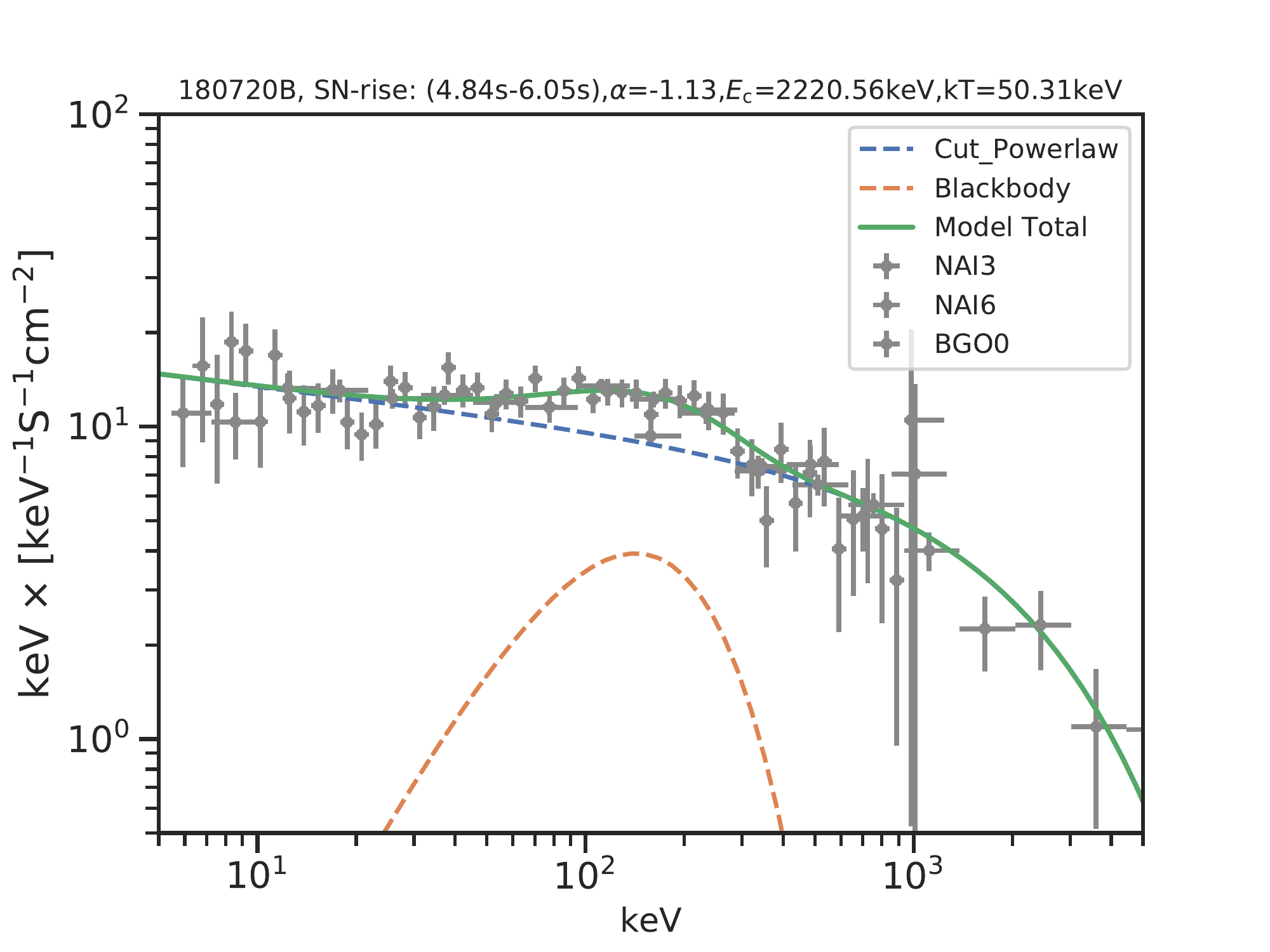}
\includegraphics[width=0.49\hsize,clip]{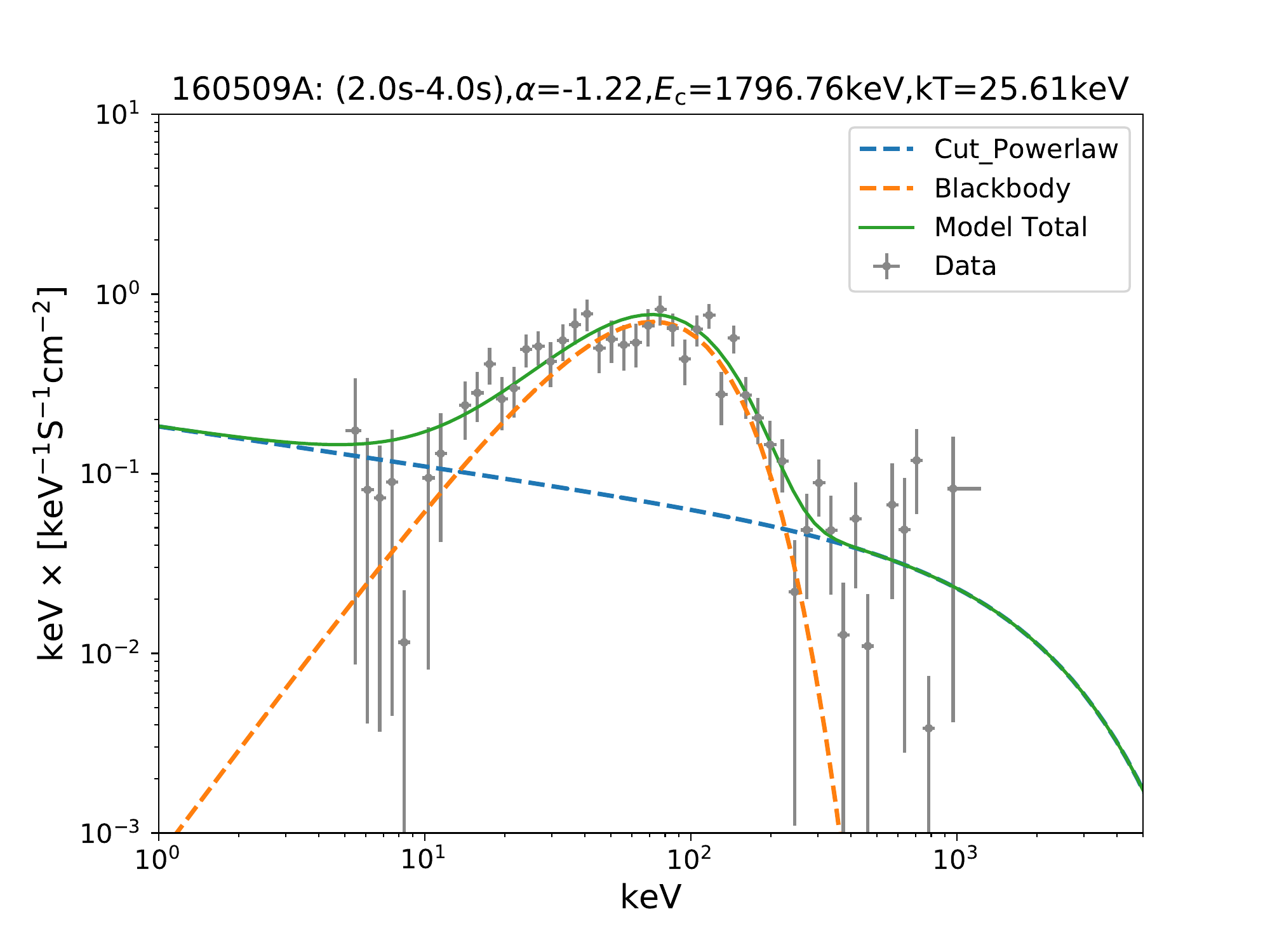}
\caption{The Spectrum of the SN-rise of GRB 160509A as observed by Fermi-GBM in the energy range of $8$--$900$~keV. \textbf{Upper left:} SN-rise spectrum of BdHN I 130427A, well fitted by a CPL+BB model, from $0$ to $0.65$s ($t_{\rm rf} \simeq 0.49$s); the spectral index $\alpha$ is -0.58, cutoff energy $E_{\rm c}$ is $547.59$ keV, and the BB temperature is $42.63$~keV in the observer's frame. \textbf{Upper right:} The spectra of SN-rise of BdHN I 190114C corresponding to $t = 1.12$~s ($t_{\rm rf} = 0.79s$) to $t=1.68$~s ($t_{\rm rf} = 1.18s$), which is best fit by a CPL+BB model, with a low-energy photon index $\alpha$ of -0.71, and a peak energy $E_{\rm c}$ of 524.7~keV, and a BB temperature $18.42$~keV. Time is reported in both the observer's frame and the rest-frame. Spectral parameters of the best fit are presented in the observer's frame. \textbf{Lower left:} SN-rise spectrum of BdHN I 180720B, well fitted by a CPL+BB model, from $4.84$ to $6.05$s ($t_{\rm rf} \simeq 0.$s); the spectral index $\alpha$ is -1.13, cutoff energy $E_{\rm c}$ is $2220.569$ keV, and the BB temperature is $50.31$~keV in the observer's frame.  \textbf{Lower right}: SN-rise spectrum of BdHN I 160509A, well fitted by a CPL+BB model, from $2.0$ to $4.0$s ($t_{\rm rf} \simeq 0.$s); the spectral index $\alpha$ is -1.22., cutoff energy $E_{\rm c}$ is $1796.76$ keV, and the BB temperature is $25.66$~keV in the observer's frame.}
\label{fig:sn-riseall}
\end{figure*}

The first list of the BdHNe I was composed of $161$ sources
spanning $12$~years of Swift/XRT observation activity till $2015$ presented in \citet{2016ApJ...833..159P} which was further extended to $173$ sources in \citet{2018ApJ...852...53R} up through the end of $2016$
which led to a total of $345$ BdHNe I within 1997-2016 observed by other satellites like Fermi and \textit{Konus}-WIND in addition to Swift. This list is further extended here to $378$ BdHN I till Dec. 2018; see Appendix.~\ref{updated}; see Table~\ref{table:taxonomy}. 

When the orbital period of the binary system is $\gtrsim 5$~min, the hypercritical accretion is not sufficient  to trigger the collapse of the NS companion into a BH: therefore no GeV emission can be produced nor be observed. Therefore, a MNS is formed. In these systems, the observed peak energy is in the range $4$~keV$<E_{\rm p,i}<300$~keV and the isotropic energy in the range of $10^{48} \lesssim E_{\rm iso}\lesssim 10^{52}$~erg, as observed by the Fermi-GBM. They have been indicated as X-ray flashes (XRF) in contrast with the
more energetic BdHN I \citep{2015ApJ...798...10R,2015ApJ...812..100B,2016ApJ...833..107B,2016ApJ...832..136R}. We here use for the XRFs the name BdHN II, according to \citet{2019ApJ...874...39W}. A canonical example has been given in \citet{2019ApJ...874...39W}; see Table.~\ref{table:taxonomy}.

BdHNe III have the same composition as BdHNe II, but the
binary is further detached. No BH is formed and no GeV radiation is produced nor observed. This subclass is characterized by binary systems widely separated and weaker energy emission with $E_{\rm iso}$ in the range of $10^{48}$--$10^{50}$~erg. 

As we will see in section~\ref{sec:9}, the most energetic BdHN I originate from extremely tight binary systems with the companion NS grazing the radius of the CO$_{\rm core}$. It is therefore conceivable that in some systems the NS companion merges with the CO$_{\rm core}$ just prior to the SN explosion leading to the possible direct formation of a BH, a concept envisaged by \citet{1993ApJ...405..273W} in the failed SN scenario. We have left such a possibility opened in an additional BdHN IV family; see Table~\ref{table:taxonomy}.

The hypercritical accretion of the SN ejecta onto the $\nu$NS leads to the pulsar-like emission which gives rise to the X-ray afterglow emission observed by \emph{Swift} \citep{2020ApJ...893..148R}. This is a property intrinsic to the nature of the model and shared by all BdHN subclasses. It is therefore natural to expect, as has been verified, that the luminosity of the X-ray afterglows of \textit{all} long GRBs, in all BdHN subclasses, follow a common decaying power-law of 
\begin{equation}\label{eq:Lx}
    L_X = A_X \left(\frac{t}{1\,\rm s}\right)^{-\alpha_X},
\end{equation}
with $\alpha_X = 1.48\pm 0.32$, including the SN-rise, when averaged over all BdHN I up to $10^{6}$~s \citep{2016ApJ...833..159P}. The different amplitudes, $ A_X$, and power-law indices, $\alpha_X$, of the X-ray afterglow luminosity can be used to determine the spin and magnetic field of the $\nu$NS \citep{2019ApJ...874...39W, 2020ApJ...893..148R}.

\begin{table*}
\small\addtolength{\tabcolsep}{-2pt}
\label{tab:shockbreakout}
\caption{The properties of the SN-rise in BdHN I: GRB 190114C, GRB 130427A, GRB 160509A, and GRB 160625B; and the properties of the SN-rise in BdHN II: GRB 180728A.}
\centering                         
\begin{tabular}{ccccccccc}       
\hline\hline    
GRB&$t_{1}\sim t_{2}$ &Duration&Flux&$E_{\rm sh}$&$E_{\rm iso}$&Temperature&redshift&Reference\\
&(s)&(s)&(erg cm$^{-2}$ s$^{-1}$)&(10$^{52}$ erg)&(erg)&(keV)&&\\
&(Observation)&(Rest)&&(SN-rise)&(Total)&(Rest)&&(For SN-rise)\\
\hline                        
190114C&1.12$\sim$1.68&0.39&$1.06^{+0.20}_{-0.20}(10^{-4})$&$2.82^{+0.13}_{-0.13}$&$(2.48\pm0.20)\times10^{53}$&27.4$^{+45.4}_{-25.6}$&0.424&\cite{GCN23983}\\
\hline
130427A&0.0$\sim$0.65&0.49&2.14$^{+0.28}_{-0.26}$(10$^{-5}$)&0.65$^{+0.17}_{-0.17}$&$\sim$1.40$\times$10$^{54}$&44.9$^{+1.5}_{-1.5}$&0.3399&\cite{2013ApJ...776...98X}\\
160509A&2.0$\sim$4.0&0.92&1.82$^{+1.23}_{-0.76}$(10$^{-6}$)&1.47$^{+0.6}_{-0.6}$&$\sim$1.06$\times$10$^{54}$&25.6$^{+4.8}_{-4.7}$&1.17&\cite{2017ApJ...844L...7T}\\
160625B&0$\sim$2.0&0.83&6.8$^{+1.6}_{-1.6}$(10$^{-7}$)&1.09$^{+0.2}_{-0.2}$&$\sim$3.00$\times$10$^{54}$&36.8$^{+1.9}_{-1.9}$&1.406&This paper\\
\hline
180728A&$-1.57\sim1.18$&$0.83$&4.82$^{+1.16}_{-0.82}(10^{-8}$)&7.98$^{+1.92}_{-1.34} \times 10^{49}$&$2.76^{+0.11}_{-0.10}\times$10$^{51}$& - &0.117& \citet{2018GCN.23142....1I}\\
\hline                                   
\end{tabular}
\end{table*}

Before leaving this topic, we  mention a few cases of X-ray afterglows in BdHN II and BdHN III. Each BdHN II and BdHN III must be also characterized by an x-ray afterglow originating from the accretion of the SN ejecta into the $\nu$NS. Their power-law index $\alpha_X$ coincides with the one of BdHN I, although the difference in the total angular momentum of the binary progenitors and its conservation leads necessarily to a smaller value of the amplitude $A_X$ in Eq.~(\ref{eq:Lx}), to a corresponding
lower value of the $\nu$NS spin and to a smaller value of the SN-rise; see Fig.\ref{fig:2bdhn2}. 

In the rest of this article, we  mainly examine the properties of BdHN I with special attention to:
\begin{enumerate}
    \item 
     their SN-rise emission;
    \item 
     the power-law decay of the X-ray emission of the afterglow observed by Swift, measured in the cosmological rest-frame of the source;
    \item 
     the corresponding GeV emission observed by Fermi-LAT, centering on the identification of the BdHN morphology to explain the absence of this emission in a subclass of BdHN I. 
\end{enumerate}

\begin{figure*}
\centering
\begin{tabular}{ccc}
\includegraphics[width=0.49\hsize,clip]{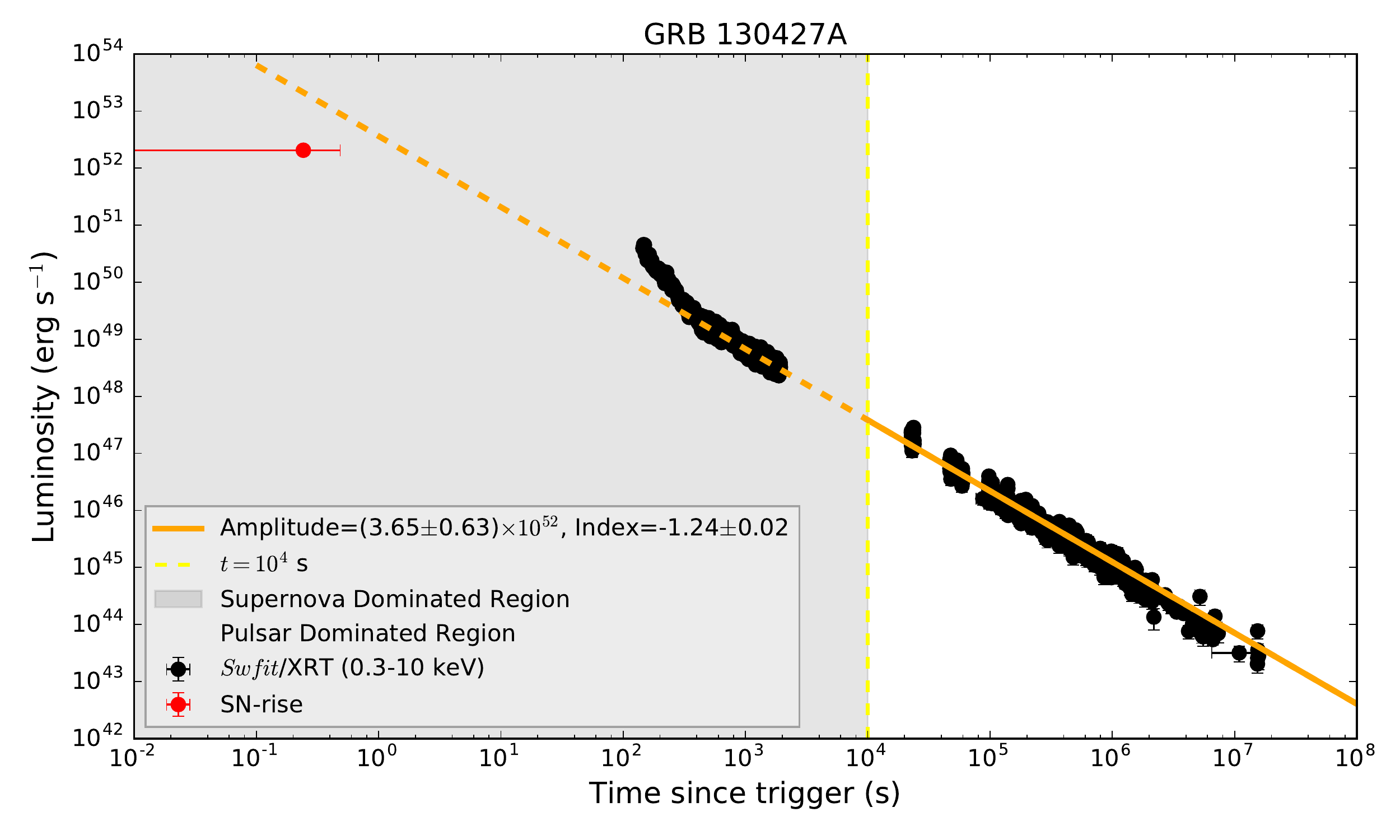} & \includegraphics[width=0.49\hsize,clip]{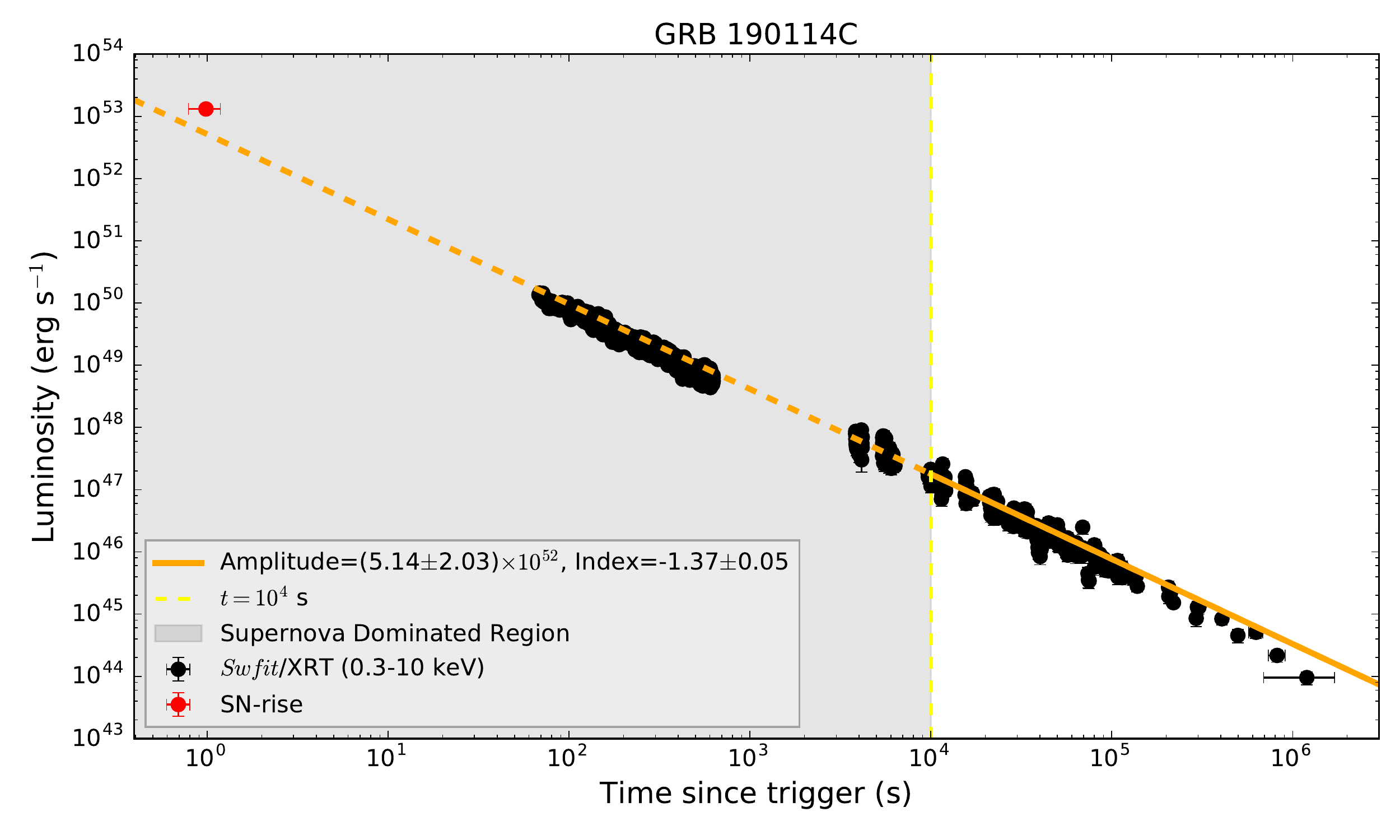}  \\  (a) &  (b)\\
\includegraphics[width=0.49\hsize,clip]{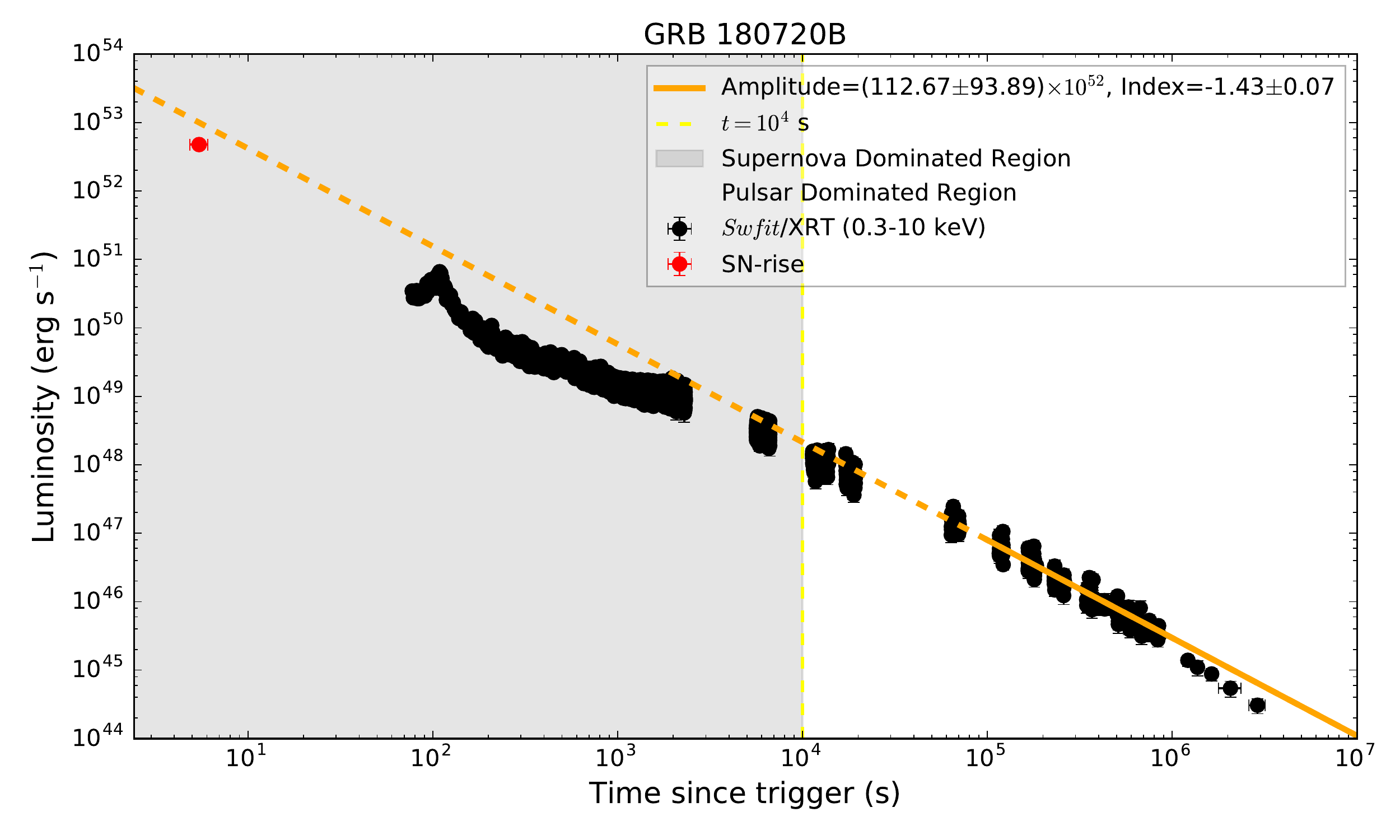} & \includegraphics[width=0.49\hsize,clip]{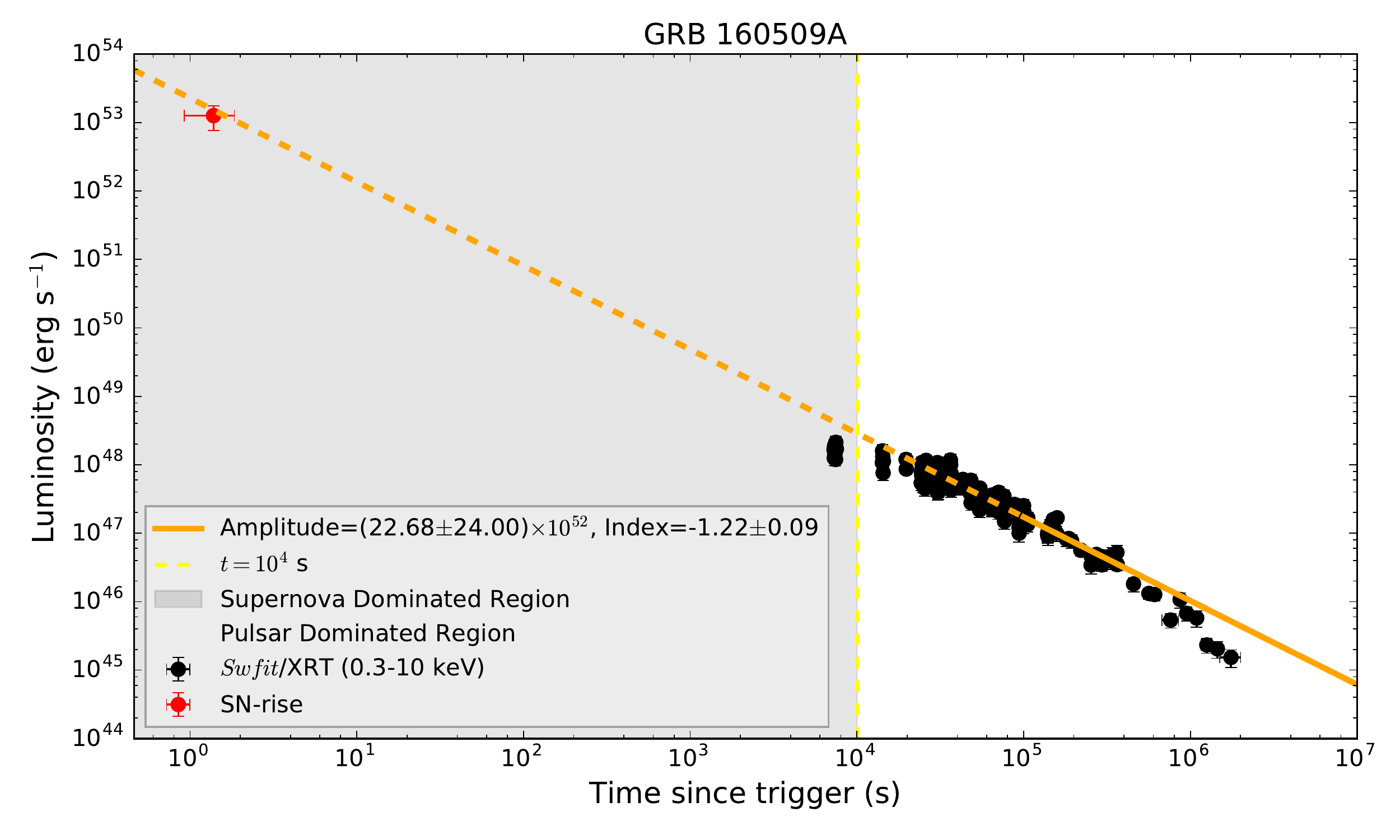}\\
    (c) &  (d)
\end{tabular}

\caption{X-ray afterglow luminosities of four BdHNe I observed by \textit{Swift}-XRT which follow a decaying power-law: \textbf{(a)}: GRB 130427A (BdHNe I) with amplitude $(3.65\pm 0.63)\times 10^{52}$~erg~s$^{-1}$ and power-law index $\alpha_X=1.24\pm 0.02$. 
\textbf{(b)}: GRB 190114C with amplitude $(5.14\pm 2.03)\times 10^{52}$~erg~s$^{-1}$ and power-law index $\alpha_X=1.37\pm 0.05$.    
\textbf{(c)}: GRB 180720B with amplitude $(112.67\pm 93.89)\times 10^{52}$~erg~s$^{-1}$ and power-law index $\alpha_X=1.43\pm 0.07$. 
\textbf{(d)}: GRB 160509A with amplitude $(22.68\pm 24.00)\times 10^{52}$~erg~s$^{-1}$ and power-law index $\alpha_X=1.22\pm 0.09$.  
The red points show the luminosity of SN-rise in each BdHN. The fallback of material from the SN onto the $\nu$NS produces this X-ray afterglow emission \citep{2020ApJ...893..148R}. As shown in \citet{2018ApJ...852...53R, 2018ApJ...869..151R}, till $\sim 10^4$~s the gamma/X-ray afterglow is mainly produced by the SN kinetic energy (\textit{SN dominated region}) and its interaction with the magnetic field of the $\nu$NS. After $ 10^4$~s, as shown by \citet{2018ApJ...869..101R}, the role of $\nu$NS becomes prominent (\textit{pulsar dominated region}).}
\label{fig:4bdhn}
\end{figure*}

\begin{figure*}
\centering
\includegraphics[width=1.0\hsize,clip]{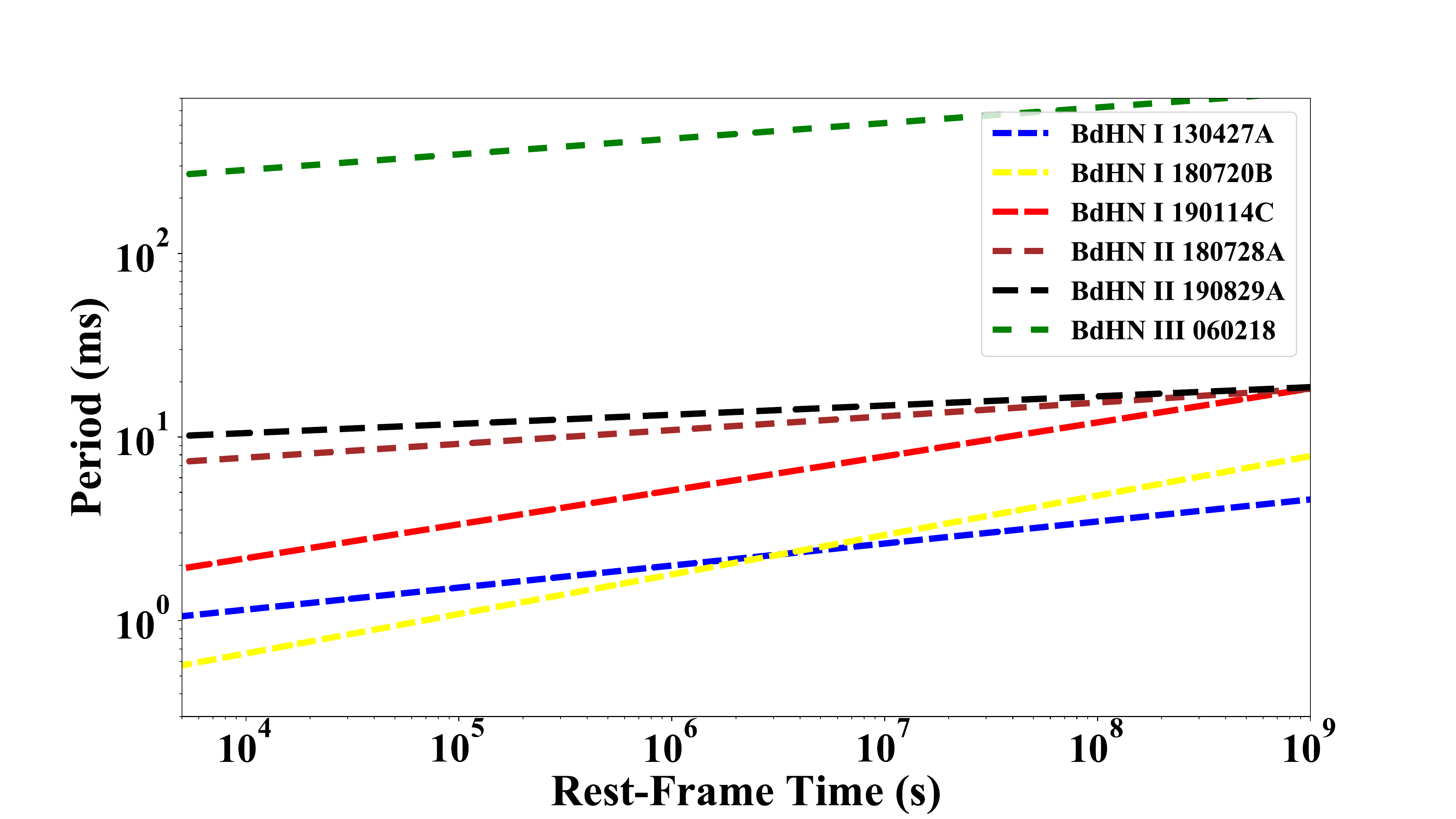}
\caption{The evolution of the $\nu$NS period of six BdHNe, as a function of rest-frame time. The values of the $\nu$NS period at $10^4$~s, namely in the pulsar dominated region of the afterglow are tabulated in Table~\ref{tab:ns_parameters}. The trend of the $\nu$NS period indicates that the rotational energy is being released due to the radiation losses in the keV band revealing itself as the X-ray afterglow luminosity.}
\label{fig:angdown} 
\end{figure*}

\section{The SN-rise in BdHN I and BdHN II: Fermi observation}\label{sec:snrise}

The trigger of all BdHNe is represented by the gravitational collapse of the CO$_{\rm core}$ which gives origin to a SN and its Fe-core collapses to form a $\nu$NS. We have indicated the first appearance of the SN as the SN-rise. In BdHN I, the SN-rise is characterized by the presence of the thermal component in the Fermi-GBM data with isotropic energy of $\sim 10^{52}~\rm erg$; see \citet{2019arXiv190404162R,2019arXiv191012615L,2014ApJ...793L..36F}. In BdHN II, the SN-rise is weaker and has no thermal component in the Fermi-GBM data with energy of $\sim 10^{50}~\rm erg$; see \citet{2019ApJ...874...39W,2019arXiv190404162R,2019arXiv191012615L}, Fig.~\ref{fig:sn-riseall} and Table~\ref{tab:shockbreakout}. In this article, we just recall the observation of the SN-rise in $4$ BdHNe I: GRB 130427A, GRB 160509A, GRB 180720B and GRB 190114C, as well as in two BdHNe II: GRB 180728A and GRB 190829A. In Fig.~\ref{fig:sn-riseall} we  show  the spectra of the SN-rise in the aforementioned sources and in Fig.~\ref{fig:4bdhn} we show their corresponding luminosity.

\section{The afterglows of BdHN I, BdHN II and BdHN III: The Swift observations}\label{sec:xrayafterglow}

Following the CO$_{\rm core}$ gravitational collapse and the appearance of the SN-rise, which characterizes all BdHN subclasses, the hypercritical accretion of the SN ejecta onto the $\nu$NS and the magnetic field of the $\nu$NS leads to the pulsar-like emission powering the X-ray afterglow observed by the Swift satellite \citep{2020ApJ...893..148R}. 

We present four afterglows of BdHN I (Fig.~\ref{fig:4bdhn}), two afterglows of BdHNe II, and one afterglow of BdHNe III (Fig.~\ref{fig:2bdhn2}). In each case, we also reproduce the SN-rise presented in the previous section; see Figs.~\ref{fig:2bdhn2} and \ref{fig:4bdhn}. 

The BdHN I in GRB 130427A, GRB 190114C, GRB 180720B and GRB 160509A follow a decaying luminosity consistent with Eq.~(\ref{eq:Lx}); see Fig.~\ref{fig:4bdhn}:
\begin{itemize}
    \item  
    GRB 130427A with  amplitude $(3.65\pm 0.63)\times 10^{52}$~erg~s$^{-1}$ and power-law index $\alpha_X=1.24\pm 0.02$;
    \item 
    GRB 160509A with amplitude $(22.68\pm 24.00)\times 10^{52}$~erg~s$^{-1}$ and power-law index $\alpha_X=1.22\pm 0.09$;
    \item 
    GRB 180720B with amplitude $(112.67\pm 93.89)\times 10^{52}$~erg~s$^{-1}$ and power-law index $\alpha_X=1.43\pm 0.07$;
    \item 
    GRB 190114C with amplitude $(5.14\pm 2.03)\times 10^{52}$~erg~s$^{-1}$ and power-law index $\alpha_X=1.37\pm 0.05$
\end{itemize}
 
The BdHNe II in GRB  180728A and GRB 190829A follow a decaying luminosity consistent with Eq.~(\ref{eq:Lx}) (see \citealp{2019ApJ...874...39W} and Wang, et al., in preparation); see Fig.~\ref{fig:2bdhn2}(a) and (b):
\begin{itemize}
\item 
GRB 180728A with amplitude $(2.19\pm 0.13)\times 10^{50}$~erg~s$^{-1}$ and power-law index $\alpha_X=1.15\pm 0.05$;
\item GRB 190829A with amplitude $(5.20\pm 0.89)\times 10^{49}$~erg~s$^{-1}$ and power-law index $\alpha_X=1.1\pm 0.1$. 
\end{itemize}

As an example of the X-ray afterglow luminosity of a BdHN III, we indicate the case of GRB 060218 where the X-ray luminosity, as in the case of BdHNe I and II, follows a decaying power-law consistent with Eq.~(\ref{eq:Lx}), with an amplitude $(2.19\pm 0.53)\times 10^{47}$~erg~s$^{-1}$ and power-law index $\alpha_X=1.17\pm 0.02$. This is consistent with $\alpha_X=1.2\pm 0.1$ obtained by \citet{2006Natur.442.1008C}; see Fig.~\ref{fig:2bdhn2}(c). 

We can then reach the following general conclusions:
\begin{enumerate}
    \item
    The X-ray afterglow is present in all three BdHN subclasses: BdHN I, BdHN II, BdHN III. 
    \item 
    The X-ray afterglow is always present in \emph{all} of the $378$ BdHNe I; see Appendix~\ref{updated}.
    \item 
    This result clearly indicates the spherical symmetry, or a very wide-angle emission of the X-ray afterglow.
\end{enumerate}

\begin{table*}[!ht]
\centering
\caption{Observational properties of three BdHN I, GRB 130427A, GRB 180720B and GRB 190114C together with two BdHNe II 180728A and 190829A as well as one BdHN III, GRB 060218 and inferred physical quantities of the $\nu$NS of the corresponding BdHN model that fits the GRB data. Column 1: GRB name; column 2: identified BdHN type; column 3: cosmological redshift ($z$); column 4: the isotropic energy released ($E_{\rm iso}$) in gamma-rays; column 5: $\nu$NS rotation period ($P_{\nu \rm NS}$) at 10$^4$~s, column 6: The isotropic energy of the  X-ray afterglow  ($E_{\rm X}$). In We assume the NS mass of $1.4 M_\odot$ and the NS radius of $10^{6}$~cm for all these cases.}
\label{tab:ns_parameters}
\small
\begin{tabular}{@{}cccccccc@{}}
\hline
GRB     & Type    & Redshift & $E_{\rm iso}$  & $P_{\nu \rm NS} @ 10^4$~s  & $E_{\rm X} (\rm after 10^4$~s) & $A_X$ & $\alpha_X$\\ 
        &         &          & (erg) & (ms) & (erg) & (erg/s)& \\ 
        \hline
130427A & BdHN I  & 0.34     & $9.2\times10^{53}$& 1.15  & $1.67\times10^{52}$ & $3.65\pm 0.63)\times10^{52}$& $1.24 \pm$ 0.02 \\
180720B & BdHN I  & 0.654     & $6.8\times10^{53}$& 0.66  & $4.99\times10^{52}$ &$(112.67\pm 93.89)\times10^{52}$& $1.43 \pm$ 0.07  \\
190114C & BdHN I  & 0.42     & $1.5\times10^{53}$& 2.19 & $4.60\times10^{51}$ &$(5.14\pm 2.03)\times10^{52}$& $1.37 \pm$ 0.05   \\
180728A & BdHN II & 0.117    & $2.3\times10^{51}$& 7.74  & $3.68\times10^{50}$ &$(2.19\pm 0.13)\times10^{50}$& $1.15 \pm$ 0.05  \\
190829A & BdHN II & 0.0785    & $2.2\times10^{50}$& 10.31  & $2.07\times10^{50}$ &$(5.20\pm 0.89)\times10^{49}$& $1.10 \pm$ 0.06  \\
060218 & BdHN III &0.033   & $5.4\times10^{49}$& 285.81  & $2.69\times10^{47}$ &$(2.19\pm 0.53)\times10^{47}$& $1.17 \pm$ 0.02  \\

\hline
\end{tabular}
\end{table*}

\subsection{The spin of the $\nu$NS}\label{sec:nuns1}

In \citet{2018ApJ...869..101R, 2020ApJ...893..148R} and \citet{2019ApJ...874...39W}, the bolometric luminosity contributing to the optical and X-ray bands by the $\nu$NS rotational energy loss by magnetic braking has been modeled for the emission at late times $t\gtrsim 10^4$~s of the ``Nousek-Zhang'' (flare-plateau-afterglow, FPA phase).  This allows the inference  of the initial rotation period of the $\nu$NS as well as its magnetic field structure. The origin of the long GRB afterglows at this phase is the interaction between the SN ejecta and the spinning magnetised $\nu$NS and their synchrotron emission \citep{2018ApJ...869..101R}.

Since the $\nu$NS is just born, it must be rapidly rotating and contains abundant rotational energy:
\begin{equation}
	E_{\rm rot} = \frac{1}{2} I \Omega^2, \label{roten}
\end{equation}
where $I$ is the moment of inertia, and $\Omega = 2\pi/P_{\nu\rm NS}$ is the angular velocity. For a $\nu$NS with a period of $P_{\nu\rm NS}$=1~ms, $M = 1.4 M_{\odot}$, $R = 10$~ km, the moment of inertia is $I \sim 10^{45}$~g~cm$^2$.  This leads to a total rotational energy of $E \sim 2 \times 10^{52}$~erg. 

We assume that the rotational energy of the $\nu$NS provides the energy budget of the X-ray radiation via synchrotron emission of the electrons \citep{2018ApJ...869..101R}:
\begin{equation}
	E_{\rm rot} = E_{\rm X}.
\end{equation}
This is reminiscent of the extraction of the BH rotational energy via synchrotron radiation of electrons radiating in the GeV energy band \citep{2019ApJ...886...82R}.

Therefore, using the Eq.~(\ref{roten}) and substituting the Eq.~(\ref{eq:Lx})
\begin{equation}
\frac{dE_X}{dt} =	L_{\text{X}}(t) = A_X \left(\frac{t}{\rm 1s}\right)^{-\alpha_x}  = -I \Omega  \dot{\Omega}. \label{eq:pulsar_luminosity}
\end{equation} 

The best fit to the X-ray luminosity of Eq.~(\ref{eq:Lx}), together with Eq.~(\ref{eq:pulsar_luminosity}), allow an estimate of the spin of the $\nu$NS in \textit{all} BdHNe, as well as their spin evolution; see Table~\ref{tab:ns_parameters} and Fig.~\ref{fig:angdown}. 

In Table~\ref{tab:ns_parameters}, we report the physical quantities of three BdHNe I, GRB 130427A, GRB 180720B and GRB 190114C, together with two BdHNe II, GRB 180728A and GRB 190829A, as well as one BdHN III, GRB 060218; assuming a $\nu$NS of mass and radius, respectively, $1.4 M_\odot$ and $10^{6}$~cm. The $\nu$NS emission is not able to explain the emission of the ``Nousek-Zhang'' phase at early times $10^2$--$10^4$~s. As it is shown in \citet{2018ApJ...869..101R,2018ApJ...869..151R}, that emission is mainly powered by the mildly-relativistic SN kinetic energy which we refer it to as the \textit{SN dominated region}.  After $ 10^4$~s, as shown by \citet{2018ApJ...869..101R}, the role of $\nu$NS  becomes prominent, referred to as the \textit{pulsar dominated region}. 

The first main results of this paper are: 1) the first identification of the SN-rise, 2) the agreement of the extrapolated luminosity of the X-ray afterglow with the luminosity of the SN-rise, and 3) the measurement of the $\nu$NS period, leading to  the power-law emission of the afterglow; see Fig.~\ref{fig:4bdhn}. The two process of the SN-rise energetics and the $\nu$NS dynamics appear to be strongly correlated. 

\section{ BdHN I: The Fermi-LAT observations}\label{sec:4}

\subsection{BdHNe I  observed by Fermi-LAT}

We now address the $378$ BdHNe I with known redshifts; see \citet{2016ApJ...833..159P,2018ApJ...852...53R} and Appendix~\ref{updated}: out of them, we are first interested in the $25$ BdHNe I emitting GeV radiation and within the boresight angle of \textit{Fermi}-LAT, i.e. $\theta < 75^{\circ}$, at the time of the trigger, since exposure drops quickly for larger angles \citep{2019ApJ...878...52A}. They have as well a TS value $>25$, which means the GeV photons are excluded at the $5$--$\sigma$ level from background sources. We follow the first and second \textit{Fermi} catalogs \citep{2013ApJS..209...11A, 2019ApJ...878...52A} for the time-resolved likelihood spectral analysis. Therefore, we divide the data into logarithmic spaced bins and, if the test statistic (TS) value of each bin is smaller than $16$, we merge the time bin with the next one and repeat the likelihood analysis. In Table~\ref{tab:cb}, we indicate in the first column the name of the BdHNe I, in the second their measured redshift, we report in the third column the $E_{\rm p,i}$ obtained from the \textit{Fermi} data, we estimate in the fourth column the $E_{\rm \gamma, iso}$, which is itself larger than the $10^{52}$~erg. In the fifth column, the \textit{Fermi} GCN numbers are shown. In the sixth column, the values of $E_{\rm LAT}$ are provided and finally we add the boresight angle of the LAT $\theta < 75^{\circ}$ and the TS values of these GRBs observed by LAT.

\begin{table*}
\centering
\begin{tabular}{llcccclr}
\hline\hline
GRB       	&  $z$     &  $E_{\rm p,i}$   &  $E_{\gamma, \rm iso}$    & Fermi GCN&  $E_{\rm LAT}$        &  $\theta$  &  TS     \\
        		&          &  (MeV)           &  ($10^{52}$~erg)  &			&  ($10^{52}$~erg)       &  (deg)     &        \\
\hline
080916C	&  $4.35$  &  $2.27\pm0.13$   &  $407\pm86$	      &$8246$&  $230\pm10$     &  $48.8$	&  $1450$  \\
090323A	&  $3.57$  &  $2.9\pm0.7$     &  $438\pm53$	      &$9021$&  $120\pm20$ &  $57.2$	&  $150$   \\
090328A	&  $0.736$ &  $1.13\pm0.08$   &  $14.2\pm1.4$     &$9044$&  $2.7\pm0.4$  &  $64.6$	&  $107$   \\
090902B	&  $1.822$ &  $2.19\pm0.03$   &  $292.0\pm29.2$   &$9867$&  $47\pm 2$	   	 &  $50.8$	&  $1832$  \\
090926A	&  $2.106$ &  $0.98\pm0.01$   &  $228\pm23$       &$9934$&  $149\pm 8$	     &  $48.1$  &  $1983$  \\
091003A	&  $0.897$ &  $0.92\pm0.04$   &  $10.7\pm1.8$	  &$9985$&  $0.8 \pm 0.3$  &  $12.3$	&  $108$   \\
091127	&  $0.49$ &  $0.05\pm 0.01$   &  $0.81\pm0.18$	  &$ 10204$&  $0.03\pm0.02$  &  $25.8$	&  $34$   \\
091208B	&  $1.063$ &  $0.25\pm0.04$   &  $2.10\pm0.11$	  &$10266$&  $\gtrsim0.41\pm0$  &  $55.6$	&  $20$   \\
100414A	&  $1.368$ &  $1.61\pm0.07$   &  $55.0\pm0.5$	  &$10594$	& $7\pm 1$  &  $69$	&  $81$	   \\
100728A	&  $1.567$ &  $1.00\pm0.45$   &  $72.5\pm2.9$	  &$11006$&  $0.9\pm0.3$  &  $59.9$	&  $32$	   \\
110731A	&  $2.83$  &  $1.21\pm0.04$   &  $49.5\pm4.9$	  &$12221$&  $15\pm 2$   &  $3.4$   &  $460$   \\
120624B &  $2.197$ &  $1.39\pm0.35$   &  $347\pm16$	      &$13377$	&  $22\pm2$  &  $70.8$	&  $312$	   \\
130427A	&  $0.334$ &  $1.11\pm0.01$   &  $92\pm13$	      &$14473$&  $8.6\pm0.4$  &  $47.3$  &  $163$   \\
130518A &  $2.488$ &  $1.43\pm0.38$   &  $193\pm1$	      &$14675$	&  $15\pm5 $  &  $41.5$	&  $50$	   \\
131108A &  $2.40$  &  $1.27\pm0.05$   &  $51.20\pm3.83$	  &$15464$	&  $37\pm 4$  &  $23.78$	&  $870$\\
131231A	&  $0.642$ &  $0.27\pm0.01$   &	 $21.50\pm0.02$	  &$15640$&  $1.6\pm0.3$  &  $38$	&  $110$   \\
141028A	&  $2.33$  &  $0.77\pm0.05$   &  $76.2\pm0.6$	  &$16969$&  $9\pm 2$  &  $27.5$	&  $104.5$ \\
150314A &  $1.758$ &  $0.86\pm0.01$   &  $70.10\pm3.25$	  &$17576$	&  $1.8 \pm0.7$  &  $47.13$	&  $27.1$  \\
150403A &  $2.06$  &  $0.95\pm0.04$   &  $87.30\pm7.74$	  &$17667$	&  $1.1 \pm 0.4$  &  $55.2$	&  $37$	 \\
150514A &  $0.807$ &  $0.13\pm0.01$   &  $1.14\pm0.03$	  &$17816$	&  $0.06\pm0.05$  &  $38.5$	&  $33.9$	 \\
160509A	&  $1.17$  &  $0.80\pm0.02$   &  $84.5\pm2.3$	  &$19403$&  $10\pm 1$ &  $32$	&  $234$   \\
160625B	&  $1.406$  &  $1.3\pm0.1$   &  $337\pm 1$	  &$19581$, $19604 $&  $17\pm 1$ &  $41.46$	&  $961.33$ \\ 
170214A	&  $2.53$  &  $0.89\pm0.04$   &  $392\pm 3$	  &$20675$, $20686 $&  $53\pm 4$ &  $33.2$	&  $1571$ \\ 
170405A	&  $3.51$  &  $1.20\pm0.42$   &  $241.01\pm 52.02  $	  &$20990$, $20986$&  $16\pm 7$ &  $52.0$	&  $56$ \\

180720B	&  $0.654$  &  $1.06\pm 0.24$   &  $68.2\pm 2.2  $	  &$22996 $, $23042$&  $2.2\pm 0.2$ &  $49.1$	&  $975$ \\
\hline
\end{tabular}
\caption{\textit{Prompt and GeV emission of the $25$ long GRBs inside the Fermi-LAT boresight angle and with GeV photons detected}. The columns list: the source name, $z$, $E_{\rm p,i}$, $E_{\gamma, \rm iso}$, $E_{\rm LAT}$ in $0.1$--$10$~GeV, the position of the source from the LAT boresight $\theta$, the likelihood TS. The $E_{\rm LAT}$ includes only the energy in the observed time duration, which does not cover the whole GeV emission period, and is different for each GRB, so we put a symbol '$\gtrsim$' to indicate that the value is the lower limit.}
\label{tab:cb}
\end{table*}

\begin{table*}
\centering
\begin{tabular}{llcclccl}
\hline\hline
GRB	& $z$       &	$E_{\rm p}$	        & $E_{\rm \gamma, iso}$     & Fermi GCN     & $\theta$  & GeV observed & comments \\
        &           &	(MeV)			        & ($\times 10^{52}$~erg)   &               & (deg)     &               & \\
\hline	          
081222  &   $2.77$  &   $0.51\pm0.03$                  &   $27.4\pm2.7$    &   8715    &   $50.0$  &   no      & \\
090424A &   $0.544$ &   $0.27\pm 0.04$                  &   $4.07\pm0.41$   &  9230    &   $71.0$  &   no      & \\
090516A	&	$4.109$	&	$0.14\pm0.03$	&	$99.6\pm16.7$	&	9415	&	$20.0$	&	no	    & Clear X-ray flare\\
100615A &   $1.398$ &   $0.21\pm 0.02$                  &   $5.81\pm0.11$   &    10851   &   $64.0$  &   no      & \\
100728B &   $2.106$ &   $0.32\pm 0.04$                  &   $3.55\pm0.36$   &    11015   &   $57.1$  &   no      & \\
110128A &   $2.339$ &   $0.46\pm 0.01$                    &   $1.58\pm0.21$   &    11628   &   $45.0$  &   no      & \\
111228A &   $0.716$ &   $0.060\pm0.007$                  &   $2.75\pm0.28$   &    12744   &   $70.0$  &   no      & \\
120119A &   $1.728$ &   $0.52\pm0.02$                  &   $27.2\pm3.6$    &    12874   &   $31.4$  &   no      & \\
120712A &   $4.175$ &   $0.64\pm 0.13$                  &   $21.2\pm2.1$    &    13469   &   $42.0$  &   no      & \\
120716A &   $2.486$ &   $0.4\pm 0.04$                  &   $30.2\pm3.0$    &    13498   &   $63.0$  &   no      & \\
120909A &   $3.93$  &   $0.87\pm 0.01$                   &   $87\pm10$       &    13737   &   $66.0$  &   no      & \\
130528A &	$1.250$	&	$0.27\pm0.18$	&	$18.01\pm2.28$	&	 14729	&	$60.0$	&	no	    & X-ray flare\\
130925A &	$0.347$	&	$0.14\pm0.04$		&	$3.23\pm0.37$	&	 15261	&	$22.0$	&	no	    & X-ray flare\\
131105A &   $1.686$ &   $0.55\pm0.08$       &   $34.7\pm1.2$    &    15455   &   $37.0$  &   no      & \\
140206A&   $2.73$  &   $1.1\pm0.03$	&	$144.24\pm19.20$&    15790   &   $46.0$  &   no   & Clear X-ray flare\\
140213A &   $1.2076$&   $0.176\pm0.004$       &   $9.93\pm0.15$   &    15833   &   $48.5$  &   no      & \\
140423A &   $3.26$  &   $0.53\pm0.04$       &   $65.3\pm3.3$    &    16152   &   $44.0$  &   no      & \\
140623A &   $1.92$  &   $1.02\pm0.64$        &   $7.69\pm0.68$   &    16450   &   $32.0$  &   no      & \\
140703A &   $4.13$  &   $0.91\pm0.07$        &   $1.72\pm0.09$   &    16512   &   $16.0$  &   no      & \\
140907A &   $1.21$  &   $0.25\pm0.02$	&	$2.29\pm0.08$   &    16798   &   $16.0$  &   no      & X-ray flare\\
141220A &   $1.3195$&   $0.42\pm0.02$       &   $2.44\pm0.07$   &    17205   &   $47.0$  &   no      & \\
150301B &   $1.5169$&   $0.45\pm0.10$   &   $2.87\pm0.42$   &    17525   &   $39.0$  &   no      & \\
150821A &   $0.755$ &   $0.57\pm0.03$   &   $14.7\pm1.1$    &    18190   &   $57.0$  &   no      & \\
151027A &   $0.81$  &   $0.62\pm0.11$	&	$3.94\pm1.33$   &    18492   &   $10.0$  &   no      & Clear X-ray flare\\
151111A &   $3.5$   &   $0.25\pm0.04$	&	$3.43\pm1.19$   &    18582   &   $50.0$  &   no      & X-ray flare observed\\
161014A &   $2.823$ &   $0.64\pm0.06$   &   $10.1\pm1.7$    &    20051   &   $69.0$  &   no      & \\

171222A &   $2.409$ &   $0.1\pm0.01$   &   $20.73\pm1.7$    &    22272, 22277   &   $43$  &   no      & \\

180703A &   $0.67$ &   $0.58\pm0.05$   &   $3.15\pm0.7$    &    23889, 22896   &   $44$  &   no      & \\

180728A &   $0.117$ &   $0.1\pm0.02$   &   $3.15\pm0.7$    &    23055, 23067   &   $35$  &   no      & \\
\hline
\end{tabular}
\caption{\textit{{List of $29$ BdHNe I inside the Fermi-LAT boresight angle and with no GeV photons detected}}: $29$ BdHNe I with redshift taken from \citep{2016ApJ...832..136R} from 2008, when \textit{Fermi} started to operate, till the end of 2016. All of them are within the boresight of Fermi-LAT, but no detected GeV photons. For each source the columns list: $z$, $E_{\rm \gamma, iso}$, $E_{\rm p}$, GCN number, position of the source from LAT boresight $\theta$, whether there was a detection by LAT, and additional information.}
\label{tab:BdHNe_No_GeV}
\end{table*}

\subsection{BdHNe I without GeV emission and geometry of the BdHNe I}\label{sec:10}

We now turn to an additional unexpected result obtained in the analysis of the BdHNe I subtended within the $75^{\circ}$ of the Fermi-LAT boresight angle: the existence of $29$ BdHNe I without observed GeV emission, see Table~\ref{tab:BdHNe_No_GeV}. Although the distribution of the boresight angle and redshift is analogous to the one of the $25$ sources considered in section~\ref{sec:4}, no GeV emission is observed.

Some BdHNe I of this group have been observed previously by \textit{Swift} and have been identified as sources of a) gamma and hard X-ray flares, b) soft X-ray flares and of c) the extended thermal emission \citep[see][for details]{2018ApJ...852...53R}. A particular example has been given by GRB 151027A in \citet{2017A&A...598A..23N} and \citet{2018ApJ...869..151R}. There, we assumed that the viewing angle of these sources lies in the equatorial plane of the progenitor system; see section~\ref{sec:1} and Fig.~\ref{fig:SPHsimulation}. As we will show in this article, in none of these sources GeV radiation can be observed due to the new morphology discovered in the BdHNe I; see next section. 
\section{Morphology of BdHN I}\label{sec:morphology}

We here assume that the $25$ sources considered in Table~\ref{tab:cb}, all emitting in the GeV have a viewing angle close to the normal of the plane. This assumption is confirmed in \citet{2019ApJ...886...82R} where indeed the high energy GeV-TeV radiations are emitted in direction close the BH rotation axis.

The remaining $29$ sources in Table~\ref{tab:BdHNe_No_GeV}  have a viewing angle in the equatorial plane of the binary progenitor and in that case only flaring activities in gamma and X-ray are observable, i.e. no GeV-TeV emission, as explicitly shown in \citet{2018ApJ...869..151R, 2019ApJ...886...82R}.  This allows us to introduce a new morphology for the BdHNe I and predict specific observational properties.

\begin{figure*}
\centering
\includegraphics[width=0.49\hsize,clip]{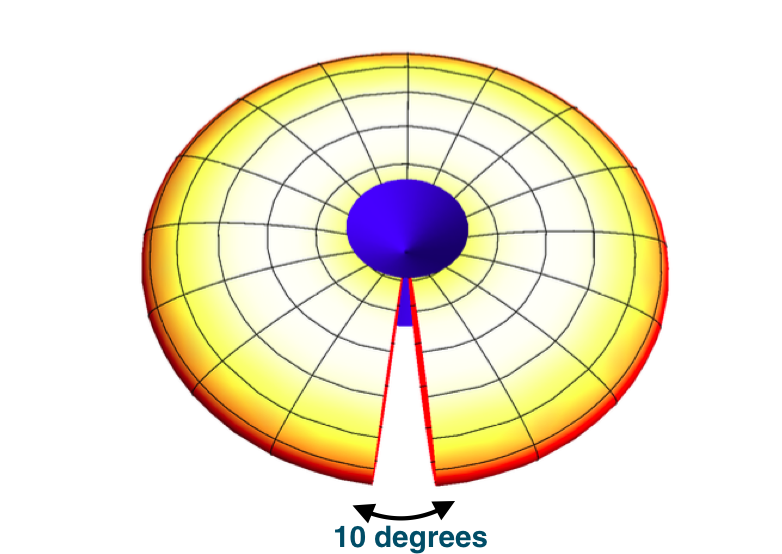}
\includegraphics[width=0.49\hsize,clip]{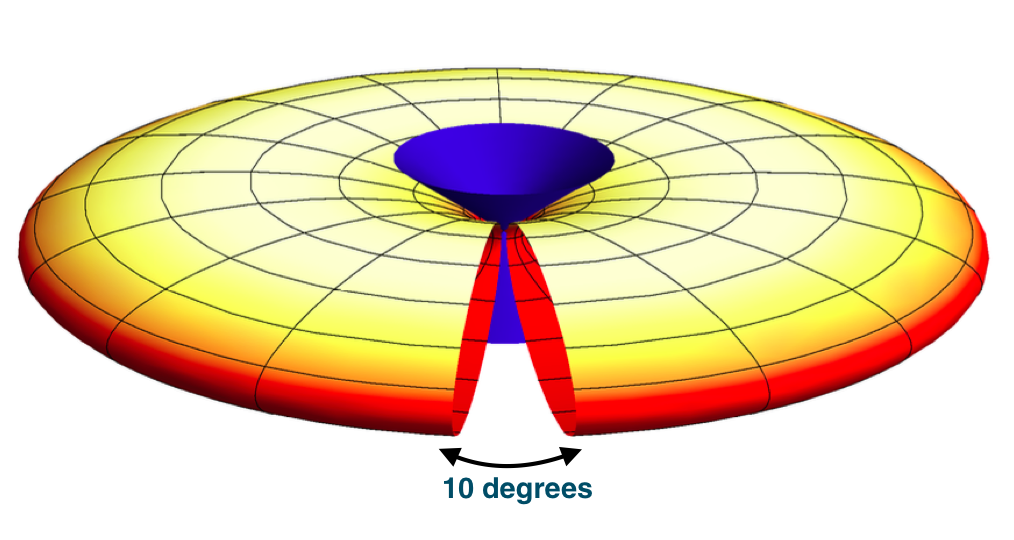}
\caption{Schematic plot for showing the morphology of the BdHNe I. The GeV emission is detectable when the viewing angle is less than the $60^{\circ}$ from the normal to the orbital plane. Left panel is the situation in which the detectors can observe {GeV and  Prompt}  emissions and the right panel is the one for which GeV emission is not detectable and only Gamma-ray and X-ray flares are detectable. The $10^{\circ}$ cuts in both figures indicate the low density region in Fig~\ref{fig:cc} through which the prompt radiation phase can be ``seen in the orbital plane''. The existence of such a $10^{\circ}$ cut was first identified by the SPH simulation quoted in \citet{2016ApJ...833..107B,2019ApJ...871...14B} and further confirmed in GRB 151027A \citep{2018ApJ...869..151R}.}
\label{fig:dd}
\end{figure*}

\begin{figure*}
\centering
\includegraphics[width=0.85\hsize,clip]{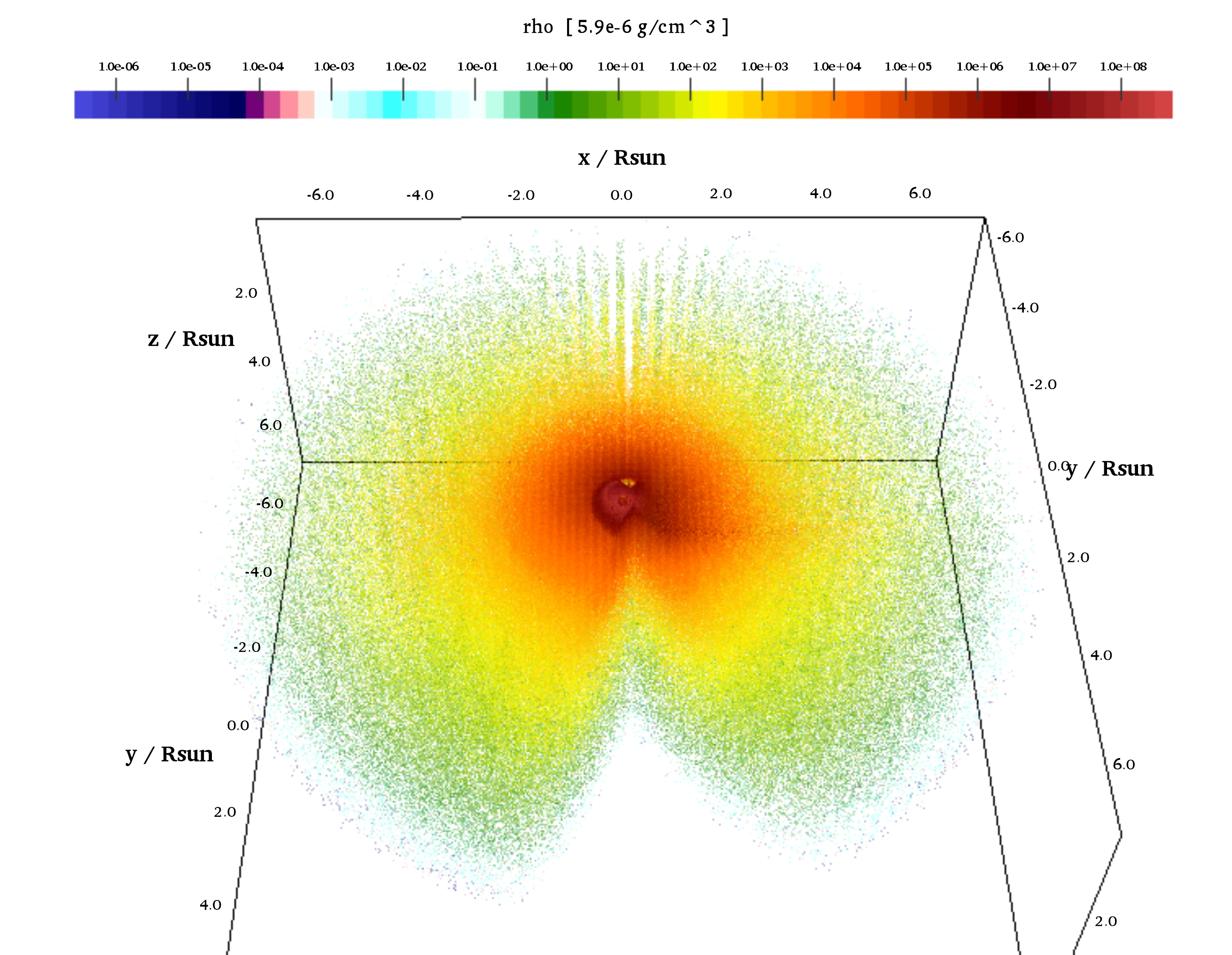}
\caption{Three-dimensional, half hemisphere views of the density distribution of the SN ejecta at the moment of BH formation in a BdHN I. The simulation is performed with a SPH code that follows the SN ejecta expansion under the influence of the NS companion gravitational field including the effects of the orbital motion and the changes in the NS gravitational mass by the hypercritical accretion process. The initial conditions of the SN ejecta are set as a homologous velocity distribution in free expansion and the mass-distribution is modeled with $16$ millions point-like particles \citep[see][for additional details]{2016ApJ...833..107B,2019ApJ...871...14B}. The binary parameters of this simulation are: the NS companion has an initial mass of $2.0~M_\odot$; the CO-star, obtained from a progenitor with zero-age main-sequence (ZAMS) mass $M_{\rm ZAMS}=30~M_\odot$, leads to a total ejecta mass of $7.94~M_\odot$ and to a $1.5~M_\odot$ $\nu$NS, the orbital period is $P\approx 5$~min (binary separation $a\approx 1.5\times 10^{10}$~cm). The distribution of the ejecta is not axially symmetric; it is strongly influenced by the rotation of the system and accretion occurring in the binary component, see Fig.~\ref{fig:SPHsimulation}. Particularly relevant for the observations is the low density region of $\approx 10^{\circ}$ which allows, for the sources with viewing angle in the equatorial plane to detect the prompt radiation phase. This has been qualitatively indicated in Fig.~\ref{fig:dd}. In these sources,  only a fraction of approximately 10$\%$ of the prompt radiation can be detectable, they are the only ones able to trigger the \textit{Fermi}-GBM and the remaining 90$\%$ will not have detectable prompt radiation, see \citet{2018ApJ...869..151R}. Figure is taken from \citet{2018ApJ...869..151R} with the kind permission of the authors.}
\label{fig:cc}
\end{figure*}

We now look at the ratio between the number of GRBs with an observed GeV radiation, $N_{\rm LAT}$, and the total number of GRBs, $N_{\rm tot}$, both within the LAT $75^{\circ}$ boresight angle. We assume that: 1) BdHNe I follow the same cosmological isotropic distribution of all GRBs first observed by the BATSE instrument onboard the CGRO satellite \citep[see, e.g.,][]{1992Natur.355..143M,1999ApJS..122..465P}; 2) all orientations of the BdHNe I with respect to the LAT detector are equally probable; 3) the GeV emitting region is a two-side cone whose opening angle is the same for all sources. Under these assumptions, we can then estimate the half-opening angle of a single cone $\vartheta$ as:
\begin{equation}
\label{angle}
1-\cos\vartheta = \frac{N_{\rm LAT}}{N_{\rm tot}}.
\end{equation}

Our search in the LAT data\footnote{\url{https://fermi.gsfc.nasa.gov/ssc/observations/types/grbs/lat_grbs/table.php}} gives $N_{\rm LAT}=25$ and $N_{\rm tot}=54$, leading to $\vartheta\approx 60^\circ$. Therefore, in BdHN I the GeV emission comes from a wide-angle emission, as it is schematically shown in Fig.~\ref{fig:dd}. This is in agreement with theory of synchrotron radiation produced around the Kerr BH along the rotation axis \citep[see details in][]{2019ApJ...886...82R}.

Therefore, we have identified a {\it new} morphology of the BdHN I (see Figs.~\ref{fig:dd} and \ref{fig:cc}). The identification of this morphology has been possible thanks to the analysis of the GeV emission in the present paper, by the soft and hard X-ray flares in \citet{2018ApJ...852...53R}, the extended thermal emission in \citet{2017A&A...598A..23N,2018ApJ...852...53R} in GRB 151027A. In this identification, we have been guided by the large number of numerical simulations describing the accretion of the SN ejected material around the NS companion; see Figs.~\ref{fig:SPHsimulation} and \ref{fig:cc}, and its idealized representation in Fig.~\ref{fig:dd}; \citep[see][for additional details]{2016ApJ...833..107B,2019ApJ...871...14B}.

What can be concluded from the above results is that in BdHNe I, the GeV emission is only detectable when the viewing angle is less than  $\approx 60^\circ$ from the normal to the plane and the BdHN I is ``seen from the top'', see the left plot in Fig.~\ref{fig:dd}. Whenever the viewing angle is within $60^\circ$ from the orbital plane, no GeV emission is observed, though X-ray and gamma-ray flares are observed, see right plot in Fig.~\ref{fig:dd}. 

Therefore, the second main result of this paper is the identification of the BdHN I morphology and its explanation within the BdHN I model.



\section{SPH simulation of BdHNe I}\label{sec:11}

The numerical simulations at the moment of BH formation in a BdHN I is presented in \citet{2016ApJ...833..107B, 2019ApJ...871...14B}. Three-dimensional (3D) views of the density distribution at the moment of the BH formation in a BdHN I are shown Fig.~\ref{fig:cc}. These plots correspond to the simulation of the SN ejecta expansion in the presence of the NS companion. The simulation is performed using an SPH code in which the SN ejecta material is evolved with $N$ point-like particles, in the present case 16 million, with different masses and  their motion is followed under the NS gravitational field. The orbital motion of the NS around the SN explosion center is also taken into account as the NS star gravitational mass changes via the hypercritical accretion process. The latter was modeled independently estimating the accretion rate onto the NS via the Bondi-Hoyle formalism. For the initial conditions of the simulation an homologous velocity distribution in free expansion was adopted and a power-law initial density profile of the SN matter was modeled by populating the inner layers with more particles \citep[see][for additional details]{2014ApJ...793L..36F, 2016ApJ...833..107B,2019ApJ...871...14B}. Figures~\ref{fig:SPHsimulation} and~\ref{fig:cc} correspond to an initial binary system formed by a $2\,M_\odot$ NS and the CO$_{\rm core}$ obtained from a progenitor with $M_{\rm ZAMS}=30\,M_\odot$. When the CO$_{\rm core}$ collapses and explodes, it ejects $7.94\, M_\odot$ and leads a $\nu$NS of $1.5\,M_\odot$ at its center. The initial binary period is about $5$~min, corresponding to a binary separation of $\approx 1.5\times 10^{10}$~cm. 

The new morphology of the BdHNe I presented here and in the previous section leads to a difference in the observed energy spectra and time variability for sources with viewing angle in the plane, or normal to the orbital plane of the binary progenitor. We infer that our $25$ BdHNe I, with viewing angles less than $ \approx 60^{\circ}$ from the normal to the orbital plane of the binary progenitor, ``seen from the top'', have larger  $E_{\gamma, \rm iso}$ than the ones with a viewing angle lying in the plane of the binary system (see Tables~\ref{tab:cb} and \ref{tab:BdHNe_No_GeV}). This explains the association/non-association of the GeV emission with bright GRBs often mentioned in the current literature (see \citealp{2011ApJ...738..138R,2011ApJ...732...29C} and Fig.~4 in \citealp{2018arXiv180401524N}).

An additional issue in the traditional approach (see e.g.~\citealp{2015MNRAS.454.1073B, 2011ApJ...738..138R} and sections 3 and 4 in \citealp{2018arXiv180401524N}) is also solvable: the sources that are seen with a viewing angle lying in the orbital plane have stronger flaring activities in the X-ray afterglow when compared to the $25$ emitting in the GeV range. Therefore, the ratio between $E_{\rm iso}$ and the luminosity in the X-ray afterglow is systematically smaller than in the $25$ with GeV emission. This offers a different explanation than the one presented in the traditional approach. However, all of these matters that have already been mentioned in \citet{2018ApJ...869..151R} need a new operational definition of $E_{\rm \gamma, iso}$, taking into due account the hard and soft X-ray flares and the extended thermal emission; see also \citet{2019ApJ...883..191R}.

Another important specific feature of the new morphology of BdHN I is the presence of the $\nu$NS formed at the center of the exploding SN (see Fig.~\ref{fig:SPHsimulation} and \citealp{2019ApJ...871...14B,2016ApJ...833..107B}). We have shown that the $\nu$NS manifests itself through the synchrotron emission by relativistic electrons injected from it into the expanding magnetized SN ejecta, as well as through its pulsar emission which explain the early and late optical and X-ray afterglow, respectively, allowing the inference of the $\nu$NS rotation period (see \citealp{2018ApJ...869..101R}). A smoking gun of this picture, namely the verification of the $\nu$NS activity following the above mechanism, both in XRFs (BdHNe II) and in BdHNe I, and the connection of the inferred rotation period of the $\nu$NS to the one of the CO-star and to the orbital period, from angular momentum conservation, has been explicitly shown in the GRB 180728A (BdHN II) and GRB 130427A (BdHN I) and GRB 190114C (BdHN I); see \citet{2019ApJ...874...39W} for details.
%

\section{The Luminosity power-law behavior in BdHNe  measured in the rest-frame}\label{sec:8}

\begin{figure*}
\centering
\includegraphics[width=1\hsize,clip]{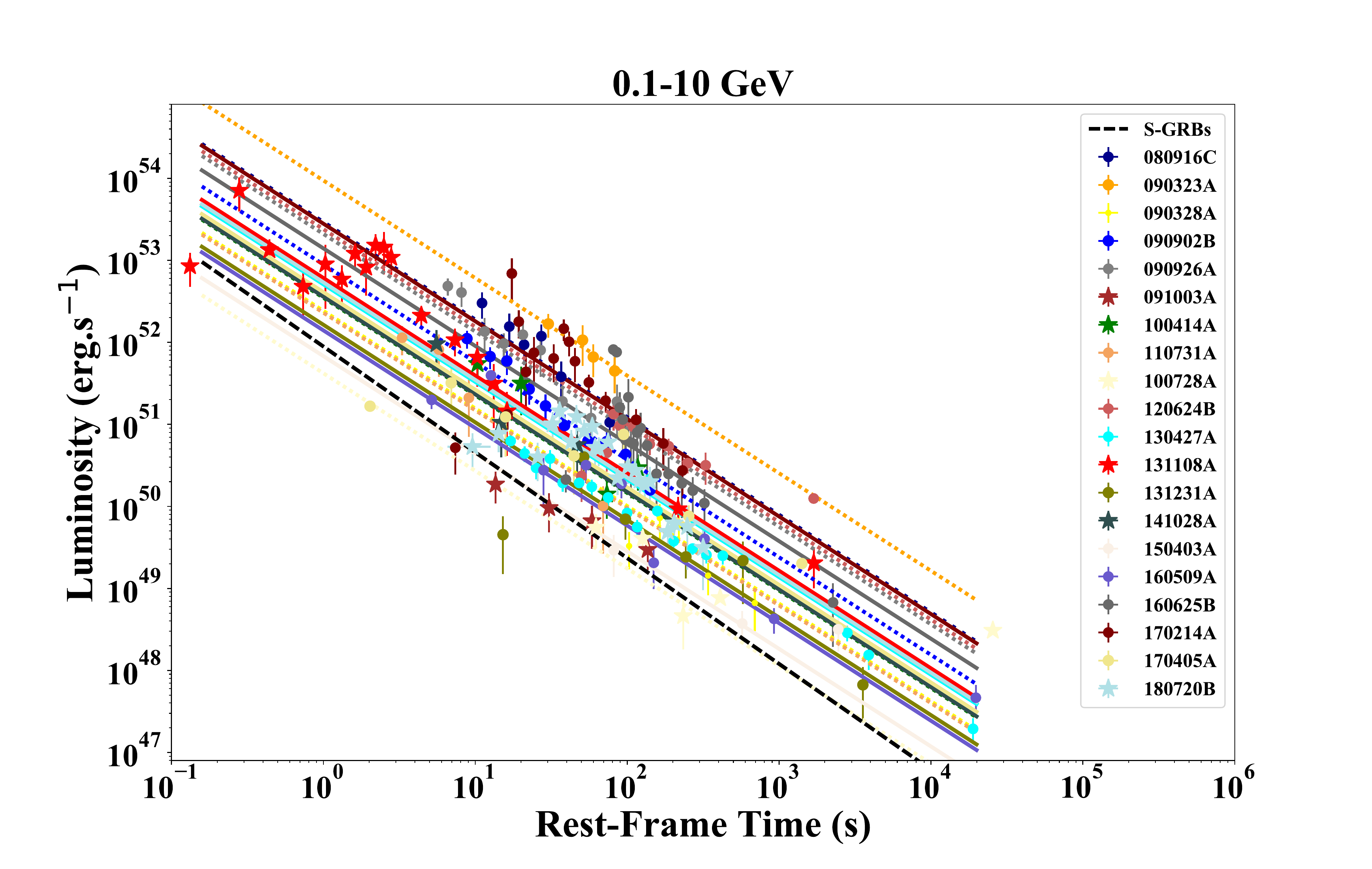}
\caption{The rest-frame $0.1$--$10$~GeV isotropic luminosity of $20$ selected BdHNe with LAT emission. The solid red line marks the common power-law behavior of the GeV emission for BdHNe with slope $\alpha_{\rm GeV}=1.19\pm0.04$; the shaded gray area encloses all the luminosity light-curves of the selected BdHNe. The dashed black line  marks the common power-law behavior of the GeV emission in S-GRBs with a slope of $\gamma=1.29 \pm0.06$.}
\label{fig:02}
\end{figure*}
\begin{table*}
\centering
\begin{tabular}{ccccc}
\hline\hline
BdHN & $A_{n}$ (Amplitude) & uncertainty of $A_{n}$  & $L_{10 \rm s}$ & uncertainty of $L_{10s}$\\
\hline
\hline
080916C & $2.9\times10^{53}$ & $^{+9.1}  _{-7.4} \times10^{52}$ & $1.88\times10^{52}$ & $^{+1.1}  _{-1.0} \times10^{52}$ \\
090323A & $9.4\times10^{53}$ & $^{+3.5}  _{-2.9} \times10^{53}$ & $6.04\times10^{52}$ & $^{+3.8}  _{-1.4} \times10^{52}$ \\
090328A & $2.4\times10^{52}$ & $^{+1.1}  _{-0.7} \times10^{52}$ & $1.5\times10^{51}$ & $^{+1.0}  _{-0.9} \times10^{51}$  \\
090902B & $8.9\times10^{52}$ & $^{+2.5}  _{-2.0} \times10^{52}$ & $5.7\times10^{51}$& $^{+3.3}  _{-3.0} \times10^{51}$\\
090926A & $2.1\times10^{53}$ & $^{+5.9}  _{-4.8} \times10^{52}$ & $1.4\times10^{52}$ & $^{+7.9}  _{-7.3} \times10^{51}$ \\
091003A & $5.7\times10^{51}$ & $^{+1.7}  _{-1.5} \times10^{51}$ & $3.7\times10^{50}$ & $^{+2.1}  _{-2.0} \times10^{50}$ \\
100414A & $3.5\times10^{52}$ & $^{+1.4}  _{-1.1} \times10^{52}$ & $2.3\times10^{51}$ & $^{+1.4}  _{-1.3} \times10^{51}$\\
100728A & $4.2\times10^{51}$ & $^{+1.9}  _{-1.5} \times10^{51}$ & $2.7\times10^{50}$ & $^{+1.9}  _{-1.6} \times10^{50}$ \\
110731A & $2.3\times10^{52}$ & $^{+0.8}  _{-0.5} \times10^{52}$ & $1.8\times10^{51}$& $^{+0.9}  _{-0.8} \times10^{51}$ \\
120624B & $2.4\times10^{53}$ & $^{+8.2}  _{-6.2} \times10^{52}$ & $1.6\times10^{52}$ & $^{+9.6}  _{-8.5} \times10^{51}$\\
130427A & $5.1\times10^{52}$ & $^{+2.1}  _{-2.0} \times10^{51}$ & $3.3\times10^{51}$ & $^{+1.3}  _{-1.3} \times10^{51}$\\
131108A & $6.1\times10^{52}$ & $^{+9.1}  _{-8.9} \times10^{51}$ & $3.9\times10^{51}$  & $^{+2.0}  _{-1.9} \times10^{51}$\\
131231A & $1.64\times10^{52}$ & $^{+7.9}  _{-5.4} \times10^{51}$ & $1.1\times10^{51}$& $^{+7.3}  _{-6.1} \times10^{50}$ \\
141028A & $3.6\times10^{52}$ & $^{+1.2}  _{-1.1} \times10^{52}$ & $2.3\times10^{51}$& $^{+1.4}  _{-1.3} \times10^{51}$ \\

150403A & $6.8\times10^{51}$ & $^{+3.0}  _{-2.3} \times10^{51}$ & $4.3\times10^{50}$ & $^{+2.9}  _{-3.0} \times10^{50}$\\
160509A & $1.4\times10^{52}$ & $^{+4.9}  _{-3.8} \times10^{51}$ & $8.9\times10^{50}$& $^{+5.4}  _{-4.1} \times10^{50}$ \\
160625B & $1.4\times10^{53}$ & $^{+4.6}  _{-3.4} \times10^{52}$ & $8.7\times10^{51}$ & $^{+5.2}  _{-4.6} \times10^{51}$\\
170214A & $2.8\times10^{53}$ & $^{+7.4}  _{-5.9} \times10^{52}$ & $1.8\times10^{52}$ & $^{+1.0}  _{-0.9} \times10^{52}$\\
170405A & $4.1\times10^{52}$ & $^{+1.1}  _{-1.0} \times10^{52}$ & $2.5\times10^{51}$ & $^{+1.5}  _{-1.4} \times10^{51}$\\
180720B & $5.4\times10^{52}$ & $^{+6.6}  _{-6.1} \times10^{51}$ & $3.5\times10^{51}$ & $^{+2.2}  _{-2.1} \times10^{50}$\\
\hline
\hline
\end{tabular}
\caption{\textit{Fitting parameters of the $0.1$--$10$~GeV power-law luminosity when measured in the rest-frame of $20$ BdHNe with GeV emission}: amplitude of the $0.1$--$10$~GeV luminosity, $A_n$, and its uncertainty, the inferred $0.1$--$10$~GeV luminosity at $10$~s from the fitting and its uncertainty. The common power-law index is $\alpha_{\rm GeV}= 1.19\pm 0.04$. Out of $25$ BdHNe emitting GeV emission, we performed the fitting for $20$ GRBs which have more than two data points in their luminosity light-curves. GRBs 091127, 091208B, 130518A, 150314A, 150514A have only two data points in their GeV luminosity light-curves.}
\label{tab:23fit}
\end{table*}

In the following, we fit simultaneously the luminosity light-curves of all the $25$ BdHNe with GeV emission expressed in their rest-frame. We assume the same power-law decay index for all of them, but allow different amplitude values. This assumption is consistent with our model, moreover, it is a benefit for those GRBs with limited data that cannot be fitted solely.

We limit our analysis of the light-curves after than the BdHN I prompt emission, when the GeV luminosity is already in the asymptotic power-law regime. We assume the power-law
\begin{equation}
	L_n(t) = A_n t^{\alpha_{\rm GeV}},
\end{equation}
describing the rest-frame $0.1$--$100$~GeV isotropic luminosity light-curve of $n$th BdHN I. In the simultaneous fitting, we perform the Levenberg-Marquardt method to perform the minimization \citep{Levenberg-Marquardt}. The basic idea of fitting is to minimize the $\chi^2$; when fitting one curve to one equation, the $\chi^2$ is minimized. To fit $N$ curves to $N$ equations simultaneously, the sum of the $\chi^2$ values should to be minimized. The related equations are:
\begin{align}
\chi^2 &= \sum_{n=1}^{N}~\chi^2 _n,\label{eq1} \\
\chi^2 _n &= \sum_{i=1}^{M} \frac{1}{\sigma_{ni}^2}(L_{ni}-L_n(t_{ni},A_n, \alpha_{\rm GeV}))^2,\label{eq2}
\end{align}
where $n$ represents each BdHN I, $i$ represents each data point in a given BdHN I, $A_n$ is the amplitude of a power-law function for the $n$th BdHN I, $\alpha_{\rm GeV}$ is the common power-law index shared for all the BdHNe I. Thus, for the $n_{\rm th}$ BdHN I, at time $t_{ni}$, the observed luminosity is $L_{ni}$, and the predicted luminosity is $L_n(t_{ni},A_n, \alpha)$. The value of $\chi^2$ represents the difference between the best power-laws fitting and all the observed data, it is a summation of individual $\chi^2 _n$, which represents the difference between the power-law fitting and the observed value of each GRB.

\begin{figure*}
\centering
\includegraphics[width=0.45\hsize,clip]{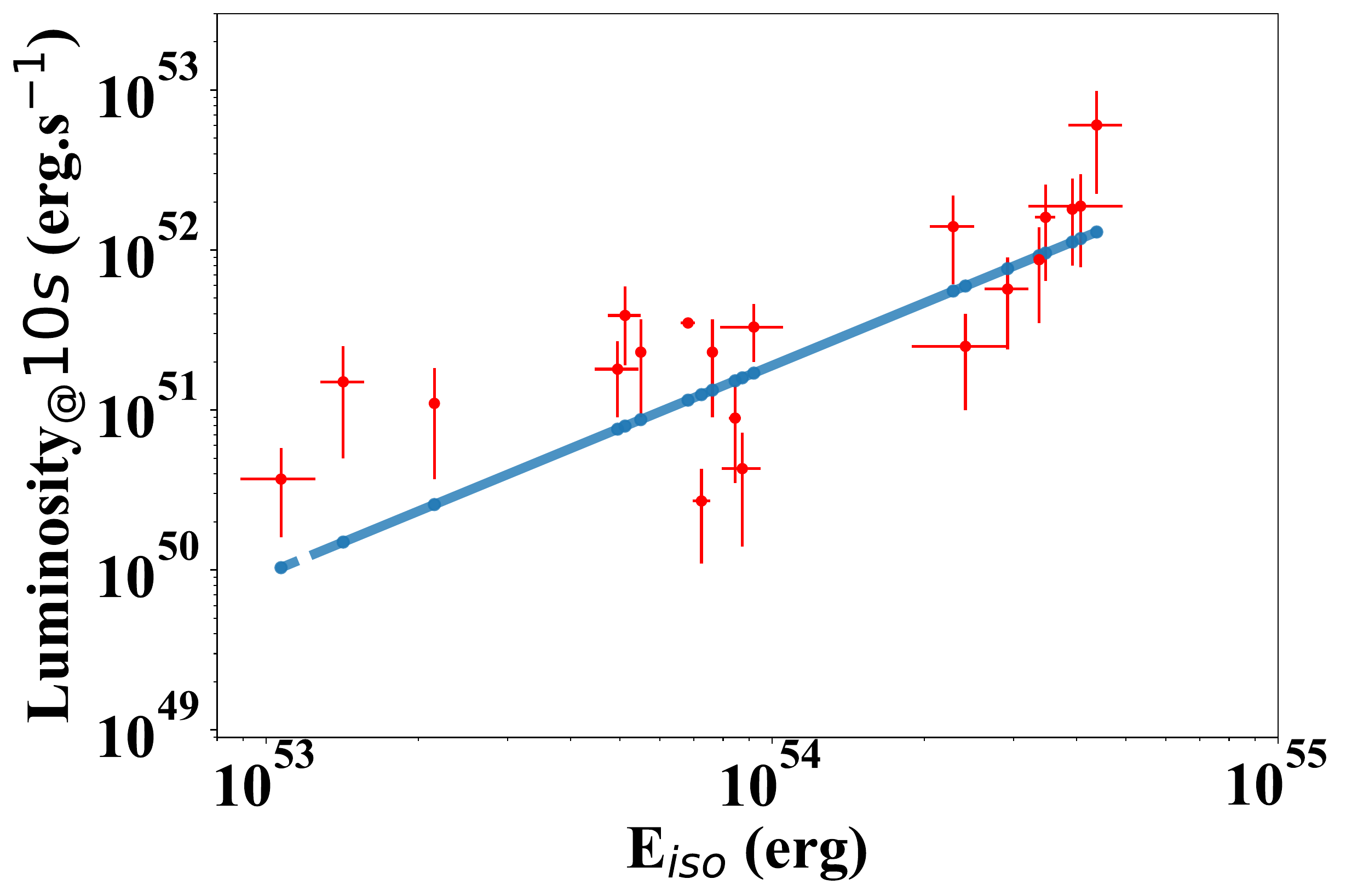}
\includegraphics[width=0.45\hsize,clip]{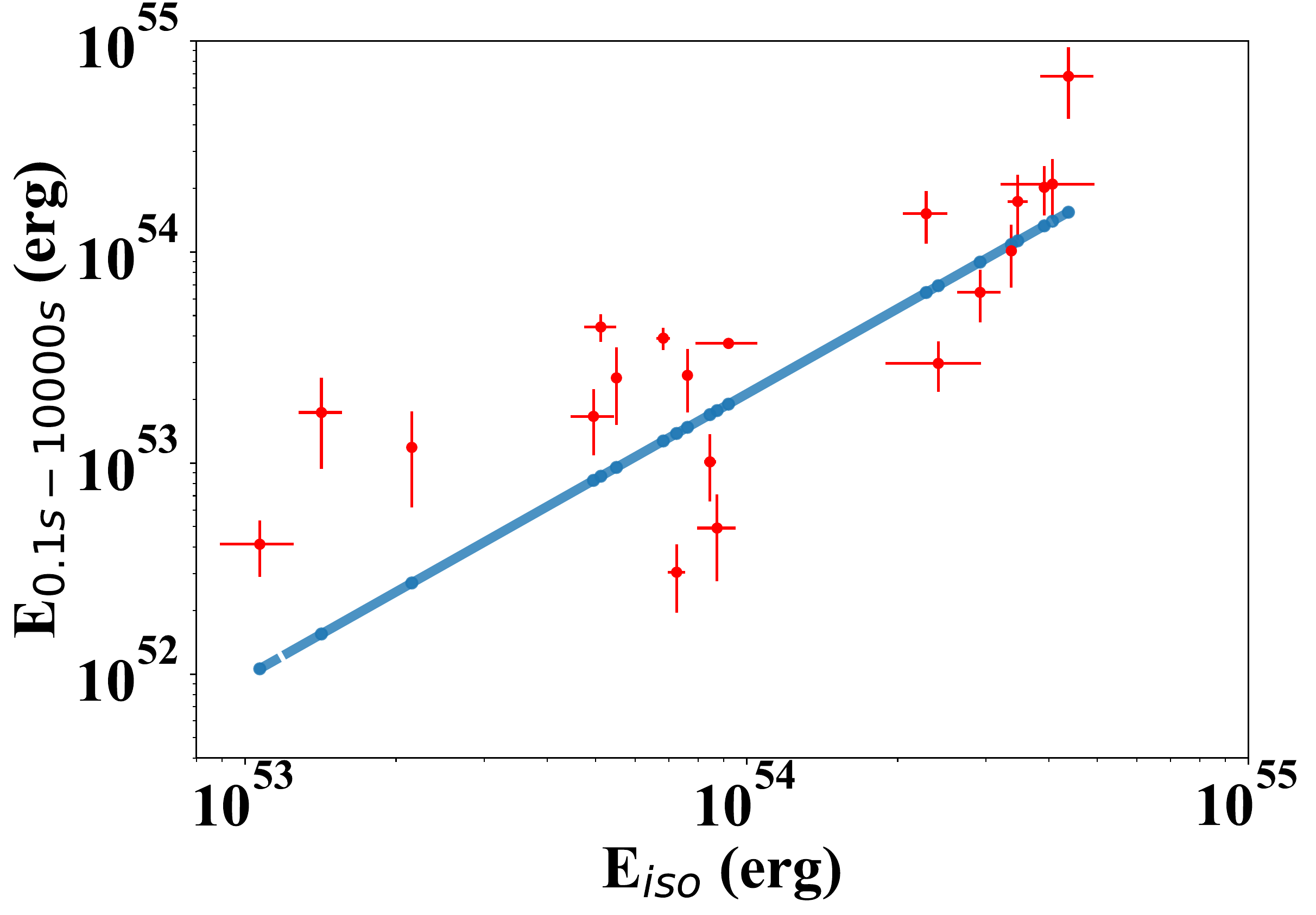}
\caption{Left: the \textit{Fermi}-LAT luminosity at $10$~s in the energy range $0.1$--$10$~GeV versus the isotropic gamma-ray energy from $1$~keV to $10$~MeV. The BdHNe are listed in Table~\ref{tab:23fit}. Right: the \textit{Fermi}-LAT energy from $0.1$ to $10^4$~s versus isotropic gamma-ray energy from $1$ keV to $10$~MeV. See the corresponding values in Table~\ref{tab:14fit}.}
\label{K10Eiso}
\label{EEiso}
\end{figure*}

\begin{table*}
\centering
\begin{tabular}{cccc}
\hline\hline
BdHN & $E_{0.1-10^{4}\rm s}$ &  uncertainty of $E_{0.1-10^{4}\rm s}$ \\
\hline
\hline
080916C & $2.1\times10^{54}$ & $6.6\times10^{53}$ \\
090323A & $6.8\times10^{54}$ & $2.5\times10^{54}$ \\ 
090328A & $1.73\times10^{53}$ & $7.9\times10^{52}$ \\
090902B & $6.4\times10^{53}$ & $1.8\times10^{53}$\\
090926A & $1.54\times10^{54}$ & $5.60\times10^{53}$\\
091003A & $4.12\times10^{52}$ & $1.58\times10^{52}$ \\
100414A & $2.53\times10^{53}$ & $1.18\times10^{53}$\\
100728A & $3.0\times10^{52}$ & $1.6\times10^{52 }$ \\
110731A & $1.6\times10^{53}$ & $5.8\times10^{52}$\\
120624B & $1.7\times10^{54}$ & $7.2\times10^{53}$  \\
130427A & $3.6\times10^{53}$ & $1.8\times10^{52}$\\
131108A & $4.4\times10^{53}$ & $1.2\times10^{53}$ \\
131231A & $1.2\times10^{53}$ & $6.3\times10^{52}$\\
141028A & $2.6\times10^{53}$ & $1.1\times10^{53}$ \\
150403A & $4.9\times10^{52}$ & $1.7\times10^{52}$ \\
160509A & $1.1\times10^{53}$ & $3.5\times10^{52}$ \\
160625B & $1.1\times10^{54}$ & $3.3\times10^{53}$ \\ 
170214A & $2.1\times10^{54}$ & $5.3\times10^{53}$ \\ 
170405A & $3.0\times10^{53}$ & $7.9\times10^{52}$ \\ 
180720B & $3.8\times10^{53}$ & $4.7\times10^{52}$\\ 
\hline
\hline
\end{tabular}
\caption{Results of $E_{0.1-10^{4}\rm s}$ and related error of $20$ BdHNe. $E_{0.1-10^{4}\rm s}$ is the total GeV energy (in erg) emitted from $0.1$ to $10^4$~s. GRBs 091127, 091208B, 130518A, 150314A, 150514A are excluded since they have only two data points in their GeV luminosity light-curves.} 
\label{tab:14fit}
\end{table*}

Out of $25$ BdHNe I presented in Table~\ref{tab:cb}, we perform the fitting for only $20$ GRBs which have more than two data points in their luminosity light-curves. Therefore, for the fitting of BdHNe I, there are $20$ bursts and each one has its power-law function. Consequently, there are in total $17$ parameters, including $20$ amplitudes, and $1$ power-law index. The fitting gives a power-law index of $\alpha_{\rm GeV} = 1.19\pm 0.04$, i.e.:
\begin{equation}\label{eq3}
L_{n}=A_{n}~t^{~-1.19\pm0.04}, 
\end{equation}
which is plotted in Fig.~\ref{fig:02} and the amplitudes of each GRB, $A_{n}$, with the uncertainty are shown in Table~\ref{tab:23fit}. This inferred power-law index is similar to the one obtained from fitting the GeV flux, $f_{\nu}(t)$, see e.g. \citep{2009MNRAS.400L..75K} and \citep{2017ApJ...837...13P}, in which the power-law index is $\alpha_{\rm GeV}=1.2 \pm 0.2$ and $\alpha_{\rm GeV}=1.2 \pm 0.4$, respectively. 

In our approach, we adopt an alternative interpretation of these power-laws: instead of using the flux expressed in arrival time, we use the luminosity expressed in the rest-frame of the source. Since the luminosity is proportional to the flux, i.e.~$L= 4 \pi d^2_L (1+z)^{\alpha_{\rm GeV} -2} f_{\nu}$, where $d_L$ is the luminosity distance, this similarity of the power-law index is not surprising. The advantage of using luminosity expressed in the rest-frame of the source, instead of flux in arrival time, is that one can determine the intrinsic energy loss of the system which produces the GeV radiation, regardless of differences in the redshift of the sources. This allows us following our recent understanding of the BdHN I 130427A \citep[see][and references therein]{2019ApJ...886...82R}, to relate the GeV radiation to the slowing down of the BH spin; see section~\ref{sec:9}.


After obtaining the best power-law parameters for the luminosity light-curve for each BdHNe I, we check the correlation between the GeV luminosity at $10$~s from Eq.~(\ref{eq3}) using the fitted parameters and the isotropic energy $E_{\gamma, \rm iso}$. The power-law fitting gives (see Fig.~\ref{K10Eiso}):
\begin{equation}\label{eq4}
L_{\mathrm{10s}}=(4.7\pm1.2) \times 10^{48}~(E_{\mathrm{iso}}/10^{52})^{~1.3\pm0.3}, 
\end{equation}
and the fitting parameters for each GRB including their uncertainties are shown in Table~\ref{tab:23fit}. Furthermore, we estimate the energy released in the GeV band by each GRB in the $0.1$--$10^{4}$~s time interval, i.e.:
\begin{equation}
E_{\mathrm{0.1-10^{4}s}}=A_{\mathrm{GRB}}~\int_{0.1}^{10000}t^{-1.19}~dt~~,
\label{eq6}
\end{equation}
and the derived $E_{\mathrm{0.1-10^{4}s}}$ are shown in Table~\ref{tab:14fit}. The parameters $E_{\mathrm{0.1-10^{4}s}}$ and $E_{\rm \gamma, iso}$ (isotropic energy of the prompt emission in $\gamma$ band) are also correlated by a power-law relation (see Fig.~\ref{EEiso}):
\begin{equation}
E_{\mathrm{0.1-10^{4}s}}=( 4.4\pm1.5) \times 10^{50}~(E_{\mathrm{iso}}/10^{52})^{1.4\pm0.3}.\label{eq7}
\end{equation}

This positive correlation indicates that the BdHNe I with higher isotropic energy are also more luminous and more energetic in the GeV emission. 

\section{The determination of the mass and spin of the BH in BdHNe I}\label{sec:5}

The theoretical progress introduced in \citet{2019ApJ...886...82R} has identified the GeV radiation as originating in the \emph{inner engine} of BdHN I. There, for the first time, it has been shown that indeed the rotational energy of a Kerr BH can be extracted for powering an astrophysical system.  The \emph{inner engine} is composed of: a) a non-stationary Kerr BH, b) a uniform magnetic field of $\sim 10^{10}$G aligned with the rotation axis, and c) the presence of a very tenuous fully ionized electron-nuclei plasma. The fundamental new conceptual breakthrough introduced by the physics of the \emph{inner engine} are developed in parallel papers; see e.g. \citet{2020EPJC...80..300R}. The main goal here is to show, using our recently published results, that the rotational energy of the Kerr BH is indeed sufficient to explain the energetics of the GeV emission. In turn, this allows us to determine here the mass and spin of the Kerr BH in each BdHN I. 

We here apply the self-consistent solution already well tested in the case of GRB 130427A \citep{2019ApJ...886...82R} and GRB 190114C \citep{2019arXiv191107552M} for determining the three parameters of the \emph{inner engine}, namely the mass and spin of the BH as well as the strength of the surrounding magnetic field $B_0$. The values are obtained satisfying three conditions:
\begin{enumerate}
\item 
The energy budget for the observed GeV luminosity is provided by the extractable rotational energy of a Kerr BH; see Eq.~(\ref{aone}); see Eq.~(34) in \citet{2019ApJ...886...82R}.
\item 
The magnetic field $B_0$ fulfills the transparency condition for the propagation of the GeV radiation imposed by the $e^+e^-$ pair production process in the \emph{inner engine}; see Eq.~(35) in \citet{2019ApJ...886...82R}. 
\item The ``quantized'' emission of the GeV radiation is determined by the density of the plasma and by the synchrotron radiation timescale \citep{2019ApJ...886...82R}; see Eq.~(36) in \citet{2019ApJ...886...82R}.
\end{enumerate}

\begin{figure}
\centering
\includegraphics[width=\hsize,clip]{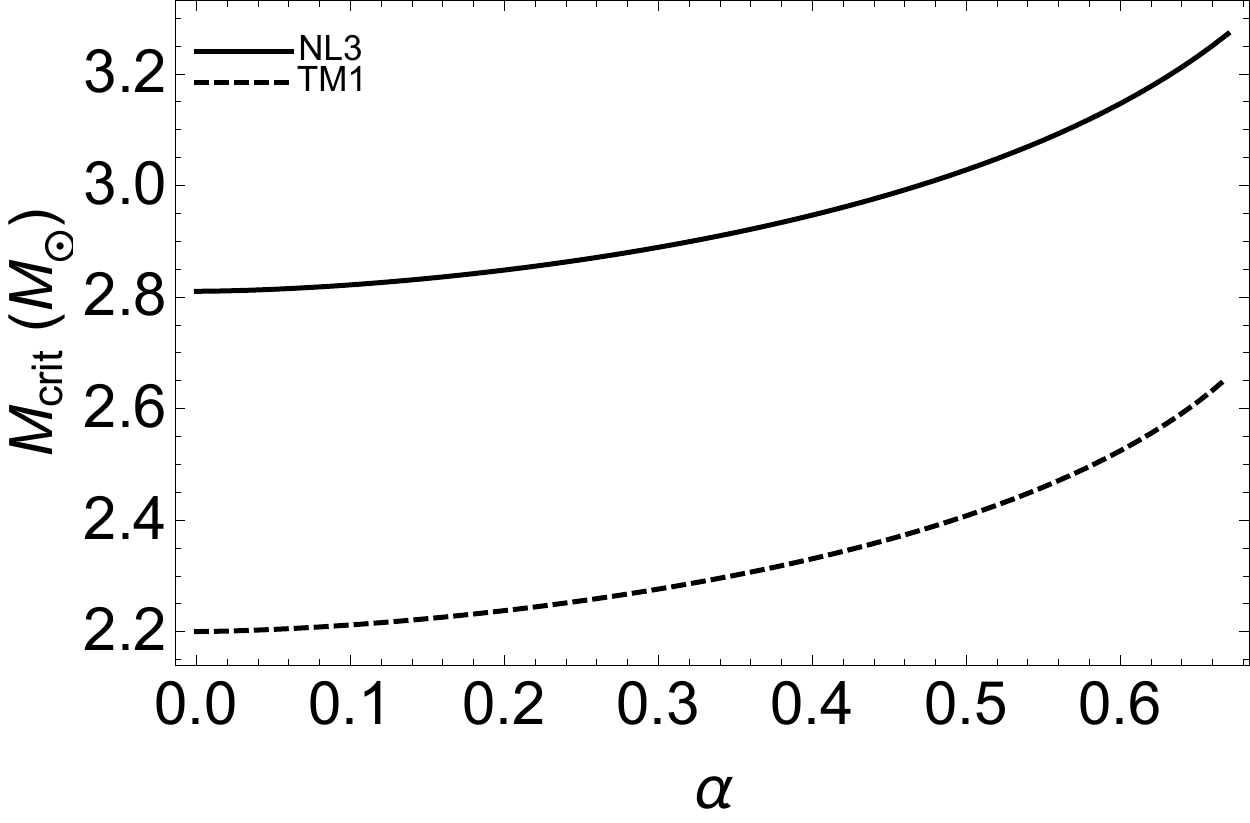}
\caption{NS critical mass as a function of the spin parameter $\alpha$ for the NL3 and TM1 EOS. We recall that the maximum spin parameter of a uniformly rotating NS is $\alpha_{\rm max}\approx0.71$, independently of the NS EOS; see e.g. \citet{2015PhRvD..92b3007C}.}
\label{fig:Mcrit}
\end{figure}

The high-quality GeV data in $11$ BdHNe I out of the $25$ long GRBs in Table~\ref{tab:cb} allow us to determine the starting point of the decreasing luminosity, by identifying the transition of the power-law dependence of the GeV luminosity from a positive to a negative slope \citep[see][for more information]{2019ApJ...886...82R}. This enables us to calculate the lower limit of the mass, $M$, spin parameter of the BH, $\alpha$, the corresponding irreducible mass of the BH, $M_{\rm irr}$; which remains constant during the energy extraction process, and finally the surrounding magnetic field strength, $B_0$; as reported in Table~\ref{tab:3b}. The values of the masses $M>2.21~M_{\odot}$ and spin parameters of $\alpha<0.71$ of the BH for BdHNe I presented in Table~\ref{tab:3b} show the consistency with the upper limit of the critical mass of the NS in \citet{1974PhRvL..32..324R} and the mass and spin of rotating NSs computed in \citet{2015PhRvD..92b3007C}; see Fig.~\ref{fig:Mcrit}. 

\begin{table*}
\centering
\begin{tabular}{lccccc}
\hline\hline
Source        	&   $\alpha$    &   $M(\alpha)$ &$M_{\rm irr}$ & $B_0$\\
        	&              &   (M$_\odot$)&(M$_\odot$)& $10^{10}$~G \\
\hline
\hline

BdHN I 080916C	&  $0.87$      &  $8.9$ & 7.6 &1.9 \\
BdHN I 090902B	&  $0.59$      &  $5.3$ & 5 &2.8   \\
BdHN I 090926A	&  $0.76$      &  $8.4$ & 7.7 &2.1  \\
BdHN I 110713A	&  $0.37$      &  $4.7$ & 4.6 &4.5   \\
BdHN I 130427A	&  $0.40$      &  $2.3$ & 2.24 &4.1   \\
BdHN I 130518A	&  $0.50$      &  $2.5$ & 2.4 &3.3   \\
BdHN I 131108A	&  $0.56$      &  $4.7$ & 4.4 &2.9   \\
BdHN I 160509A	&  $0.41$      &  $2.4$ & 2.3 &4   \\
BdHN I 170214A	&  $0.80$      &  $2.8$ & 2.5 &2.1   \\
BdHN I 170405A	&  $0.45$      &  $3.4$ & 3.3 &3.7  \\
BdHN I 180720B	&  $0.27$      &  $2.3$ & 2.29 &6   \\
\hline
\end{tabular}
\caption{The mass, $M$, the spin parameter, $\alpha=J/M^2$, and surrounding magnetic field, $B_0$ in $11$ BdHNe I, out of the $25$ long GRBs in Table~\ref{tab:cb}. The high-quality GeV data of this sample allows for a measurement of the lower limit of their ``inner engine'' parameters; see Eq.~(\ref{aone}).}
\label{tab:3b}
\end{table*}

This has indeed been addressed in  recent works \citep{2019ApJ...886...82R}, where we have developed a complementary theory and its related analysis to identify the physical conditions which have to be enforced in order to extract the rotational energy of a Kerr BH. We have there addressed an approach of considering a Kerr BH placed in a uniform magnetic field of $10^{10}$~G aligned along the BH symmetry axis, fulfilling the Einstein-Maxwell equations via the Papapetrou-Wald solution \citep{1974PhRvD..10.1680W, 1966AIHPA...4...83P} modeling the \emph{inner engine} which produces the MeV, GeV, TeV radiation and UHECRs as well \citep{2020EPJC...80..300R}. 

\section{Spin down of the BH in BdHNe I}\label{sec:9}

Following our previous work \citep{2019ApJ...886...82R}, we can turn now from the luminosity expressed in the rest-frame of the sources, see Eq.~(\ref{eq3}), and from the initial values of the spin and mass of the BH expressed in Section~\ref{sec:5}, to derive the slowing down of the BH due to the energy loss in the GeV emission. 

The relation of the luminosity and the extractable rotational energy is \citep[see Eq. (39) in][]{2019ApJ...886...82R}
\begin{equation}
\label{sdown1}
L=-\frac{dE_{extr}}{dt}=-\frac{dM}{dt},
\end{equation}
 For each BH during the GeV emission process the $M_{\rm irr}$ is constant.  Utilizing the best fit obtained for the GeV luminosity $L_{\rm GeV}=A_{\rm GeV}~t^{-1.2}$~erg/s, we obtain a relation for the loss of mass-energy of the BH by integrating Eq.~(\ref{sdown1}):
\begin{equation}
\label{sdown2}
M= M_0 + 5 A t^{-0.2}-5A t_0^{-0.2},
\end{equation}
where $M_0$ is the initial mass of the newborn BH tabulated in Table~\ref{tab:3b}. From the mass-energy formula of the BH we have \citep{2019ApJ...886...82R}
\begin{equation}
\label{sdown4}
a=\frac{J}{M}= 2 M_{\rm irr} \sqrt{1-\frac{M^2_{\rm irr}}{(M_0 + 5 A t^{-0.2}-5A t_0^{-0.2})^2}}.
\end{equation}
where $M_0$ is the initial mass of the BH presented in Table~\ref{tab:3b} as M$_{\alpha}$ at time $t_0$ at which the decaying part of GeV luminosity begins.

\begin{figure*}
\centering
\includegraphics[width=0.9\hsize,clip]{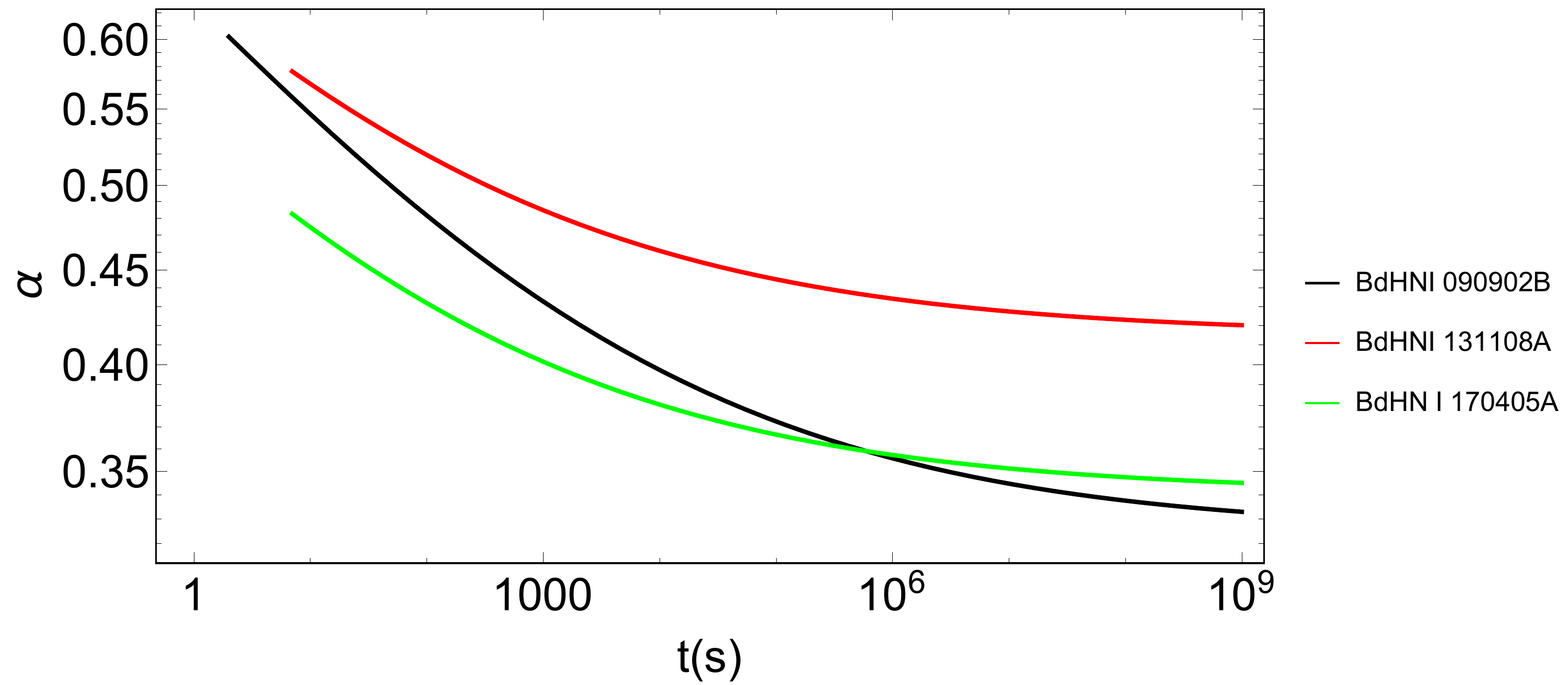}
\caption{{The  BH spin, as a function of rest-frame time. The initial values of the spin and mass of the BH for GRB 090902B are $\alpha=0.59$ and $M(\alpha)=5.3 M_{\odot}$, for 131108A: $\alpha=0.56$ and $M(\alpha)=4.7 M_{\odot}$
and for 170405A: $\alpha=0.45$ and $M(\alpha)=3.4 M_{\odot}$. This behavior of the spin parameter indicates that the rotational energy of the BH is decreasing due to the radiation losses in the GeV energy band.}}
\label{fig:sdown}
\end{figure*}

As indicative examples, we show in Fig.~\ref{fig:sdown} the decrease of the BH spin, $\alpha=a/M=J/M^2$, as a function of time in GRBs 090902B, 131108A and 170405A.

The third main results of this paper are: the identification of the rotational energy extraction from a Kerr BH and the consequent measure of the BH mass and spin.

\section{Conclusions}\label{sec:12}

The unprecedented observations of GRBs, pioneered by the Beppo-Sax satellite, have developed into the largest ever observational multi-wavelength effort in astrophysics: starting with the \textit{Swift}, BAT and XRT instruments in the X-ray band; see Fig.~\ref{fig:130427AF}, progressing with the \textit{AGILE} and with \textit{Fermi}-GBM in the MeV-GeV bands. These have worked in synergy with hundreds of optical, radio and VHE telescopes worldwide including MAGIC (see Fig.~\ref{fig:tev190114C}) and H.E.S.S. (see Fig.~\ref{fig:hess}).

This unprecedented observational effort assisted by parallel theoretical developments has allowed in this article the achievement of a new understanding of three new basic properties of the BdHNe: the first appearance of the SN triggering the entire BdHN process, the SN-rise; the presence of a mildly-relativistic afterglow in the X-ray in \textit{all} BdHN; the identification in \textit{all} BdHN of the origin of the high-energy emission in an \textit{inner engine} driven by a newborn BH; the description of their morphology. We show, for the first time, the extractable energy of a Kerr BH as an astrophysical energy source, which has allowed the inference of the BH mass and spin. 

In section~\ref{sec:2}, we first recall that binary systems have an important role in understanding both short and long GRBs and we report the progress in the classification of GRBs in nine different subclasses \citep[see e.g.][and references therein]{2019ApJ...874...39W}. We then focus on the BdHNe: long GRB model with progenitors composed of CO$_{\rm core}$ and the binary NS companion. The CO$_{\rm core}$ undergoes gravitational collapses that gives origin to a SN and the collapse of its Fe-core produces a $\nu$NS. 

We also there recall the fundamental role of the hypercritical accretion of the SN into the companion binary NS and into the $\nu$NS determine the BdHNe further evolution; see Fig.~\ref{fig:SPHsimulation} and \citet{2016ApJ...833..159P,2016ApJ...832..136R,2018ApJ...869..151R,2019ApJ...874...39W} for further details. The SN accretion onto the $\nu$NS gives origin to the X-ray afterglow emission, while the SN accretion onto the companion NS leads to different outcomes as a function of the binary period. For periods shorter than $5$~min, the hypercritical accretion onto the companion NS is sufficient for the NS to overcome its critical mass and gravitationally collapse to a BH. The BH formation characterizes a BdHN I with an isotropic energy in the range of $10^{52} \lesssim E_{\rm is} \lesssim 10^{54}$~erg. We here show that it gives origin, \textit{only in some} of them, to the GeV emission observed by Fermi-LAT. For larger binary periods, no BH is formed and consequently no GeV radiation is observed, the hypercritical SN accretion leads to a MNS with an isotropic energy in the range of $10^{50} \lesssim E_{\rm is} \lesssim 10^{52}$~erg. We refer to these binaries as BdHN II paradigm. The same occurs for more detached binary systems leading to a BdHN III, where the isotropic energy is in the range of $10^{48} \lesssim E_{\rm iso} \lesssim 10^{50}$~erg.

In section~\ref{sec:snrise}, we have given the spectral properties of the first appearance of the SN-rise in BdHN I and in BdHN II and also differentiate their energetics. 

In section~\ref{sec:xrayafterglow}, we have related the SN-rise luminosity to the X-ray luminosity of the afterglow in all three BdHNe types. It is a fortunate coincidence that we have recently understood the origin of the afterglow as a consequence of the SN hypercritical accretion on the $\nu$NS. This process is dominated by a mildly-relativistic synchrotron pulsar-like emission with Lorentz factor $\Gamma \sim 2$ that gives rise to the X-ray afterglow \citep{2018ApJ...869..101R,2019ApJ...874...39W,2020ApJ...893..148R}, and we have also related their X-ray luminosity to the NS spin. This has allowed us to represent in Fig.~\ref{fig:angdown} the afterglows for two BdHNe I, for two BdHNe II as well as one BdHN III and estimate in Table~\ref{tab:ns_parameters} the initial spin value of the $\nu$NS. What is the most remarkable, is that the X-ray afterglow is present in \textit{all} BdHN types which implies that, unlike the GeV emission, which as we show in this article to be necessarily beamed, the X-ray afterglow emission is necessarily isotropic. What is equally relevant is that independently of the differences among these four subclasses of BdHN, the X-ray afterglow luminosity emission is consistent with a power-law index of $-1.48\pm0.32$ as measured from the Swift observations \citep{2016ApJ...833..159P}, and a common energy source well explained by the rotational energy of the $\nu$NS.

The first identification of the SN-rise and of the measurement of the $\nu$NS mass originating power-law emission of the afterglow are the first main result of this paper.

The first main results of this paper are: 
\begin{enumerate}
    \item 
    the first identification of the SN-rise;
    \item 
    the agreement of the extrapolated luminosity of the X-ray afterglow with the luminosity of the SN-rise;
    \item 
    the measurement of the $\nu$NS period, originating the power-law emission of the afterglow; see Figs.~\ref{fig:4bdhn} and {\ref{fig:angdown}}.
\end{enumerate}

The two process of the SN-rise energetics and the $\nu$NS dynamics appear to be strongly correlated. 

We then turn in section~\ref{sec:4} to consider only the case of BdHN I and their Fermi-GBM and LAT observations. In Appendix~\ref{updated}, we update our previous classification of BdHN I following \citet{2016ApJ...833..159P,2016ApJ...832..136R,2018ApJ...869..151R} reaching the total number of $378$ BdHN I, \textit{all} of them are characterized by:
\begin{enumerate}
    \item 
    a measured cosmological redshift; 
    \item 
    a prompt emission of $T_{90}>2$s, measured by Fermi-GBM, with isotropic energy larger than $10^{52}$~erg;
    \item 
    a decaying X-ray afterglow, measured by Swift-XRT, characterized by a luminosity decreasing with a mean power-law with index of $\alpha_X=-1.48\pm 0.32$.
\end{enumerate}

Contrary to the case of the X-ray afterglow, universally present in \textit{all} BdHN types, the GeV radiation is present \textit{only in some} BdHN I. No GeV emission occurs in BdHN II and BdHN III. We first explore the possibility that the non-detection of GeV radiation in \textit{some} of BdHNe I could be due to the observational limitation of the LAT field of view, i.e. because of the boresight angle smaller than $75^{\circ}$. Indeed, we find that only $N_{\rm tot}=54$ out of the $378$ BdHNe I are inside the boresight angle of Fermi-LAT.
What is unexpected is that only N$_{\rm LAT}=25$ out of these $54$ BdHNe I exhibit the GeV emission observed by Fermi-LAT. For each of these $25$ sources, we have given the basic parameters in Table~\ref{tab:cb}. The corresponding data of the remaining $29$ BdHN I, without observed GeV radiation, are given in Table~\ref{tab:BdHNe_No_GeV}.

In section~\ref{sec:morphology}, we have assumed that \textit{all} BdHNe I, like all GRBs are homogeneously distributed in space \citep[see, e.g.,][]{1992Natur.355..143M,1999ApJS..122..465P}, we have inferred that the emission of the GeV radiation occurs in two opposite cones each of half opening angle of  $\sim 60^{\circ}$ from the normal to the binary plane.

We duly recall as well that the visualization of the morphology has been made possible thanks to a close collaboration with LANL \citep[see][for additional details]{2016ApJ...833..107B,2019ApJ...871...14B}, leading to the results well illustrated in the simulation presented
in Figs.~\ref{fig:SPHsimulation} and \ref{fig:cc}.
We then conclude from this simulation that all of the $25$ LAT sources are actually ``seen from the top'' which allows us to fully observe the conical emission of $60^{\circ}$ half-opening angle. For the remaining $29$ BdHN I without an observed GeV emission, we evidence that when the Swift data are available, gamma-ray flares and hard and soft X-ray flares as well as extended thermal emissions are observed in these systems \citep{2018ApJ...869..151R, 2018ApJ...852...53R}, and that these sources have a viewing angle laying in the ``orbital plane'' of the binary progenitor system.

We conclude that we are faced with a new morphology of the BdHN I which depends significantly on the viewing angle, ``seen from the top'', normal to the binary orbital plane when the GeV emission is observed, or seen ``in the plane'' of the binary when the observation of the GeV radiation is impeded by the accreting binary material; see Figs.~\ref{fig:SPHsimulation}, \ref{fig:dd} and \ref{fig:cc}. This is reminiscent of the morphology encountered in some AGNs; see e.g. the AGN IC 310 in \citet{2014Sci...346.1080A}. 

The second main result of this paper is the identification of the BdHN I morphology and its explanation within the BdHN I model.


We then recall some theoretical progresses in understanding
the origin of the GeV emission: 

a) The identification of the three components of the GRB \textit{inner engine} in GRB 130427A \citep{2019ApJ...886...82R}, composed of a Kerr BH with a magnetic field $B_0$ aligned with the BH rotation axis, both embedded in a tenuous ionized plasma composed of electrons and ions, has represented a turning point in the study of BdHN I. The electrodynamics of this \textit{inner engine}, based on the Papapetrou-Wald solution \citep{1974PhRvD..10.1680W,1966AIHPA...4...83P,2019ApJ...886...82R}, leads to a high energy emission in two opposite lobes in the MeV, GeV, and TeV radiation as well as narrowly beamed UHECR
along the BH polar axis \citep{2019arXiv191107552M};

b) This high-energy emission, unlike the traditional models which implies ultrarelativistic baryonic motion with $\Gamma \sim 10^3$ at $10^{16}$~cm--$10^{18}$~cm, occurs very close to the BH horizon; 

c) The energy source is the extractable energy of the BH \citep{1970PhRvL..25.1596C, 1971PhRvD...4.3552C,1971PhRvL..26.1344H, Hawking:1971vc}, see Eq.~(\ref{aone}), and is emitted in a sequence of impulsive process, the ``\textit{blackholic quanta}'', occurring on a timescale of $10^{-14}$~s \citep{2020EPJC...80..300R}.

On the basis of these results, we have examined the 
physical origin of the GeV emission observed by Fermi-LAT
both in BdHN I. We find that the luminosity of
the GeV emission as a function of time in the rest-frame of the
source fulfills a universal decaying power-law dependence with index of $-1.19\pm0.04$; see Fig~\ref{fig:02}. This has allowed: 1) to verify that indeed the entire GeV radiation observed by Fermi-LAT can be energetically expressed in terms of the rotational energy of the Kerr BH; 2) following the procedures in \citet{2019ApJ...886...82R} to evaluate the mass and spin of the BH; see Table~\ref{tab:3b}; and 3) to explicitly compute the slowing down rate of the BH spin due to the GeV emission; see Fig.~\ref{fig:sdown}.

It has been possible for some of the $25$ sources, with the best data:
\begin{itemize}
    \item [a)] to compute the lower limit of the initial value of the BH masses, $M$, and show their consistency with the absolute upper limit of the NS critical mass \citep{1974PhRvL..32..324R}, and the upper limit of the NS mass of $M=2.21 M_{\odot}$ and spin parameter of $\alpha<0.71$ computed in \citet{2015PhRvD..92b3007C};
    \item [b)] to evaluate the value of the spin, $a$, and show the consistency with the canonical upper limit $\alpha=a/M \leq 1$;
    \item [c)] by combining the value of the spin of the $\nu$NS observed from the afterglow (see Table~\ref{tab:ns_parameters}), the time intervening between the SN-rise and the UPE phase, the mass estimate of the BH in GRB 190114C and in GRB 090926A and in GRB 180720B, we infer that necessarily in these system we are observing the presence of a BdHN precursor with a companion NS grazing the surface of the CO$_{\rm core}$.
\end{itemize}

The third main results of this paper are: 
\begin{enumerate}
    \item the identification of the rotational energy extraction from a Kerr BH and,
    \item the consequent measure of the BH mass and spin.
\end{enumerate} 
 
All the above three main results are important: the underlying proof that indeed we can use the extractable rotational energy of a Kerr BH for explaining the high-energy jetted emissions of GRBs and AGNs stands alone. Even more subtle is the fact that the jetted emission does not originate from massive ultra-relativistic jetted emissions, but from very special energy-saving ultra-relativistic quantum and classical electrodynamical processes originating in the high-energy jetted emission. This brings up the issue of \emph{blackholic} energy \citep{2020EPJC...80..300R}, which is not treated here. We were waiting for this result for $49$ years, since writing  Eq.~(\ref{aone}).

\section*{Acknowledgements}

We acknowledge the protracted discussion with Roy Kerr. {We are thankful to the referee for the interesting report and suggestions}. We also acknowledge the continuous support of the MAECI and the Italian Space Agency (ASI). Y.~A. is supported by the Erasmus Mundus Joint Doctorate Program Grant N.~2014-0707 from EACEA of the European Commission. Y.~A. acknowledges funding by the Science Committee of the Ministry of Education and Science of the Republic of Kazakhstan (Grant No. AP08855631) and also partial support from targeted financial program No. BR05336383 by Aerospace Committee of the Ministry of Digital Development, Innovations and Aerospace Industry of the Republic of Kazakhstan. G.~J.~M is supported by the U.S. Department of Energy under Nuclear Theory Grant DE-FG02-95-ER40934. This work made use of data from \textit{Fermi} space observatory. This research has made use of data and software provided by the High Energy Astrophysics Science Archive Research Center (HEASARC), which is a service of the Astrophysics Science Division at NASA/GSFC and the High Energy Astrophysics Division of the Smithsonian Astrophysical Observatory.

\section*{Data Availability}

The data underlying this article are available in the appendix \ref{updated} of the article.


\appendix

\section{Updated BdHNe I list: BdHNe I within 2017--2018}\label{updated}

The  list of BdHNe I consists of $345$ sources\footnote{~\url{https://iopscience.iop.org/0004-637X/852/1/53/suppdata/apjaa9e8bt9_mrt.txt}} within 1997--2016 published in \citet{2018ApJ...852...53R}. Therein $152$ bursts occurred before the launch of \textit{Fermi} space mission. The first event detected by \textit{Fermi}-GBM was the object GRB~080810 reported in \texttt{GCN\,8100} \cite{Meegan...2008GCN...8100...1M}. Consequently, the further $193$ events of the list occurred within operational era of \textit{Fermi} observatory.

BdHNe I are all the GRBs which satisfy the following criteria \cite{Pisani...2018EPJWC...16804002P}:
\begin{itemize}
	\item measured redshift z;
	\item GRB rest-frame duration larger than 2 s;
	\item isotropic energy $E_{\rm \gamma, iso}$ larger than $\sim 10^{52}$~erg;
	\item presence of associated \textit{Swift}-XRT data.
\end{itemize}
 For fitting the X-ray afterglow in \citet{2016ApJ...833..159P} only the ones having \textit{Swift}-XRT data lasting at least up to $t_{\rm rf}=10^4$~s have been considered, which shows a distribution of power-law index follows a Gaussian behavior
with a  value of $\alpha_{X}=-1.48\pm 0.32$.

In order to update the BdHNe I list with ones occurred within 2017--2018 we go through the all triggered bursts. To have a comprehensive list we use several sources of information\footnote{~\url{https://gcn.gsfc.nasa.gov/gcn3_archive.html}}~\footnote{~\url{https://heasarc.gsfc.nasa.gov/W3Browse/fermi/fermigbrst.html}}~\footnote{~\url{https://heasarc.gsfc.nasa.gov/W3Browse/fermi/fermilgrb.html}}~\footnote{~\url{http://www.mpe.mpg.de/~jcg/grbgen.html}} and cross-correlate data on each GRB.

There are $197$ (2017) and $164$ (2018) events in total classified as GRBs and reported in \texttt{GCN} circulars. Among them $17$ (2017) and $16$ (2018) bursts have the measured redshift values. These $33$ GRBs are represented by $31$ long and $2$ short duration bursts. In total of $31$ long events the \textit{Fermi}-GBM triggered in $16$ cases while the rest objects were observed by different instruments---\textit{Konus}-WIND and \textit{Swift}-BAT, etc.

In Table~\ref{tab:newbdhne} we present GRB list with measured redshift values within January~2017--December~2018. The values of redshift $z$ and duration $T_{90}$ are retrieved from \texttt{GCN} and literature. The values of isotropic-equivalent energy $E_{\rm \gamma, iso}$ are calculated using the spectral parameters ($\alpha$, $\beta$, $E_{\rm p,i}$, etc.) of the model best-fitting the $T_{90}$ interval defined within [$50$--$300$]~keV band and reported in appropriate \texttt{GCN}, preferentially by \textit{Fermi}-GBM team. Note that the preference for duration info and best-fit was given to \textit{Fermi}-GBM (denoted with full name in parenthesis), then if absent---to \textit{Konus}-WIND (denoted as *KW), then if absent---to \textit{Swift}-BAT (denoted as *SW), and the motivation for such a gradation stays on different energy bands used to define $T_{90}$ value. The subsequent data on spectral fit should be understood as attributed to the analysis on substitute instruments when indicated. The boresight angle $\theta$ with respect to \textit{Fermi}-LAT instrument is an off-axis angle defined at the trigger moment.
\newcolumntype{g}{>{\columncolor{gray2}}c}
\newcolumntype{f}{>{\columncolor{gray2}}r}

\begin{table*}

	\centering\scriptsize
	\begin{tabular}{rlgfccccfrl}
	\hline\hline
\# & GRB & $z$ & $T_{90}$ & Best-fit & $\alpha$ & $\beta$ & $E_{\rm p,i}$ & $E_{\rm \gamma,iso}$          & $\theta$\cellcolor{gray3} & References\\
   &  &     & (s)      & (model)  &          &         & (keV)         & ($\times 10^{52}$~erg) & (deg\cellcolor{gray3}) & \\
\hline
1 & 170113A\,(420) & $1.968$ & $49.2$ & PL & $-1.95\pm 0.08$ & --- & --- & $19.91$ & $145.0$ & 20452, 20458\\
2 & 170202A\,*KW & $3.645$ & $30.0$ & CPL & $-1.16_{-0.34}^{+0.59}$ & --- & $247.0_{-86.0}^{+166.0}$ & $17.0$ & --- & 20584, 20588, 20590, 20604 *KW\\
3 & 170214A\,(649) & $2.53$ & $122.9$ & SBPL & $-1.063\pm 0.008$ & $-2.246\pm 0.038$ & $253.4\pm 10.4$ & $\cdots$ & $33.2$\cellcolor{gray3} & 20675, 20686\\
4 & 170405A\,(777) & $3.51$ & $78.6$ & Band & $-0.799\pm 0.020$ & $-2.354\pm 0.089$ & $267.0\pm 9.3$ & $241.01$ & $52.0$\cellcolor{gray3} & 20990, 20986\\
5 & \textbf{170428A}\,*KW & $0.454$ & $0.14$\cellcolor{white} & Band & $-0.47_{-0.21}^{+0.28}$ & $-2.46_{-7.54}^{+0.52}$ & $982.0_{-355.0}^{+394.0}$ & $\cdots$\cellcolor{white} & --- & 21059, 21045 *KW\\
6 & 170519A\,*SW & $0.818$ & $216.4$ & PL & $-1.94\pm 0.26$ & --- & --- & $\cdots$ & --- & 21119, 21112 *SW\\
7 & 170531B\,*SW & $2.366$ & $164.1$ & PL & $-1.95\pm 0.14$ & --- & --- & $\cdots$ & --- & 21177, 21209, 21186 *SW\\
8 & 170604A\,*KW & $1.329$ & $30.0$ & CPL & $-1.31_{-0.20}^{+0.26}$ & --- & $220.0_{-48.0}^{+72.0}$ & $4.7\pm 0.6$ & --- & 21197, 21247 *KW\\
9 & 170607A\,(971) & $0.557$ & $20.9$ & CPL & $-1.401\pm 0.044$ & --- & $145.2\pm 11.9$ & $1.13$ & $100.0$ & 21240, 21218\\
10 & 170705A\,(115) & $2.01$ & $22.8$ & Band & $-0.991\pm 0.069$ & $-2.303\pm 0.108$ & $97.9\pm 7.6$ & $14.95$ & $118.0$ & 21298, 21297\\
11 & 170714A\,*SW & $0.793$ & $537.3$ & PL & $-1.76\pm 0.17$ & --- & --- & $\cdots$ & --- & 21359, 21347 *SW\\
12 & \textbf{170817A}\,(529) & $0.0093$ & $2.0$\cellcolor{white} & PL & $-1.428\pm 0.094$ & --- & --- & $\cdots$\cellcolor{white} & $91.0$ & \\
13 & 170903A\,(534) & $0.886$  & $25.6$ & CPL & $-1.316\pm 0.108$ & --- & $95.6\pm 13.4$ & $1.15$ & $93.0$ & 21799, 21812\\
14 & 171010A\,(792) & $0.3285$ & $107.3$ & Band & $-1.089\pm 0.006$ & $-2.191\pm 0.009$ & $137.7\pm 1.4$ & $14.49$ & $114.7$ & 22002, 22096, 21992\\
15 & 171020A\,*SW & $1.87$ & $41.9$ & PL & $-1.04\pm 0.20$ & --- & --- & $\cdots$ & --- & 22039, 22038 *SW\\
16 & 171205A\,*KW & $0.0368$ & $145.0$ & PL & $-2.0\pm 0.18$ & --- & --- & $0.0024$\cellcolor{white} & --- & 22180, 22227 *KW\\
17 & 171222A\,(684) & $2.409$ & $80.4$ & PL & $-2.068\pm 0.059$ & --- & --- & $20.73$ & $43.0$\cellcolor{gray3} & 22272, 22277\\
\hline
18 & 180115A\,*SW & $2.487$ & $40.9$ & PL & $-1.66\pm 0.22$ & --- & --- & $\cdots$ & --- & 22346, 22348 *SW\\
19 & 180205A\,(184) & $1.409$ & $15.4$ & PL & $-1.887\pm 0.052$ & --- & --- & $3.06$ & $94.0$ & 22384, 22386\\
20 & 180314A\,(030) & $1.445$ & $22.0$ & CPL & $-0.445\pm 0.055$ & --- & $106.2\pm 2.9$ & $6.58$ & $99.0$ & 22484, 22485\\
21 & 180325A\,*KW & $2.248$ & $10.0$ & Band & $-0.50_{-0.19}^{+0.21}$ & $-2.65_{-1.06}^{+0.31}$ & $306.0_{-39.0}^{+50.0}$ & $23.0$ & --- & 22535, 22555, 22546 *KW\\
22 & 180329B*SW & $1.998$ & $210.0$ & CPL & $-0.97\pm 0.56$ & --- & $48.6\pm 9.1$ & $\cdots$ & --- & 22567, 22566 *SW\\
23 & 180404A\,*SW & $1.000$ & $35.2$ & PL & $-1.95\pm 0.12$ & --- & --- & $\cdots$ & --- & 22591, 22599 *SW\\
24 & 180510B\,*SW & $1.305$ & $134.3$ & PL & $-2.00\pm 0.17$ & --- & --- & $\cdots$ & --- & 22702, 22705 *SW\\
25 & 180620B\,(660) & $1.1175$ & $46.7$ & Band & $-1.206\pm 0.116$ & $-1.660\pm 0.035$ & $175.6\pm 49.8$ & $16.30$ & $137.0$ & 22823, 22813\\
26 & 180624A\,*SW & $2.855$ & $486.4$ & PL & $-1.91\pm 0.10$ & --- & --- & $\cdots$ & --- & 22845, 22848 *SW\\
27 & 180703A\,(876) & $0.6678$ & $20.7$ & Band & $-0.776\pm 0.041$ & $-1.967\pm 0.103$ & $350.8\pm 32.2$ & $3.15$ & $44.0$\cellcolor{gray3} & 23889, 22896\\
28 & 180720B\,(598) & $0.654$ & $48.9$ & Band & $-1.171\pm 0.005$ & $-2.490\pm 0.071$ & $636.0\pm 15.4$ & $19.08$ & $49.1$\cellcolor{gray3} & 22996, 22981\\
29 & 180728A\,(728) & $0.117$ & $6.4$ & Band & $-1.54\pm 0.01$ & $-2.46\pm 0.02$ & $79.2\pm 1.4$ & $\cdots$ & $35.0$\cellcolor{gray3} & 23055, 23067, 23053\\ 
30 & 180914B\,*KW & $1.096$ & $280.0$ & Band & $-0.81_{-0.04}^{+0.04}$ & $-2.12_{-0.7}^{+0.08}$ & $466.0_{-27.0}^{+29.0}$ & $360.0$ & $94.0$ & 23246, 23240 *KW\\
31 & 181010A\,(247) & $1.39$ & $9.7$ & CPL & $-0.8\pm 0.2$ & --- & $280.0\pm 80.0$ & $\cdots$ & $48.1$\cellcolor{gray3} & 23315, 23320\\
32 & 181020A\,(792) & $2.938$ & $15.1$ & Band & $-0.70\pm 0.02$ & $-2.06\pm 0.07$ & $367.0\pm 17.0$ & $\cdots$ & $50.0$\cellcolor{gray3} & 23356, 23352\\
33 & 181110A\,*KW & $1.505$ & $140.0$ & CPL & $-1.63_{-0.26}^{+0.34}$ & --- & $48.0_{-27.0}^{+14.0}$ & $11.0$ & --- & 23421, 23424 *KW\\
	\hline
	\end{tabular}
	\caption[List of $33$~GRBs with measured redshift values within January~2017--December~2018.]{List of $33$~GRBs with measured redshift values within January~2017--December~2018. Four criteria were applied (columns from left to right) to the sample and highlighted elements are consistent with their requirements.}
	\label{tab:newbdhne}
\end{table*}

In Table~\ref{tab:newbdhnelat} we present the above list narrowed to the BdHNe I which fell within \textit{Fermi}-LAT FoV at the trigger. Additional information includes whether the high-energy photons were detected, and if so, the calculated value of energy $E_{\rm LAT}$ is given together with test statistic (TS) value of the signal to be associated with GRB. The bursts observed by other instruments lack the information on \textit{Fermi}-LAT boresight angle and consequently no observations in high-energy domain were carried out. Therefore, consideration of that events is not possible even their energy values are sufficient ($E_{\rm \gamma,iso}\gtrsim10^{52}$~erg) and GRBs are classified as BdHNe I.
\newcolumntype{e}{>{\columncolor{gray2}}l}
\begin{table*}
	\centering\normalsize
	\begin{tabular}{ceffgcccl}
		\hline\hline
		BdHN & $z$ & $T_{90}$ & $E_{\rm \gamma,iso}$   & $\theta$ & GeV & $E_{\rm LAT}$ & TS & References\\
		     &     & (s)      & ($\times 10^{52}$~erg) & (deg)    & photons & ($\times 10^{52}$~erg) & & \\
		\hline
		170214A\,(649) & $2.53$ & $122.9$ & $\cdots$ & $33.2$ & Yes\cellcolor{gray3} & \cellcolor{gray3} & \cellcolor{gray3}& 20675 (z), 20686 (GBM)\\
		170405A\,(777) & $3.51$ & $78.6$ & $241.01$ & $52.0$ & Yes\cellcolor{gray3} & \cellcolor{gray3} & \cellcolor{gray3} & 20990 (z), 20986 (GBM)\\
		171222A\,(684) & $2.409$ & $80.4$ & $20.73$ & $43.0$ & No & --- & --- & 22272 (z), 22277 (GBM)\\
		180703A\,(876) & $0.6678$ & $20.7$ & $3.15$ & $44.0$ & No & --- & --- & 23889 (z), 22896 (GBM)\\
		180720B\,(598) & $0.654$ & $48.9$ & $19.08$ & $49.1$ & Yes\cellcolor{gray3} & \cellcolor{gray3} & \cellcolor{gray3} & 22996 (z), 22981 (GBM)\\
		180728A\,(728) & $0.117$ & $6.4$ & $\cdots$ & $35.0$ & & & & 23055 (z), 23067 (z), 23053 (GBM)\\ 
		181010A\,(247) & $1.39$ & $9.7$ & $\cdots$ & $48.1$ & & & & 23315 (z), 23320 (GBM)\\
		181020A\,(792) & $2.938$ & $15.1$ & $\cdots$ & $50.0$ & & & & 23356 (z), 23352 (GBM)\\
		\hline
	\end{tabular}
	\caption[List of BdHN I within January~2017--December~2018 with boresight angle $\theta \lesssim 75^{\circ}$ at the trigger.]{List of BdHNe I within January~2017--December~2018 with boresight angle $\theta \lesssim 75^{\circ}$ at the trigger.}
	\label{tab:newbdhnelat}
\end{table*}

\newpage

We present here in Table~\ref{tab:bdhne} the complete list of the $378$ BdHNe observed up through the end of $2018$, which includes the $161$ BdHNe already presented in \citet{2016ApJ...833..159P} and $345$ BdHNe already presented in \citet{2018ApJ...852...53R}.
\onecolumn
\begin{longtable}[c]{cc|c|c|c|c|c|c|c|cc}
\caption{List of the $378$ BdHNe during $14$ years of Swift/XRT observation activity, updated in this work. We report: the redshift $z$, the isotropic energy $E_{\rm iso}$, the instrument which detected the sources in the gamma-ray band, and the reference from which the gamma-ray spectral parameters are taken in order to evaluate the $E_{\rm iso}$.\\
$^{(a)}$: in units of $10^{52}$ erg.\\
$^{(b)}$: ``LX'' indicates the sources with {Swift}/XRT data observed up to times larger than $t_{\rm rf}\sim 10^4$~s after the trigger time.\\
$^{(c)}$: ``C'' and ``E'' represent the sources with early flares in their {Swift}/XRT, and they stand for ``confirmed'' and ``excluded'' respectively; see \citet{2018ApJ...852...53R}.\\
$^{(d)}$: ``UL'' indicates the sources with $T_{90} \gtrsim 1000$ s known as ultra-long GRBs.\\
$^{(e)}$: observed $T_{90}$ (s).\\
$^{(f)}$: ``B-SAX'' stands for {Beppo-SAX}/GRBM; ``BATSE'' stands for {Compton-GRO}-BATSE; ``Ulysses'' stands for {Ulysses}-GRB; ``KW'' stands for Konus-WIND; ``HETE'' stands for {HETE-2}-FREGATE; ``Swift'' stands for {Swift}-BAT; ``Fermi'' stands for {Fermi}-GBM.\\
$^{(g)}$: (1) \citet{1998ApJ...493L..67F}; (2) \citet{2015ARep...59..626R}; (3) \citet{1998ApJ...505L.119I}; (4) \citet{2000Sci...290..953A}; (5) \citet{2000ApJ...534L..23H}; (6) \citet{2001ApJ...559..710I}; (7) \citet{2003A&A...400.1021B}; (8) \citet{2008PASJ...60..919S}; (9) \citet{2006ApJ...652..490C}.
\label{tab:bdhne}}\\
\hline\hline
GRB & z & $E_{\rm iso}^{(a)} $  & LX$^{(b)}$ & Early flare$^{(c)}$  & UL$^{(d)}$ & $T_{90}^{(e)}$ & Instrument$^{(f)}$ & Reference$^{(g)}$ \\
\hline
\endfirsthead
\caption{continued.}\\
\hline\hline
GRB & z & $E_{\rm iso}^{(a)} $  & LX$^{(b)}$ & Early flare$^{(c)}$  & UL$^{(d)}$ & $T_{90}^{(e)}$ & Instrument$^{(f)}$ & Reference$^{(g)}$ \\
\hline
\endhead
\hline
\endfoot
970228	&	$0.695$	&	$1.65\pm0.16$	&		&		&		&	$80$	&	B-SAX	&	(1) 	\\
970828	&	$0.958$	&	$30.4\pm3.6$	&		&		&		&	$90$	&	BATSE	&	(2) 	\\
971214	&	$3.42$	&	$22.1\pm2.7$	&		&		&		&	$40$	&	BATSE	&	IAUC 6789 	\\
980329	&	$3.5$	&	$267\pm53$	&		&		&		&	$54$	&	B-SAX	&	(3) 	\\
980703	&	$0.966$	&	$7.42\pm0.74$	&		&		&		&	$400$	&	BATSE	&	GCN 143	\\
990123	&	$1.6$	&	$241\pm39$	&		&		&		&	$63.3$	&	BATSE	&	GCN 224	\\
990506	&	$1.3$	&	$98.1\pm9.9$	&		&		&		&	$131.33$	&	BATSE	&	GCN 306	\\
990510	&	$1.619$	&	$18.1\pm2.7$	&		&		&		&	$75$	&	BATSE	&	GCN 322	\\
990705	&	$0.842$	&	$18.7\pm2.7$	&		&		&		&	$42$	&	B-SAX	&	(4)	\\
991208	&	$0.706$	&	$23.0\pm2.3$	&		&		&		&	$68$	&	Ulysses	&	(5)	\\
991216	&	$1.02$	&	$69.8\pm7.2$	&		&		&		&	$15.17$	&	BATSE	&	GCN 504 	\\
\hline
000131	&	$4.5$	&	$184\pm32$	&		&		&		&	$50$	&	KW+Ulysses	&	GCN 529	\\
000210	&	$0.846$	&	$15.4\pm1.7$	&		&		&		&	$12.3$	&	BATSE	&	GCN 540	\\
000301C	&	$2.0335$	&	$4.96\pm0.50$	&		&		&		&	$10$	&	Ulysses	&	GCN 568	\\
000418	&	$1.12$	&	$9.5\pm1.8$	&		&		&		&	$30$	&	KW+Ulysses	&	GCN 642 	\\
000911	&	$1.06$	&	$70\pm14$	&		&		&		&	$500$	&	KW+Ulysses	&	GCN 791	\\
000926	&	$2.07$	&	$28.6\pm6.2$	&		&		&		&	$25$	&	KW+Ulysses	&	GCN 801	\\
\hline
010222	&	$1.48$	&	$84.9\pm9.0$	&		&		&		&	$170$	&	B-SAX	&	(6)	\\
010921	&	$0.45$	&	$0.97\pm0.10$	&		&		&		&	$12$	&	HETE	&	GCN 1096	\\
011121	&	$0.36$	&	$8.0\pm2.2$	&		&		&		&	$28$	&	Ulysses	&	GCN 1148	\\
011211	&	$2.14$	&	$5.74\pm0.64$	&		&		&		&	$270$	&	B-SAX	&	GCN 1215	\\
\hline
020124	&	$3.2$	&	$28.5\pm2.8$	&		&		&		&	$78.6$	&	HETE	&	(7)	\\
020127	&	$1.9$	&	$3.73\pm0.37$	&		&		&		&	$9.3$	&	HETE	&	(7)	\\
020405	&	$0.69$	&	$10.6\pm1.1$	&		&		&		&	$40$	&	KW+Ulysses	&	GCN 1325	\\
020813	&	$1.25$	&	$68\pm17$	&		&		&		&	$90$	&	HETE	&	(7)	\\
021004	&	$2.3$	&	$3.47\pm0.46$	&		&		&		&	$57.7$	&	HETE	&	(7)	\\
021211	&	$1.01$	&	$1.16\pm0.13$	&		&		&		&	$5.7$	&	HETE	&	GCN 1734	\\
\hline
030226	&	$1.98$	&	$12.7\pm1.4$	&		&		&		&	$100$	&	HETE	&	GCN 1888	\\
030323	&	$3.37$	&	$2.94\pm0.92$	&		&		&		&	$26$	&	HETE	&	GCN 1956	\\
030328	&	$1.52$	&	$38.9\pm3.9$	&		&		&		&	$100$	&	HETE	&	GCN 1978 	\\
030329	&	$0.169$	&	$1.62\pm0.16$	&		&		&		&	$50$	&	HETE	&	IAUC 8101	\\
030429	&	$2.65$	&	$2.29\pm0.27$	&		&		&		&	$14$	&	HETE	&	GCN 2211	\\
030528	&	$0.78$	&	$2.22\pm0.27$	&		&		&		&	$21.6$	&	HETE	&	GCN 2256 	\\
\hline
040912	&	$1.563$	&	$1.36\pm0.36$	&		&		&		&	$122$	&	HETE	&	GCN 2723	\\
040924	&	$0.859$	&	$0.98\pm0.10$	&		&		&		&	$2.4$	&	KW	&	GCN 2754	\\
041006	&	$0.716$	&	$3.11\pm0.89$	&		&		&		&	$27.3$	&	HETE	&	(8) 	\\
041219A	&	$0.31$	&	$10.0\pm1.0$	&		&		&		&	$520$	&	Swift	&	GCN 2874	\\
\hline
050126	&	$1.29$ 	&	$2.47\pm0.25$	&		&		&		&	$26$	&	Swift	&	GCN 2987 	\\
050315	&	$1.95$	&	$6.15\pm0.30$	&	LX	&		&		&	$96$	&	Swift	&	GCN 3099	\\
050318	&	$1.444$	&	$2.30\pm0.23$	&	LX	&		&		&	$32$	&	Swift	&	GCN 3134	\\
050319	&	$3.243$	&	$4.63\pm.0.56$	&	LX	&		&		&	$10$	&	Swift	&	GCN 3119	\\
050401	&	$2.898$	&	$37.6\pm7.3$	&	LX	&		&		&	$33$	&	KW	&	GCN 3179	\\
050408	&	$1.2357$	&	$2.48\pm0.25$	&	LX	&		&		&	$34$	&	HETE	&	GCN 3188	\\
050502B	&	$5.2$	&	$2.66\pm0.22$	&		&		&		&	$17.5$	&	Swift	&	GCN 3339 	\\
050505	&	$4.27$	&	$16.0\pm1.1$	&	LX	&		&		&	$60$	&	Swift	&	GCN 3364 	\\
050525A	&	$0.606$	&	$2.30\pm0.49$	&	LX	&		&		&	$5.2$	&	KW	&	GCN 3479	\\
050603	&	$2.821$	&	$64.1\pm6.4$	&		&		&		&	$6$	&	KW	&	GCN 3518	\\
050714B	&	$2.4383$	&	$4.99\pm0.85$	&		&		&		&	$46.7$	&	Swift	&	GCN 3615	\\
050730	&	$3.969$	&	$11.8\pm0.8$	&	LX	&		&		&	$155$	&	Swift	&	GCN 3715	\\
050802	&	$1.71$	&	$5.66\pm0.47$ 	&	LX	&		&		&	$13$	&	Swift	&	GCN 3737	\\
050803	&	$4.3$	&	$1.16\pm0.12$	&	LX	&	E	&		&	$85$	&	Swift	&	GCN 3757	\\
050814	&	$5.3$	&	$9.9\pm1.1$	&	LX	&		&		&	$65$	&	Swift	&	GCN 3783	\\
050819	&	$2.5043$	&	$3.60\pm0.55$	&		&		&		&	$36$	&	Swift	&	GCN 3828	\\
050820	&	$2.615$	&	$103\pm10$	&	LX	&		&		&	$549.2$	&	KW	&	(9)	\\
050822	&	$1.434$	&	$10.8\pm1.1$	&	LX	&	E	&		&	$102$	&	Swift	&	GCN 3856	\\
050904	&	$6.295$	&	$133\pm14$	&		&		&		&	$225$	&	Swift	&	GCN 3938	\\
050908	&	$3.347$	&	$1.54\pm0.16$	&		&		&		&	$20$	&	Swift	&	GCN 3951	\\
050915	&	$2.5273$	&	$1.8\pm1.3$	&		&		&		&	$53$	&	Swift	&	GCN 3982	\\
050922B	&	$4.9$	&	$46.4\pm4.6$	&	LX	&	E	&		&	$980$	&	Swift	&	GCN 4019 	\\
050922C	&	$2.199$	&	$5.6\pm1.8$	&	LX	&		&		&	$8.4$	&	KW	&	GCN 4030	\\
051001	&	$2.4296$	&	$2.3\pm1.7$	&		&		&		&	$190$	&	Swift	&	GCN 4052	\\
051006	&	$1.059$	&	$1.02\pm0.56$	&		&		&		&	$26$	&	Swift	&	GCN 4063	\\
051008	&	$2.77$	&	$115\pm20$	&		&		&		&	$280$	&	KW	&	GCN 4078	\\
051022	&	$0.8$	&	$56.0\pm5.6$	&		&		&		&	$200$	&	KW	&	GCN 4150	\\
051109A	&	$2.346$	&	$6.85\pm0.73$	&	LX	&		&		&	$130$	&	KW	&	GCN 4238	\\
051111	&	$1.55$	&	$15.4\pm1.9$	&		&		&		&	$47$	&	KW	&	GCN 4260	\\
\hline
060108	&	$2.03$	&	$1.51\pm1.33$	&	LX	&		&		&	$14.4$	&	Swift	&	GCN 4445	\\
060111	&	$2.32$	&	$1.62\pm0.08$	&	LX	&	E	&		&	$13$	&	Swift	&	GCN 4486	\\
060115	&	$3.533$	&	$5.9\pm3.8$	&	LX	&		&		&	$142$	&	Swift	&	GCN 4518	\\
060124	&	$2.296$	&	$43.8\pm6.4$	&	LX	&		&		&	$300$	&	KW	&	GCN 4599	\\
060202	&	$0.785$	&	$1.20\pm0.09$	&	LX	&		&		&	$203.7$	&	Swift	&	GCN 4635	\\
060204B	&	$2.3393$	&	$29.3\pm6.0$	&	LX	&	C	&		&	$134$	&	Swift	&	GCN 4671	\\
060206	&	$4.056$	&	$4.1\pm1.9$	&	LX	&		&		&	$7$	&	Swift	&	GCN 4697	\\
060210	&	$3.91$	&	$32.2\pm3.2$	&	LX	&		&		&	$255$	&	Swift	&	GCN 4748	\\
060223	&	$4.41$	&	$9.73\pm0.72$	&		&		&		&	$11$	&	Swift	&	GCN 4820	\\
060306	&	$3.5$	&	$7.6\pm1.0$	&		&		&		&	$61$	&	Swift	&	GCN 4851	\\
060418	&	$1.489$	&	$13.5\pm2.7$	&	LX	&		&		&	$52$	&	Swift	&	GCN 4975	\\
060502A	&	$1.51$	&	$10.57\pm0.48$	&	LX	&		&		&	$33$	&	Swift	&	GCN 5053	\\
060510B	&	$4.9$	&	$19.1\pm0.8$	&	LX	&		&		&	$276$	&	Swift	&	GCN 5107	\\
060512	&	$2.1$	&	$2.38\pm2.70$	&	LX	&		&		&	$8.6$	&	Swift	&	GCN 5124	\\
060522	&	$5.11$	&	$6.47\pm0.63$	&		&		&		&	$69$	&	Swift	&	GCN 5153	\\
060526	&	$3.22$	&	$2.75\pm0.37$	&	LX	&		&		&	$298$	&	Swift	&	GCN 5174	\\
060602A	&	$0.787$	&	$6.63\pm0.41$	&		&		&		&	$60$	&	Swift	&	GCN 5206	\\
060605	&	$3.773$	&	$4.23\pm0.61$	&	LX	&		&		&	$15$	&	Swift	&	GCN 5231	\\
060607A	&	$3.082$	&	$21.4\pm11.9$	&	LX	&	C	&		&	$100$	&	Swift	&	GCN 5242	\\
060707	&	$3.424$	&	$4.3\pm1.1$	&	LX	&		&		&	$68$	&	Swift	&	GCN 5289	\\
060708	&	$1.92$	&	$1.06\pm0.08$	&	LX	&		&		&	$9.8$	&	Swift	&	GCN 5295	\\
060714	&	$2.7108$	&	$7.67\pm0.44$	&	LX	&		&		&	$115$	&	Swift	&	GCN 5334	\\
060719	&	$1.532$	&	$1.4\pm1.3$	&		&		&		&	$55$	&	Swift	&	GCN 5349	\\
060729	&	$0.54$	&	$1.20\pm0.53$	&	LX	&	E	&		&	$116$	&	Swift	&	GCN 5370	\\
060814	&	$1.923$	&	$56.7\pm5.7$	&	LX	&		&		&	$40$	&	KW	&	GCN 5460	\\
060906	&	$3.6856$	&	$7.81\pm0.51$	&	LX	&		&		&	$43.6$	&	Swift	&	GCN 5538	\\
060908	&	$1.884$	&	$7.2\pm1.9$	&		&		&		&	$19.3$	&	Swift	&	GCN 5551	\\
060923B	&	$1.5094$	&	$2.71\pm0.34$	&		&		&		&	$8.8$	&	Swift	&	GCN 5595 	\\
060926	&	$3.2086$	&	$2.29\pm0.37$	&		&		&		&	$8$	&	Swift	&	GCN 5621	\\
060927	&	$5.46$	&	$12.0\pm2.8$	&		&		&		&	$22.6$	&	Swift	&	GCN 5639	\\
061007	&	$1.262$	&	$90.0\pm9.0$	&	LX	&		&		&	$75$	&	KW	&	GCN 5722	\\
061110B	&	$3.4344$	&	$17.9\pm1.6$	&		&		&		&	$128$	&	Swift	&	GCN 5810 	\\
061121	&	$1.314$	&	$23.5\pm2.7$	&	LX	&		&		&	$81$	&	Swift	&	GCN 5831	\\
061126	&	$1.1588$	&	$31.4\pm3.6$	&	LX	&		&		&	$191$	&	Swift	&	GCN 5860	\\
061202	&	$2.2543$	&	$21.99\pm0.63$	&		&		&		&	$91$	&	Swift	&	GCN 5887	\\
061222A	&	$2.088$	&	$30.0\pm6.4$	&	LX	&		&		&	$72$	&	Swift	&	GCN 5964	\\
061222B	&	$3.355$	&	$8.1\pm1.5$	&		&		&		&	$40$	&	Swift	&	GCN 5974	\\
\hline
070110	&	$2.3521$	&	$4.98\pm0.30$	&	LX	&		&		&	$85$	&	Swift	&	GCN 6007	\\
070125	&	$1.547$	&	$84.1\pm8.4$	&		&		&		&	$75$	&	Swift	&	GCN 6049	\\
070129	&	$2.3384$	&	$16.8\pm1.7$	&	LX	&	E	&		&	$460$	&	Swift	&	GCN 6058 	\\
070223	&	$1.6295$	&	$4.73\pm0.28$	&		&		&		&	$89$	&	Swift	&	GCN 6132	\\
070224	&	$1.9922$	&	$2.37\pm0.28$	&		&		&		&	$34$	&	Swift	&	GCN 6141	\\
070306	&	$1.4959$	&	$8.26\pm0.41$	&	LX	&		&		&	$210$	&	Swift	&	GCN 6173	\\
070318	&	$0.84$	&	$3.41\pm2.14$	&	LX	&	C	&		&	$63$	&	Swift	&	GCN 6212 	\\
070328	&	$2.0627$	&	$56.7\pm7.7$	&		&		&		&	$45$	&	KW	&	GCN 6230	\\
070411	&	$2.954$	&	$8.31\pm0.45$	&		&		&		&	$101$	&	Swift	&	GCN 6274 	\\
070419B	&	$1.959$	&	$12.1\pm1.7$	&		&		&		&	$236.5$	&	Swift	&	GCN 6327	\\
070508	&	$0.82$	&	$7.74\pm0.29$	&	LX	&		&		&	$40$	&	KW	&	GCN 6403	\\
070521	&	$1.35$	&	$10.8\pm1.8$	&		&		&		&	$55$	&	KW	&	GCN 6459	\\
070529	&	$2.4996$	&	$12.8\pm1.1$	&	LX	&		&		&	$109$	&	Swift	&	GCN 6468	\\
070611	&	$2.0394$	&	$0.92\pm0.13$	&		&		&		&	$12$	&	Swift	&	GCN 6502	\\
070612A	&	$0.617$	&	$1.96\pm0.40$	&		&		&		&	$370$	&	Swift	&	GCN 6522	\\
070721B	&	$3.6298$	&	$24.2\pm1.4$	&		&		&		&	$340$	&	Swift	&	GCN 6649	\\
070802A	&	$2.45$	&	$1.65\pm2.78$	&	LX	&		&		&	$16.4$	&	Swift	&	GCN 6699 	\\
070810A	&	$2.17$	&	$91.5\pm1.1$	&		&		&		&	$11$	&	Swift	&	GCN 6748	\\
071003	&	$1.604$	&	$38.3\pm4.5$	&	LX	&		&		&	$30$	&	KW	&	GCN 6849	\\
071010B	&	$0.947$	&	$2.32\pm0.40$	&		&		&		&	$16.6$	&	KW	&	GCN 6879	\\
071020	&	$2.145$	&	$10.0\pm4.6$	&		&		&		&	$8.45$	&	KW	&	GCN 6960	\\
071021	&	$2.452$	&	$8.18\pm0.82$	&	LX	&	E	&		&	$225$	&	Swift	&	GCN 6966	\\
071025	&	$5.2$	&	$115\pm4$	&		&		&		&	$109$	&	Swift	&	GCN 6996	\\
071031	&	$2.6918$	&	$4.99\pm0.97$	&		&		&		&	$180$	&	Swift	&	GCN 7029	\\
071112C	&	$0.823$	&	$15.7\pm2.1$	&		&		&		&	$15$	&	Swift	&	GCN 7081	\\
071117	&	$1.331$	&	$5.86\pm2.7$	&		&		&		&	$5$	&	KW	&	GCN 7114	\\
\hline
080129	&	$4.349$	&	$7.7\pm3.5$	&		&		&		&	$48$	&	Swift	&	GCN 7235	\\
080205	&	$2.72$	&	$15.21\pm0.72$	&		&		&		&	$106.5$	&	Swift	&	GCN 7257 	\\
080207	&	$2.0858$	&	$16.4\pm1.8$	&		&		&		&	$340$	&	Swift	&	GCN 7272	\\
080210	&	$2.6419$	&	$4.77\pm0.29$	&	LX	&		&		&	$45$	&	Swift	&	GCN 7289 	\\
080310	&	$2.4274$	&	$20.9\pm2.1$	&	LX	&	E	&		&	$365$	&	Swift	&	GCN 7402 	\\
080319A	&	$2.0265$	&	$27.0\pm2.2$	&		&		&		&	$64$	&	Swift	&	GCN 7447	\\
080319B	&	$0.937$	&	$118\pm12$	&	LX	&		&		&	$50$	&	KW	&	GCN 7482	\\
080319C	&	$1.95$	&	$14.9\pm3.0$	&	LX	&		&		&	$15$	&	KW	&	GCN 7487	\\
080325	&	$1.78$	&	$9.55\pm0.84$	&		&		&		&	$128.4$	&	Swift	&	GCN 7531	\\
080411	&	$1.03$	&	$16.2\pm1.6$	&		&		&		&	$70$	&	KW	&	GCN 7589	\\
080413A	&	$2.433$	&	$8.6\pm2.1$	&		&		&		&	$46$	&	Swift	&	GCN 7604	\\
080413B	&	$1.1$	&	$1.61\pm0.27$	&		&		&		&	$8$	&	Swift	&	GCN 7606	\\
080514B	&	$1.8$	&	$18.1\pm3.6$	&		&		&		&	$7$	&	KW	&	GCN 7751	\\
080515	&	$2.47$	&	$5.11\pm0.77$	&		&		&		&	$21$	&	Swift	&	GCN 7726	\\
080602	&	$1.8204$	&	$6.08\pm0.38$	&		&		&		&	$74$	&	Swift	&	GCN 7786	\\
080603B	&	$2.69$	&	$6.0\pm3.1$	&		&		&		&	$70$	&	KW	&	GCN 7812 	\\
080604	&	$1.4171$	&	$1.05\pm0.12$	&		&		&		&	$82$	&	Swift	&	GCN 7817	\\
080605	&	$1.64$	&	$28\pm14$	&	LX	&		&		&	$20$	&	Swift	&	GCN 7854	\\
080607	&	$3.04$	&	$187\pm11$	&	LX	&	C	&		&	$85$	&	KW	&	GCN 7862	\\
080710	&	$0.8454$	&	$1.68\pm0.22$	&		&		&		&	$120$	&	Swift	&	GCN 7969	\\
080721	&	$2.591$	&	$134\pm23$	&	LX	&		&		&	$30$	&	KW	&	GCN 7995	\\
080804	&	$2.205$	&	$12.0\pm1.2$	&	LX	&		&		&	$34$	&	Swift	&	GCN 8067	\\
080805	&	$1.51$	&	$7.16\pm1.90$	&	LX	&	C	&		&	$78$	&	Swift	&	GCN 8068	\\
080810	&	$3.35$	&	$50.0\pm4.4$	&	LX	&	C	&		&	$79.4$	&	KW	&	GCN 8101	\\
080825B	&	$4.3$	&	$38.4\pm3.8$	&		&		&		&	$110$	&	KW	&	GCN 8142 	\\
080905B	&	$2.3739$	&	$4.55\pm0.37$	&	LX	&		&		&	$128$	&	Swift	&	GCN 8188	\\
080906	&	$2.1$	&	$21.2\pm1.2$	&		&		&		&	$147$	&	Swift	&	GCN 8196	\\
080913	&	$6.695$	&	$9.2\pm2.7$	&		&		&		&	$8.8$	&	KW	&	GCN 8280	\\
080916A	&	$0.689$	&	$0.98\pm0.10$	&		&		&		&	$40$	&	KW	&	GCN 8259	\\
080916C	&	$4.35$	&	$407\pm86$	&	LX	&		&		&	$60$	&	Fermi	&	GCN 8263	\\
080928	&	$1.692$	&	$3.99\pm0.91$	&	LX	&		&		&	$66$	&	Fermi	&	GCN 8278	\\
081008	&	$1.967$	&	$13.5\pm6.6$	&	LX	&	C	&		&	$185.5$	&	Swift	&	GCN 8351	\\
081028	&	$3.038$	&	$18.3\pm1.8$	&	LX	&		&		&	$260$	&	Swift	&	GCN 8428	\\
081029	&	$3.8479$	&	$12.1\pm1.4$	&		&		&		&	$270$	&	Swift	&	GCN 8447 	\\
081109	&	$0.9787$	&	$1.81\pm0.12$	&	LX	&		&		&	$45$	&	Fermi	&	GCN 8505	\\
081118	&	$2.58$	&	$12.2\pm1.2$	&		&		&		&	$20$	&	Fermi	&	GCN 8550	\\
081121	&	$2.512$	&	$32.4\pm3.7$	&	LX	&		&		&	$18$	&	KW	&	GCN 8548	\\
081203A	&	$2.05$	&	$32\pm12$	&	LX	&		&		&	$213$	&	KW	&	GCN 8611	\\
081210	&	$2.0631$	&	$15.6\pm5.4$	&	LX	&	C	&		&	$146$	&	Swift	&	GCN 8649 	\\
081221	&	$2.26$	&	$31.9\pm3.2$	&	LX	&		&		&	$40$	&	Fermi	&	GCN 8704	\\
081222	&	$2.77$	&	$27.4\pm2.7$	&	LX	&		&		&	$30$	&	Fermi	&	GCN 8715	\\
081228	&	$3.44$	&	$9.9\pm2.0$	&		&		&		&	$3$	&	Swift	&	GCN 8749 	\\
081230	&	$2.0$	&	$3.21\pm0.31$	&		&		&		&	$60.7$	&	Swift	&	GCN 8759	\\
\hline
090102A	&	$1.547$	&	$22.6\pm2.7$	&	LX	&		&		&	$30$	&	KW	&	GCN 8776	\\
090113A	&	$1.7493$	&	$1.00\pm0.17$	&		&		&		&	$9.1$	&	Swift	&	GCN 8808	\\
090201A	&	$2.1$	&	$93.4\pm8.1$	&		&		&		&	$110$	&	KW	&	GCN 8878	\\
090205A	&	$4.6497$	&	$1.12\pm0.16$	&		&		&		&	$8.8$	&	Swift	&	GCN 8886	\\
090313A	&	$3.375$	&	$4.42\pm0.79$	&	LX	&		&		&	$78$	&	Swift	&	GCN 8986	\\
090323A	&	$3.57$	&	$438\pm53$	&		&		&		&	$150$	&	Fermi	&	GCN 9035	\\
090328A	&	$0.736$	&	$14.2\pm1.4$	&	LX	&		&		&	$80$	&	Fermi	&	GCN 9057	\\
090404A	&	$3$	&	$59.2\pm6.1$	&	LX	&	E	&		&	$84$	&	Swift	&	GCN 9089	\\
090418A	&	$1.608$	&	$17.2\pm2.7$	&	LX	&		&		&	$64.8$	&	KW+Swift	&	GCN 9196	\\
090423A	&	$8.26$	&	$8.8\pm2.1$	&	LX	&		&		&	$12$	&	Fermi	&	GCN 9229	\\
090424A	&	$0.544$	&	$4.07\pm0.41$	&	LX	&		&		&	$52$	&	Fermi	&	GCN 9230	\\
090429B	&	$9.3$	&	$6.7\pm1.3$	&		&		&		&	$5.5$	&	Swift	&	GCN 9290	\\
090516A	&	$4.109$	&	$99.6\pm16.7$	&	LX	&	C	&		&	$350$	&	Fermi	&	GCN 9415	\\
090519A	&	$3.85$	&	$24.7\pm2.8$	&		&		&		&	$64$	&	Swift	&	GCN 9406	\\
090529A	&	$2.625$	&	$2.56\pm0.30$	&		&		&		&	$100$	&	Swift	&	GCN 9434	\\
090530A	&	$1.266$	&	$1.73\pm0.19$	&		&		&		&	$48$	&	Swift	&	GCN 9443	\\
090618A	&	$0.54$	&	$28.6\pm2.9$	&	LX	&		&		&	$113.2$	&	Fermi	&	GCN 9535	\\
090715B	&	$3.$	&	$63.9\pm3.7$	&	LX	&	E	&		&	$100$	&	KW	&	GCN 9679	\\
090726A	&	$2.71$	&	$1.82\pm0.40$	&		&		&		&	$67$	&	Swift	&	GCN 9716	\\
090809A	&	$2.737$	&	$1.88\pm0.26$	&	LX	&		&		&	$5.4$	&	Swift	&	GCN 9756	\\
090812A	&	$2.452$	&	$44.0\pm6.5$	&	LX	&	C	&		&	$64.8$	&	KW+Swift	&	GCN 9821	\\
090902B	&	$1.822$	&	$292\pm29.2$	&	LX	&		&		&	$21$	&	Fermi	&	GCN 9866	\\
090926A	&	$2.106$	&	$228\pm23$	&	LX	&		&		&	$20$	&	Fermi	&	GCN 9933	\\
090926B	&	$1.24$	&	$4.14\pm0.45$	&		&		&		&	$81$	&	Fermi	&	GCN 9957	\\
091003A	&	$0.897$	&	$10.7\pm1.8$	&	LX	&		&		&	$21.1$	&	Fermi	&	GCN 9983	\\
091020A	&	$1.71$	&	$8.4\pm1.1$	&	LX	&		&		&	$37$	&	Fermi	&	GCN 10095	\\
091024A	&	$1.092$	&	$18.4\pm2.0$	&		&		&	UL	&	$1250$	&	KW	&	GCN 10083	\\
091029A	&	$2.752$	&	$7.97\pm0.82$	&	LX	&		&		&	$39.2$	&	Swift	&	GCN 10103	\\
091109A	&	$3.076$	&	$10.6\pm1.4$	&		&		&		&	$48$	&	Swift	&	GCN 10141	\\
091127A	&	$0.49$	&	$1.64\pm0.18$	&	LX	&		&		&	$9$	&	Fermi	&	GCN 10204	\\
091208B	&	$1.063$	&	$2.06\pm0.21$	&	LX	&		&		&	$15$	&	Fermi	&	GCN 10266	\\
\hline
100219A	&	$4.6667$	&	$3.93\pm0.61$	&		&		&		&	$18.8$	&	Swift	&	GCN 10434	\\
100302A	&	$4.813$	&	$1.33\pm0.17$	&	LX	&		&		&	$17.9$	&	Swift	&	GCN 10462	\\
100414A	&	$1.368$	&	$55.0\pm5.5$	&		&		&		&	$26.4$	&	Fermi	&	GCN 10595	\\
100424A	&	$2.465$	&	$3.05\pm0.53$	&		&		&		&	$104$	&	Swift	&	GCN 10670	\\
100425A	&	$1.755$	&	$2.76\pm3.45$	&	LX	&		&		&	$37$	&	Swift	&	GCN 10685	\\
100513A	&	$4.8$	&	$6.75\pm0.53$	&	LX	&		&		&	$84$	&	Swift	&	GCN 10753	\\
100615A	&	$1.398$	&	$5.81\pm0.11$	&		&		&		&	$37.7$	&	Fermi	&	GCN 10851	\\
100621A	&	$0.542$	&	$2.82\pm0.35$	&	LX	&		&		&	$80$	&	KW	&	GCN 10882	\\
100728A	&	$1.567$	&	$86.8\pm8.7$	&		&		&		&	$162.9$	&	Fermi	&	GCN 11006	\\
100728B	&	$2.106$	&	$3.55\pm0.36$	&		&		&		&	$11.8$	&	Fermi	&	GCN 11015	\\
100814A	&	$1.44$	&	$15.3\pm1.8$	&	LX	&		&		&	$149$	&	Fermi	&	GCN 11099	\\
100816A	&	$0.8049$	&	$0.75\pm0.10$	&	LX	&		&		&	$2$	&	Fermi	&	GCN 11124	\\
100901A	&	$1.408$	&	$4.22\pm0.50$	&	LX	&		&		&	$439$	&	Swift	&	GCN 11169	\\
100906A	&	$1.727$	&	$29.9\pm2.9$	&	LX	&		&		&	$105$	&	Fermi	&	GCN 11248	\\
\hline
101213A	&	$0.414$	&	$2.72\pm0.53$	&		&		&		&	$45$	&	Fermi	&	GCN 11454	\\
110128A	&	$2.339$	&	$1.58\pm0.21$	&	LX	&		&		&	$12$	&	Fermi	&	GCN 11628	\\
110205A	&	$2.22$	&	$48.3\pm6.4$	&	LX	&		&		&	$330$	&	KW	&	GCN 11659	\\
110213A	&	$1.46$	&	$5.78\pm0.81$	&	LX	&		&		&	$33$	&	Fermi	&	GCN 11727	\\
110213B	&	$1.083$	&	$8.3\pm1.3$	&		&		&		&	$50$	&	KW	&	GCN 11722	\\
110422A	&	$1.77$	&	$79.8\pm8.2$	&	LX	&		&		&	$40$	&	KW	&	GCN 11971	\\
110503A	&	$1.613$	&	$20.8\pm2.1$	&	LX	&		&		&	$12$	&	KW	&	GCN 12008	\\
110715A	&	$0.82$	&	$4.36\pm0.45$	&	LX	&		&		&	$20$	&	KW	&	GCN 12166	\\
110731A	&	$2.83$	&	$49.5\pm4.9$	&	LX	&		&		&	$7.3$	&	Fermi	&	GCN 12221	\\
110801A	&	$1.858$	&	$10.9\pm2.7$	&		&		&		&	$415.1$	&	KW+Swift	&	GCN 12276	\\
110808A	&	$1.348$	&	$6.09\pm4.83$	&	LX	&		&		&	$48$	&	KW	&	GCN 12270	\\
110818A	&	$3.36$	&	$26.6\pm2.8$	&		&		&		&	$75$	&	Fermi	&	GCN 12287	\\
110918A	&	$0.982$	&	$185\pm5$	&	LX	&		&		&	$22$	&	KW	&	GCN 12362	\\
111008A	&	$4.9898$	&	$24.7\pm1.2$	&	LX	&		&		&	$40$	&	KW	&	GCN 12433	\\
111107A	&	$2.893$	&	$3.76\pm0.55$	&		&		&		&	$12$	&	Fermi	&	GCN 12545	\\
111123A	&	$3.1516$	&	$24\pm14$	&	LX	&		&		&	$290$	&	Swift	&	GCN 12598	\\
111209A	&	$0.677$	&	$5.14\pm0.62$	&	LX	&		&	UL	&	$11900$	&	KW	&	GCN 12663	\\
111215A	&	$2.06$	&	$22.1\pm2.5$	&		&		&		&	$796$	&	Swift	&	GCN 12689	\\
111228A	&	$0.716$	&	$2.75\pm0.28$	&	LX	&		&		&	$101.2$	&	Fermi	&	GCN 12744	\\
\hline
120118B	&	$2.943$	&	$6.24\pm0.55$	&		&		&		&	$23.26$	&	Swift	&	GCN 12873	\\
120119A	&	$1.728$	&	$27.2\pm3.6$	&	LX	&		&		&	$55$	&	Fermi	&	GCN 12874	\\
120211A	&	$2.4$	&	$7.1\pm1.0$	&		&		&		&	$61.7$	&	Swift	&	GCN 12924	\\
120326A	&	$1.798$	&	$3.27\pm0.33$	&	LX	&		&		&	$12$	&	Fermi	&	GCN 13145	\\
120327A	&	$2.813$	&	$14.42\pm0.46$	&	LX	&		&		&	$62.9$	&	Swift	&	GCN 13137	\\
120404A	&	$2.876$	&	$4.18\pm0.34$	&		&		&		&	$38.7$	&	Swift	&	GCN 13220	\\
120521C	&	$6.01$	&	$11.9\pm1.9$	&		&		&		&	$26.7$	&	Swift	&	GCN 13333	\\
120624B	&	$2.197$	&	$319\pm32$	&		&		&		&	$271$	&	Fermi	&	GCN 13377	\\
120711A	&	$1.405$	&	$180\pm18$	&	LX	&		&		&	$44$	&	Fermi	&	GCN 13437	\\
120712A	&	$4.175$	&	$21.2\pm2.1$	&	LX	&		&		&	$23$	&	Fermi	&	GCN 13469	\\
120716A	&	$2.486$	&	$30.2\pm3.0$	&		&		&		&	$234$	&	Fermi	&	GCN 13498	\\
120802A	&	$3.796$	&	$12.9\pm2.8$	&		&		&		&	$50$	&	Swift	&	GCN 13559	\\
120805A	&	$3.1$	&	$19.0\pm3.2$	&		&		&		&	$48$	&	Swift	&	GCN 13594	\\
120811C	&	$2.671$	&	$6.41\pm0.64$	&		&		&		&	$26.8$	&	Swift	&	GCN 13634	\\
120815A	&	$2.358$	&	$1.65\pm0.27$	&		&		&		&	$9.7$	&	Swift	&	GCN 13652	\\
120909A	&	$3.93$	&	$87\pm10$	&	LX	&		&		&	$112$	&	Fermi	&	GCN 13737	\\
120922A	&	$3.1$	&	$22.4\pm1.4$	&	LX	&		&		&	$180$	&	Fermi	&	GCN 13789	\\
121024A	&	$2.298$	&	$4.61\pm0.55$	&	LX	&		&		&	$69$	&	Swift	&	GCN 13899	\\
121027A	&	$1.773$	&	$1.50\pm0.17$	&	LX	&	E	&	UL	&	$6000$	&	Swift	&	GCN 13910	\\
121128A	&	$2.2$	&	$8.66\pm0.87$	&	LX	&		&		&	$17$	&	Fermi	&	GCN 14012	\\
121201A	&	$3.385$	&	$2.52\pm0.34$	&		&		&		&	$85$	&	Swift	&	GCN 14028	\\
121209A	&	$2.1$	&	$24.31\pm0.84$	&		&		&		&	$42.7$	&	Swift	&	GCN 14052	\\
121217A	&	$3.1$	&	$25.9\pm19.7$	&	LX	&		&		&	$780$	&	Fermi	&	GCN 14094	\\
121229A	&	$2.707$	&	$3.7\pm1.1$	&		&		&		&	$100$	&	Swift	&	GCN 14123	\\
\hline
130408A	&	$3.758$	&	$35.0\pm6.4$	&		&		&		&	$15$	&	KW	&	GCN 14368	\\
130131B	&	$2.539$	&	$7.15\pm0.84$	&		&		&		&	$4.3$	&	Swift	&	GCN 14164	\\
130215A	&	$0.597$	&	$4.45\pm0.11$	&		&		&		&	$140$	&	Fermi	&	GCN 14219	\\
130408A	&	$3.757$	&	$35.4\pm5.9$	&		&		&		&	$15$	&	KW	&	GCN 14368	\\
130418A	&	$1.218$	&	$9.9\pm1.6$	&	LX	&		&		&	$120$	&	KW	&	GCN 14417	\\
130420A	&	$1.297$	&	$7.74\pm0.77$	&	LX	&		&		&	$102$	&	Fermi	&	GCN 14429	\\
130427A	&	$0.334$	&	$92\pm13$	&	LX	&		&		&	$162.8$	&	Fermi	&	GCN 14473	\\
130427B	&	$2.78$	&	$13.3\pm0.5$	&	LX	&	E	&		&	$27$	&	Swift	&	GCN 14469	\\
130505A	&	$2.27$	&	$347\pm35$	&	LX	&		&		&	$21$	&	KW	&	GCN 14575	\\
130514A	&	$3.6$	&	$49.5\pm9.2$	&	LX	&	E	&		&	$204$	&	Swift	&	GCN 14636	\\
130518A	&	$2.488$	&	$193\pm19$	&		&		&		&	$48$	&	Fermi	&	GCN 14674	\\
130528A	&	$1.25$	&	$18.0\pm2.3$	&	LX	&	E	&		&	$55$	&	Fermi	&	GCN 14729	\\
130606A	&	$5.91$	&	$28.3\pm5.1$	&	LX	&	E	&		&	$165$	&	KW	&	GCN 14808	\\
130610A	&	$2.092$	&	$6.99\pm0.46$	&	LX	&		&		&	$28$	&	Fermi	&	GCN 14858	\\
130701A	&	$1.155$	&	$2.60\pm0.09$	&	LX	&		&		&	$5.5$	&	KW	&	GCN 14958	\\
130907A	&	$1.238$	&	$304\pm19$	&	LX	&		&		&	$214$	&	KW	&	GCN 15203	\\
130925A	&	$0.347$	&	$3.23\pm0.37$	&	LX	&	E	&	UL	&	$4500$	&	Fermi	&	GCN 15261	\\
131011A	&	$1.874$	&	$86.67\pm0.39$	&		&		&		&	$77$	&	Fermi	&	GCN 15331	\\
131030A	&	$1.293$	&	$30.0\pm2.0$	&	LX	&	C	&		&	$28$	&	KW	&	GCN 15413	\\
131105A	&	$1.686$	&	$34.7\pm1.2$	&	LX	&		&		&	$112$	&	Fermi	&	GCN 15455	\\
131108A	&	$2.4$	&	$70.87\pm0.97$	&	LX	&		&		&	$19$	&	Fermi	&	GCN 15477	\\
131117A	&	$4.042$	&	$1.02\pm0.16$	&	LX	&		&		&	$11$	&	Swift	&	GCN 15499	\\
131227A	&	$5.3$	&	$24.2\pm1.7$	&		&		&		&	$18$	&	Swift	&	GCN 15620	\\
\hline
140114A	&	$3.0$	&	$27.6\pm0.8$	&	LX	&	E	&		&	$139.7$	&	Swift	&	GCN 15738	\\
140206A	&	$2.73$	&	$35.8\pm7.9$	&	LX	&	C	&		&	$27$	&	Fermi	&	GCN 15796	\\
140213A	&	$1.2076$	&	$9.93\pm0.15$	&	LX	&		&		&	$18.6$	&	Fermi	&	GCN 15833	\\
140226A	&	$1.98$	&	$5.8\pm1.1$	&	LX	&		&		&	$15$	&	KW	&	GCN 15889	\\
140301A	&	$1.416$	&	$0.95\pm0.18$	&	LX	&	C	&		&	$31$	&	Swift	&	GCN 15906	\\
140304A	&	$5.283$	&	$15.3\pm1.1$	&	LX	&	E	&		&	$32$	&	Fermi	&	GCN 15923	\\
140311A	&	$4.954$	&	$11.6\pm1.5$	&	LX	&		&		&	$71.4$	&	Swift	&	GCN 15962	\\
140419A	&	$3.956$	&	$185\pm77$	&	LX	&	C	&		&	$80$	&	KW	&	GCN 16134	\\
140423A	&	$3.26$	&	$65.3\pm3.3$	&	LX	&		&		&	$95$	&	Fermi	&	GCN 16152	\\
140428A	&	$4.7$	&	$1.88\pm0.31$	&		&		&		&	$17.42$	&	Swift	&	GCN 16186	\\
140430A	&	$1.6$	&	$1.54\pm0.23$	&		&		&		&	$173.6$	&	Swift	&	GCN 16200	\\
140506A	&	$0.889$	&	$7.75\pm0.80$	&	LX	&	E	&		&	$64$	&	Fermi	&	GCN 16220	\\
140508A	&	$1.027$	&	$23.24\pm0.26$	&	LX	&		&		&	$44.3$	&	Fermi	&	GCN 16224	\\
140509A	&	$2.4$	&	$3.77\pm0.44$	&	LX	&		&		&	$23.2$	&	Swift	&	GCN 16240	\\
140512A	&	$0.725$	&	$7.76\pm0.18$	&	LX	&		&		&	$148$	&	Fermi	&	GCN 16262	\\
140515A	&	$6.32$	&	$5.41\pm0.55$	&		&		&		&	$23.4$	&	Swift	&	GCN 16284	\\
140518A	&	$4.707$	&	$5.89\pm0.59$	&		&		&		&	$60.5$	&	Swift	&	GCN 16306	\\
140614A	&	$4.233$	&	$7.3\pm2.1$	&	LX	&		&		&	$720$	&	Swift	&	GCN 16402	\\
140620A	&	$2.04$	&	$6.28\pm0.24$	&	LX	&		&		&	$46$	&	Fermi	&	GCN 16426	\\
140623A	&	$1.92$	&	$7.69\pm0.68$	&		&		&		&	$110$	&	Fermi	&	GCN 16450	\\
140629A	&	$2.275$	&	$6.15\pm0.90$	&	LX	&		&		&	$26$	&	KW	&	GCN 16495	\\
140703A	&	$3.14$	&	$1.72\pm0.09$	&	LX	&		&		&	$84$	&	Fermi	&	GCN 16512	\\
140801A	&	$1.32$	&	$5.69\pm0.05$	&		&		&		&	$7$	&	Fermi	&	GCN 16658	\\
140808A	&	$3.29$	&	$11.93\pm0.75$	&		&		&		&	$4.7$	&	Fermi	&	GCN 16669	\\
140907A	&	$1.21$	&	$2.29\pm0.08$	&	LX	&		&		&	$35$	&	Fermi	&	GCN 16798	\\
141026A	&	$3.35$	&	$7.17\pm0.90$	&	LX	&		&		&	$146$	&	Swift	&	GCN 16960	\\
141028A	&	$2.33$	&	$68.9\pm0.02$	&		&		&		&	$31.5$	&	Fermi	&	GCN 16971	\\
141109A	&	$2.993$	&	$33.1\pm6.9$	&	LX	&		&		&	$94$	&	KW	&	GCN 17055	\\
141121A	&	$1.47$	&	$14.2\pm1.1$	&	LX	&		&	UL	&	$1200$	&	KW	&	GCN 17108	\\
141220A	&	$1.3195$	&	$2.44\pm0.07$	&		&		&		&	$7.6$	&	Fermi	&	GCN 17205	\\
141221A	&	$1.47$	&	$6.99\pm1.98$	&	LX	&	C	&		&	$23.8$	&	Fermi	&	GCN 17216	\\
141225A	&	$0.915$	&	$2.29\pm0.11$	&		&		&		&	$56$	&	Fermi	&	GCN 17241	\\
\hline
150120B	&	$3.5$	&	$7.37\pm1.09$	&	LX	&		&		&	$24.3$	&	Swift	&	GCN 17330	\\
150206A	&	$2.087$	&	$55.6\pm20.1$	&	LX	&		&		&	$60$	&	KW	&	GCN 17427	\\
150301B	&	$1.5169$	&	$2.87\pm0.42$	&	LX	&		&		&	$13$	&	Fermi	&	GCN 17525	\\
150314A	&	$1.758$	&	$95.2\pm3.1$	&	LX	&		&		&	$10.7$	&	Fermi	&	GCN 17579	\\
150323A	&	$0.593$	&	$1.30\pm0.30$	&		&		&		&	$38$	&	KW	&	GCN 17640	\\
150403A	&	$2.06$	&	$98.1\pm6.3$	&	LX	&		&		&	$22.3$	&	Fermi	&	GCN 17674	\\
150413A	&	$3.139$	&	$49.80\pm7.01$	&		&		&		&	$263.6$	&	KW+Swift	&	GCN 17731	\\
150821A	&	$0.755$	&	$14.7\pm1.1$	&	LX	&		&		&	$103$	&	Fermi	&	GCN 18190	\\
150910A	&	$1.359$	&	$21.6\pm1.8$	&	LX	&		&		&	$112.2$	&	Swift	&	GCN 18268	\\
151021A	&	$2.33$	&	$112.2\pm35$	&	LX	&		&		&	$100$	&	KW	&	GCN 18433	\\
151027A	&	$0.81$	&	$3.94\pm1.33$	&	LX	&	C	&		&	$124$	&	Fermi	&	GCN 18492	\\
151027B	&	$4.063$	&	$18.6\pm3.7$	&	LX	&		&		&	$80$	&	Swift	&	GCN 18514	\\
151111A	&	$3.5$	&	$3.43\pm1.19$	&	LX	&	E	&		&	$40$	&	Fermi	&	GCN 18582	\\
151112A	&	$4.1$	&	$12.1\pm1.5$	&	LX	&		&		&	$19.32$	&	Swift	&	GCN 18593	\\
151215A	&	$2.59$	&	$1.89\pm0.43$	&	LX	&		&		&	$17.8$	&	Swift	&	GCN 18699	\\
\hline
160121A	&	$1.96$	&	$2.54\pm0.21$	&	LX	&		&		&	$12$	&	Swift	&	GCN 18919	\\
160131A	&	$0.972$	&	$58.7\pm32.7$	&	LX	&		&		&	$200$	&	KW	&	GCN 18974	\\
160203A	&	$3.52$	&	$12.0\pm1.0$	&	LX	&		&		&	$20.2$	&	Swift	&	GCN 18998	\\
160227A	&	$2.38$	&	$5.52\pm2.38$	&	LX	&		&		&	$316.5$	&	Swift	&	GCN 19106	\\
160228A	&	$1.64$	&	$15.98\pm0.80$	&		&		&		&	$98.36$	&	Swift	&	GCN 19113	\\
160509A	&	$1.17$	&	$84.5\pm2.3$	&	LX	&		&		&	$371$	&	Fermi	&	GCN 19411	\\
160623A	&	$0.367$	&	$22.4\pm1.5$	&	LX	&		&		&	$38.9$	&	KW	&	GCN 19554	\\
160625B	&	$1.406$	&	$419.0\pm4.8$	&	LX	&		&		&	$460$	&	Fermi	&	GCN 19587	\\
160629A	&	$3.332$	&	$48.8\pm9.9$	&		&		&		&	$66.6$	&	Fermi	&	GCN 19628	\\
160804A	&	$0.736$	&	$2.46\pm0.51$	&	LX	&		&		&	$130$	&	Fermi	&	GCN 19769	\\
161014A	&	$2.823$	&	$10.1\pm1.7$	&		&		&		&	$37$	&	Fermi	&	GCN 20051	\\
161017A	&	$2.013$	&	$7.56\pm1.55$	&	LX	&		&		&	$32$	&	Fermi	&	GCN 20068	\\
161023A	&	$2.708$	&	$73.9\pm27.5$	&	LX	&		&		&	$50$	&	KW	&	GCN 20111	\\
161108A	&	$1.159$	&	$1.66\pm0.15$	&	LX	&		&		&	$105.1$	&	Swift	&	GCN 20151	\\
161117A	&	$1.549$	&	$31.2\pm5.5$	&	LX	&		&		&	$122$	&	Fermi	&	GCN 20192\\
\hline
170113A & $1.968$   & $19.91$           &   LX    &  C    &      & $49.2 $     &   Swift      &  GCN20452, GCN20458\\
170202A & $3.645$   & $17.0$           &    LX   &      &      & $30.0 $     &   Swift      &  GCN20584, GCN20604\\
170214A & $2.53$   & $392\pm 3$           &       &      &      & $122.9 $     &   Fermi     &  GCN20675, GCN20686\\
170405A & $3.51$   & $241\pm 52$           &  LX      &  C    &      & $78.6 $     &   Swift      &  GCN20986, GCN20990\\
170519A &   $0.818$  &                    & LX     &   C  &      &  $216.4$   & Swift &  GCN21112, GCN21119\\          
170531B &  $2.366$ &                    &  LX   &   C &      &  $164.1$ & Swift  &  GCN21177, GCN21209\\
170604A &  $1.329$ &  $4.7\pm 0.5$    &     LX   &   C  &      &   30 & Swift   &  GCN21197, GCN21247\\
170607A &  0.557 &  $1.113 $   &       LX   &   C   &      &  29.9   & Swift &  GCN21218, GCN21240\\                    170705A & 2.01  &   14.95     &      LX   &   C     &      & 22.8  &  Swift &  GCN21297, GCN21298\\
170903A & 0.886         &                    &      LX    &     &      & 25.6   &Swift &  GCN21799, GCN21812\\          171010A & 0.3285     & 14.49          &       LX   &     &      &   107.3 &  Fermi &  GCN22002, GCN22096\\
171222A  & 2.409  &  20.73      &         LX   &   C     &      &  80.4  & Swift &  GCN22272, GCN22277\\ 
\hline
180115A &2.487          &                    &      LX    &     &      &   40.9  & Swift&  GCN22346, GCN22348\\
180205A & 1.409  &                    &       LX   &     &      & 15.4  & Swift &  GCN22384, GCN22386\\
180314A & 1.445  &     6.58     &  LX  &     &      &   22.0  & Swift  &  GCN22484, GCN22485\\                          180325A &2.248   &    23.0    &    LX   &     &      & 10.0  & Swift &  GCN22535, GCN22555\\
180329B & 1.998   &                    &   LX &    C &      &  210.0  &  Swift &  GCN22566, GCN22567\\
180404A &  1.000        &                    &     LX     &     &      &  35.2        &  Swift &  GCN22591, GCN22599\\  180510B & 1.305         &                    &   LX   &     &      &      134.3    & Swift     &  GCN22702, GCN22705\\
180620B & 1.1175         &       16.3     &    LX &  C   &      & 46.7  & Swift        &  GCN22813, GCN22823\\
180624A & 2.855         &                    &      LX    &   C  &      & 486.4 &  Swift       &  GCN22845, GCN22848\\  180703A  &  0.6678   &       3.15      &          &     &      &  20.7   & Fermi  &  GCN22889, GCN22896\\
180720B &  0.654 &     $68.2\pm2.2$     &  LX    &   C  &      &  48.9  & Swift  &  GCN22981, GCN22996\\
180728A & 0.117   &     $3.15\pm0.7$     & LX    &     &      &  6.4  &   Swift &  GCN23055, GCN23067\\                 180914B & 1.096   &  360  &     &      &  & 280.0    &  AGILE    &  GCN23246, GCN23240\\        
181010A & 1.39   &                    &   LX   &     &      &  9.7 &  Swift  &  GCN23315, GCN23320\\
181020A & 2.938     &                    &  LX    &   C  &      &  15.1        & Swift &  GCN23452, GCN23456\\                    181110A & 1.505    &        11.0            &   LX   & C    &      &      140.0    &  Swift       &  GCN23421, GCN23424\\
	\hline
\end{longtable}

\end{document}